\documentclass[aps,pra,amsmath,amssymb,floatfix,twocolumn,amsmath,superscriptaddress,nofootinbib,tighten,letterpaper,eqsecnum]{revtex4}
\usepackage{multirow}
\usepackage{bbold}
\usepackage{subfigure}
\usepackage{color}
\usepackage{xcolor}
\usepackage{mathrsfs}
\usepackage{hyperref}
\usepackage{bm}
\usepackage{stackrel}

\usepackage{amsfonts, relsize, color}
\usepackage{graphics}
\usepackage{graphicx}
\usepackage{amssymb}
\usepackage{amsbsy}
\usepackage{amsmath}
\usepackage{setspace}
\usepackage{array}
\usepackage{fontenc}
\usepackage{textcomp}
\usepackage{rotating}
\usepackage{comment}

\hypersetup{pdfpagemode=UseNone}
\newcommand{\bsub}{\begin{subequations}}
\newcommand{\esub}{\end{subequations}}
\newcommand{\T}{\mathsf{T}}

\DeclareMathOperator{\sech}{sech}
\DeclareMathOperator{\sgn}{sgn}
\DeclareMathOperator{\diag}{diag}

\DeclareMathOperator{\Tr}{Tr}

\newcommand{\Mp}{M_{\mathsf{P}}}

\newcommand{\Ms}{M_{\mathsf{S}}}

\newcommand{\Mps}{\hat{M}^{\pupsf{S}}_{\mathsf{P}}}
\newcommand{\Mss}{\hat{M}^{\pupsf{S}}_{\mathsf{S}}}

\newcommand{\Mpb}{M_{\mathsf{P}}^\pup{b}}
\newcommand{\Mtb}{M_{\mathsf{T}}^\pup{b}}
\newcommand{\Msb}{M_{\mathsf{S}}^\pup{b}}
\newcommand{\MpW}{M_{\mathsf{P}}^\pupsf{W}}
\newcommand{\MtW}{M_{\mathsf{T}}^\pupsf{W}}
\newcommand{\MsW}{M_{\mathsf{S}}^\pupsf{W}}
\newcommand{\MpWe}{M_{\mathsf{P}}^\pupsf{Eff}}

\newcommand{\ts}[1]{{\textstyle{#1}}}

\newcommand{\av}[1]{\langle #1 \rangle}
\newcommand{\kk}{\mathbf{k}}
\newcommand{\ud}{\mathrm{d}}

\newcommand{\be}{\begin{equation}}
\newcommand{\ee}{\end{equation}}
\newcommand{\bea}{\begin{eqnarray}}
\newcommand{\eea}{\end{eqnarray}}

\newcommand{\kc}{\msf{k}}
\newcommand{\kb}{\bar{\msf{k}}}
\newcommand{\e}{\varepsilon}

\newcommand{\vex}[1]{\bm{\mathrm{#1}}}
\newcommand{\Nabla}{\bm{\nabla}}

\newcommand{\pup}[1]{{\scriptscriptstyle{({#1})}}}

\newcommand{\pupsf}[1]{{\scriptscriptstyle{(\mathsf{{#1}})}}}

\newcommand{\msf}[1]{\mathsf{#1}}
\newcommand{\ket}[1]{| {#1} \rangle}

\newcommand{\braless}[1]{\left\langle {#1} \right.}

\newcommand{\hs}{\hat{h}_{\scriptscriptstyle{\mathsf{S}}}}
\newcommand{\hsnu}{\hat{h}_{\scriptscriptstyle{\mathsf{S}}}^\pup{\nu}}

\newcommand{\Hs}{H_0^\pupsf{S}}
\newcommand{\His}{H_I^\pupsf{S}}
\newcommand{\Hds}{H_D^\pupsf{S}}

\newcommand{\parr}{\partial}
\newcommand{\parb}{\bar{\partial}}

\newcommand{\hppbandloc}{\hat{h}^{\stackrel[\scriptstyle{band}]{\scriptstyle{local}}{}}_{pp}}
\newcommand{\hsth}{\hat{h}_{3/2}^{\pupsf{S}}}
\newcommand{\rhos}{\varrho_{\mathsf{S}}}
\newcommand{\Ws}{W_{\mathsf{S}}}

\begin{document}

\title{Topological superconductivity of spin-3/2 carriers in a three-dimensional doped Luttinger semimetal}

\author{Bitan Roy}
\affiliation{Department of Physics and Astronomy, Rice University, Houston, Texas 77005, USA}
\affiliation{Max-Planck-Institut f\"{u}r Physik komplexer Systeme, N\"{o}thnitzer Stra. 38, 01187 Dresden, Germany}

\author{Sayed Ali Akbar Ghorashi}
\affiliation{Texas Center for Superconductivity and Department of Physics, University of Houston, Houston, Texas 77204, USA}
\affiliation{Department of Physics, William \& Mary, Williamsburg, Virginia 23187, USA}

\author{Matthew S. Foster}
\affiliation{Department of Physics and Astronomy, Rice University, Houston, Texas 77005, USA}
\affiliation{Rice Center for Quantum Materials, Rice University, Houston, Texas 77005, USA}

\author{Andriy H. Nevidomskyy}
\affiliation{Department of Physics and Astronomy, Rice University, Houston, Texas 77005, USA}
\affiliation{Rice Center for Quantum Materials, Rice University, Houston, Texas 77005, USA}

\date{\today}

\begin{abstract}
We investigate topological Cooper pairing, including gapless Weyl and fully gapped class DIII superconductivity, in a three-dimensional doped Luttinger semimetal. The latter describes effective spin-3/2 carriers near a quadratic band touching and captures the normal-state properties of the 227 pyrochlore iridates and half-Heusler alloys. 
Electron-electron interactions may favor non-$s$-wave pairing in such systems, including even-parity $d$-wave pairing. We argue that the lowest energy $d$-wave pairings are always of complex (e.g., $d + i d$) type, with nodal Weyl quasiparticles. This implies $\varrho(E) \sim |E|^2$ scaling of the density of states (DoS) at low energies in the clean limit, or $\varrho(E) \sim |E|$ over a wide critical region in the presence of disorder. The latter is consistent with the $T$-dependence of the penetration depth in the half-Heusler compound YPtBi. We enumerate routes for experimental verification, including specific heat, thermal conductivity, NMR relaxation time, and topological Fermi arcs. Nucleation of any $d$-wave pairing also causes a small lattice distortion and induces an $s$-wave component; this gives a route to strain-engineer exotic $s+d$ pairings. We also consider odd-parity, fully gapped $p$-wave superconductivity. For hole doping, a gapless Majorana fluid with cubic dispersion appears at the surface. We invent a generalized surface model with $\nu$-fold dispersion to simulate a bulk with winding number $\nu$. Using exact diagonalization, we show that disorder drives the surface into a critically delocalized phase, with universal DoS and multifractal scaling consistent with the conformal field theory (CFT) SO($n$)${}_\nu$, where $n \rightarrow 0$ counts replicas. This is contrary to the naive expectation of a surface thermal metal, and implies that the topology tunes the surface renormalization group to the CFT in the presence of disorder.
\end{abstract}

\maketitle

\section{Introduction}

One of the most useful concepts of modern day condensed matter physics is the topological classification of quantum phases, which at the coarsest level divides into two categories: topological and trivial. A hallmark signature of a topologically non-trivial system is the existence of robust gapless states at an interface with the trivial vacuum, exposing the information about the bulk topological invariant to the external world. This classification encompasses insulators, semimetals and superconductors (both gapped and gapless)~\cite{kane-review, zhang-review, ryu-review-2, ryu-review-1,  hasan-review, armitage-review, altland, fu-kane, SRFL2008, juricic, new-fermions, ben-kane, bansil-review, brydon-schnyder, Volovik2003}. In this paper we establish that a doped three-dimensional Luttinger semimetal (LSM), describing a quadratic touching of Kramers degenerate valence and conduction bands of $j=3/2$ (effective) fermions~\cite{luttinger, murakami}, can harbor myriad exotic gapless and gapped topological superconductors.

The LSM provides the low-energy normal-state description for a plethora of both strongly and weakly correlated compounds, such as the 227 pyrochlore iridates (Ln$_2$Ir$_2$O$_7$, with Ln being a lanthanide element)~\cite{Savrasov, Balents1, yamaji, Exp:Nakatsuji-1, Exp:Nakatsuji-2, goswami-roy-dassarma}, half-Heusler compounds (ternary alloys such as LnPtBi, LnPdBi)~\cite{Exp:cava, Exp:felser, binghai}, HgTe~\cite{hgte}, and gray-tin~\cite{gray-sn-1,gray-sn-2}. Among these materials, the 227 pyrochlore iridates might support only non-Fermi liquid or excitonic (particle-hole channel) orders~\cite{kohn, Abrikosov1, Abrikosov2, balents-kim, Herbut1, Balents3, lai-roy-goswami, arago, kharitonov} (most likely magnetic such as the all-in all-out~\cite{Savrasov, yamaji, Balents3}, spin-ice~\cite{goswami-roy-dassarma} orders), since the chemical potential lies extremely close to the band touching point~\cite{Exp:Nakatsuji-1, Exp:Nakatsuji-2}. Nevertheless, it is possible 
to move the chemical potential away from charge neutrality 
(e.g.\ via 
chemical doping), which 
can be conducive for superconductivity. 
While
Cooper pairing has not yet been found in HgTe or gray-tin, some half-Heusler compounds (such as YPtBi, LaPtBi, LuPdBi, LuPtBi) become superconducting below a few Kelvin~\cite{Exp:Takabatake, Exp:Paglione-1a, Exp:Bay2012, Exp:visser-1, Exp:Taillefer, Exp:Zhang, Exp:visser-2, Exp:Paglione-1, Exp:Paglione-2}. This has led to a surge of theoretical works recently~\cite{Fang2015, Yang16, boettcher-herbut, brydon, GDF17, brydon-2, brydon-3, herbut-2, savary, conjun-new}. Despite half-Heuslers standing as fertile ground for topological phases of matter, the nature of the actual pairing remains elusive so far and therefore demands comprehensive theoretical and experimental investigations.

In this paper, we study various experimental signatures of superconducting states that could arise in a three-dimensional LSM. Since the superconducting order parameter is formed from spin-3/2 band electrons, the SU(2) angular momentum addition rule
$
	(3/2) \otimes (3/2) = 0 \oplus 1 \oplus 2 \oplus 3
$
implies that simple paired states reside in two broad categories: 
(a) even-parity, such as local or intra-unit cell pairing (with order parameter spin $j \in \{0,2\}$), 
and 
(b) odd-parity, momentum-dependent pairing (with order parameter spin $j \in \{1,3\}$) \cite{brydon,savary}.  
We consider these two cases separately. 
The unpaired conduction and valence bands are each two-fold degenerate in the absence of inversion 
symmetry breaking; degenerate states can be labeled by a band pseudospin index.  
All spin-$j$ pairings can be classified according to their transformation under pseudospin rotations.
The $j=0$ and $2$ channels transform as 
pseudospin singlets (respectively $s$- and $d$-wave pairings), while $j=1$ represents a pseudospin triplet $p$-wave
pairing.    
We next provide a synopsis of our main findings for even- and odd-parity pairings.

\subsection{Even parity pairing: scenarios and main results}

Even-parity local pairings are represented by anomalous local bilinears of the spin-$3/2$ fermion field. 
Local or intra-unit cell pairings can be mediated by short-range interactions, such as spin exchange scattering. 
For superconductivity at low densities in an LSM, 
momentum-dependent pairing interactions 
can be strongly suppressed relative to 
local ones. 
The mechanism for this is virtual renormalization from higher energies,
as may also occur in bilayer graphene (a ``two-dimensional LSM'')~\cite{LemonikAleiner10,LemonikAleiner12,Roy13,Vafek14} 
or structurally similar bilayer silicene~\cite{Liu2013}. 
The local pairing amplitudes couple to $j = 0$ and $j = 2$ spin SU(2) tensor operators.

Although the even-parity pairings are local bilinears in the LSM, we focus
on the situation where the superconductivity itself manifests mainly near the Fermi surface.
The Fermi surface is assumed to reside at finite carrier density away from charge neutrality. 
The projection of the $j = 0$ ($j = 2$) pairing onto the conduction or valence bands give rise
to momentum-independent (dependent) $s$-wave ($d$-wave) superconductivity on the Fermi surface~\cite{brydon,savary}.
Both channels are \emph{band-pseudospin} singlets due to Kramers degeneracy. 
In this context we point out an important role played by the normal state band structure.
We note that in a Dirac semiconductor six local pairing bilinears, when projected onto the Fermi surface, 
transform into two $s$-wave pairings and four $p$-wave pairings 
(including the analogous paired state of the $B$-phase, and three polar pairings of $^3$He), see Appendix~\ref{Dirac:bandprojection}.
Thus while $d$-wave pairings in a doped LSM can be realized from an extended Hubbard model 
(containing only local or momentum-independent interactions), 
in Dirac semiconductors they may require non-local interactions.
The same logic was employed by Berg and Fu to argue for the naturalness of $p$-wave pairings from
local interactions in a (topological) Dirac insulator~\cite{fu-berg}~\footnote{
{The reason behind obtaining the $d$-wave character of the SC gap when projecting the \textit{local} pairing onto the Fermi level is reminiscent of how Fu and Berg obtain an effective $p$-wave pairing by projecting the initially local (inter-band) pairing onto the Fermi surface of a doped Bi$_2$Se$_3$. In the latter case, the electron dispersion is linear and the Hamiltonian has a form of a dot-product $(\mathbf{p}\cdot\mathbf{L})$ of momentum $\mathbf{p}$ and $L=1$ orbital momentum. In the case of the Luttinger metal, on the other hand, the Hamiltonian is a dot-product of a 5-dimensional vector formed from cubic harmonics and the 5 components of the symmetric traceless tensors (Dirac's $\Gamma$ matrices) that transform like $L=2$ under the SU(2). Thus, the $L=2$ character results in the $d$-wave pairing when projected onto the Luttinger Fermi surface, compared to $L=1$ in the case of 
Ref.~\cite{fu-berg}.}}

In this paper we carefully catalog the bulk structure of the nodal loops, 
single or double Weyl nodes that
arise via combinations of the $d$-wave pairings (at the cost of the time-reversal symmetry breaking). 
We show how strain can promote particular $d$-wave pairings, whilst simultaneously inducing a 
parasitic $s$-wave component. We also
consider the effects of quenched disorder on the bulk quasiparticle density of states (DoS).
In addition, we determine the anomalous spin and/or thermal Hall conductivities expected 
from possible time-reversal symmetry-breaking orders.  
We now highlight our main results.

1. We determine the transformation of various local pairings under the cubic point 
group symmetry and the spectra of BdG quasiparticles inside various paired phases.
We show that while pseudospin singlet $s$-wave pairing 
(transforming under the $A_{1g}$ representation) induces a full gap, each of the five 
simple $d$-wave pairings 
(belonging to $T_{2g}$ and $E_g$ representations of the $O_h$ point group) 
produces two nodal loops on the Fermi surface (see Sec.~\ref{pairing:luttinger} and Table.~\ref{table_projection}).  
However, due to the underlying cubic symmetry it is always possible to find a specific phase locking among the $d$-wave 
components that eliminates the nodal loops
from the spectra and yields only few isolated simple Weyl nodes (characterized by monopole charges $W_n=\pm 1$). 
See Secs.~\ref{phaselocking:eg} and \ref{Phaselock:T2g}. 
The DoS around each Weyl node vanishes as $\varrho(E) \sim |E|^2$. 
Within a weak coupling pairing picture, complex (e.g.\ $d + i d$) 
Weyl superconductors are therefore energetically favored over the simple $d$-wave nodal-loop pairings, since the former cause an additional power-law suppression of the DoS [$\varrho(E) \sim |E|$ for nodal loop $\to |E|^2$ for Weyl].
Nodal superconductivity can also be realized when the pairing interactions in the $E_g$ and $T_{2g}$
channels are comparable, discussed in Sec~\ref{EgT2g:Competition}, typically supporting simple Weyl nodes with $W_n=\pm 1$.
By contrast, double-Weyl nodes (with monopoles charges $W_n=\pm 2$) can only be found inside the 
$d_{x^2-y^2}+id_{xy}$ phase, which results from a competition between $E_g$ and $T_{2g}$ pairings.
This gives $\varrho(E) \sim |E|$ at low energies. 

2. The emergent topology of BdG-Weyl quasiparticles 
(and therefore the symmetry of the underlying paired state) 
can be probed from the measurement of the anomalous pseudospin and thermal Hall conductivities, discussed in 
Sec.~\ref{hall:thermal-spin}.
We show that despite possessing Weyl nodes the net anomalous pseudospin and thermal Hall conductivities inside the $E_g$ paired state 
are precisely zero, while these are finite in any high symmetry plane inside the $T_{2g}$ paired states. 
On the other hand, when pairing interactions in the $E_g$ and $T_{2g}$ channels are of comparable strength, 
only the $d_{x^2-y^2}+id_{xy}$ paired state supports non-trivial anomalous pseudospin and thermal Hall conductivities 
(see Sec.~\ref{hall:thermal-spin}). 
These results
stem from the momentum-space distribution of 
the Abelian Berry curvature, shown in Figs.~\ref{Berry:Eg} and ~\ref{Berry:T2g}.

3. In strongly correlated quantum materials, a recurring question is the coexistence of otherwise competing orders.
This includes the coexistence of charge-density-wave and superconductivity in the cuprates, 
as well as magnetic and superconducting phases in heavy-fermion compounds.
Here we demonstrate that the formation of any $d$-wave pairing in an LSM 
breaks the cubic symmetry and causes a small lattice distortion or nematicity 
that in turn induces an even smaller $s$-wave component. 
Thus any $d$-wave paired state is always accompanied by a parasitic $s$-wave 
counterpart. Such non-trivial coupling between $d$-wave and $s$-wave superconductivity with the lattice distortion can be exploited to 
strain engineer various $d+s$ paired states, by applying a weak external strain 
in 
particular
directions (see Sec.~\ref{externalstrain}).
Specifically, strain applied along the 
$[0,0,1]$, $[1,1,1]$ and $[1,1,0]$ directions 
respectively leads to the formation of 
$s+d_{3z^2-r^2}$, $s+d_{xy}+d_{yz}+d_{xz}$ and $s+d_{3z^2-r^2}+d_{xy}$ pairing. 
External strain along these three directions therefore induces time-reversal-symmetric mixing of $s$- and $d$-wave pairing.

4. Impurities and quenched disorder can be particularly important in low-carrier systems. 
We investigate the stability of various nodal topological superconductors against the onslaught of randomness or disorder.
Using renormalization group and the $\epsilon$-expansion,
we find that Weyl superconductors, 
comprised of Weyl nodes with monopole charges $\pm 1$, remain stable for sufficiently weak disorder, while at stronger disorder 
the system can undergo a continuous quantum phase transition into a thermal metallic phase where $\varrho(0)$ is finite. 
The disorder-controlled quantum critical point is accompanied by a wide quantum critical regime, where $\varrho(E) \sim |E|$, 
as long as $|E| \ll T_c$ (the superconducting transition temperature). By contrast, both double-Weyl and nodal-loop paired 
states enter into a diffusive thermal metallic phase for arbitrarily weak strength of disorder (see Sec.~\ref{Weyl:disorder}).
~\footnote{ In the presence of strong inter-band coupling due to pairing interactions nodal Fermi points gets replaced by BdG-Fermi surface at lowest energy~\cite{brydon-2,brydon-3}. Our conclusions remain valid above the scale of BdG Fermi energy. Presently the stregth or importance of such inter-band coupling in real materials is unknown. }        

5. In this work, we make an independent attempt to understand the peculiar power-law suppression of the penetration depth 
($\Delta \lambda$) in YPtBi~\cite{Exp:Paglione-2} by combining a power-law contribution (arising from gapless quasiparticles 
in the $d$-wave paired state, for example), with an exponential one, stemming from an $s$-wave component (due to its inevitable coexistence 
with any $d$-wave pairing). We find that even though both $T$-linear and $T^2$ fitting give qualitative agreement, 
the former one yields a better fit over a larger window of temperature (see Fig.~\ref{fitlambda:expthr}). 
A $T$-linear dependence may arise from either double-Weyl nodes or nodal loop(s) in a clean system, 
but it might also represent BdG-Weyl quasiparticles in the presence of quenched disorder. 
The dirty BdG-Weyl system can exhibit $\varrho(E) \sim |E|$ scaling throughout a wide quantum critical fan. 
We propose future experiments to determine the scaling of
specific heat, 
thermal conductivity, 
NMR relaxation time, 
STM measurements of surface Andreev bound states, and
anomalous thermal Hall conductivity 
that can pin down the nature of the pairing in this class of materials (see Sec.~\ref{experiment}).  Finally, we also discuss the consequences for superconductivity of the lack of spatial inversion symmetry, which is  broken in the half-Heusler family of materials, in Sec.~\ref{experiment} and in Appendix~\ref{Append:Inversionassymetry}.

\subsection{Odd parity pairing: robust surface states and topological protection}

Odd-parity pairing can arise in the LSM via $j = 1$ or $j = 3$ spin SU(2) tensor operators,
coupled to odd powers of momentum to satisfy the Pauli principle. 
The $j = 3$ operator [Eq.~(\ref{T3}) in Appendix~\ref{Luttinger:details}] plays the key role
in proposals for $p$-wave, ``septet'' pairing that has been extensively discussed in the context of YPtBi \cite{brydon,Exp:Paglione-2,brydon-2}.
It could also arise in an exotic, isotropic $f$-wave pairing scenario \cite{Yang16}.

In this paper,
we instead focus on simple isotropic $p$-wave pairing, 
different from the gapless septet scenario proposed by other authors~\cite{brydon,Exp:Paglione-2,brydon-2}.
This simpler odd-parity pairing is nevertheless very rich, and can be viewed as the spin-3/2 generalization of the $B$ phase of $^3$He~\cite{Volovik2003}, giving rise to fully gapped, strong class DIII topological superconductivity~\cite{SRFL2008,Fang2015,Yang16,GDF17}.~\footnote{ This outcome is insensitive to the magnitude of chemical doping away from the band touching point. As long the pairing takes place in the vicinity of the Fermi surface (realized either in the valence or conduction band), i.e. when the Fermi momentum $k^\pm_F=\sqrt{2 m^\pm_\ast \mu}$ is a real quantity or equivalently $\mu>0$(see Sec.~\ref{pairing:luttinger} for details), it is topological in nature.}
Unlike model spin-1/2 topological superconductors, the gapless surface Majorana fluid that arises from 
a higher-spin bulk can exhibit nonrelativistic dispersion \cite{Fang2015,Yang16}. The robustness of 
``topological protection'' for such a 2D surface fluid has not been generally established, and
we have shown previously that interactions can destabilize such states \cite{GDF17}. In the same work, however,
we demonstrated that topological protection can be \emph{enhanced} by quenched surface disorder.

The motivation for studying strong topological superconductivity in the LSM is twofold. 
We 
seek to define topological protection 
for surface states of 
higher-spin superconductors,
since this is an ingredient expected to arise in candidate materials with strong spin-orbit coupling.  
At the same time, the Eliashberg calculations in Ref.~\cite{savary} suggest that 
isotropic $p$-wave pairing gives the dominant non-$s$-wave channel in a hole-doped LSM
due to optical-phonon--mediated pairing interactions.

For isotropic $p$-wave pairing in the LSM, we show that the bulk winding number $\nu = 3$ 
describes superconductivity arising from either the $|m_s| = 3/2$ valence or $|m_s| = 1/2$ conduction band;
here $m_s$ denotes the spin projection.
Unlike the winding number, the dispersion of the surface Majorana fluid does depend on $|m_s|$.
We investigate the effects of quenched disorder on the cubic-dispersing fluid that obtains in the
hole-doped scenario.

The surface states of spin-1/2 TSCs with disorder are by now well-understood, 
thanks in large part to their exact solvability near zero energy using methods of 
conformal field theory (CFT) \cite{WZWP2,WZWP4}.	 CFT predicts universal,
disorder-indepedent statistical properties of the surface states, including power-law
scaling of the average 
low-energy 
\emph{surface}
density of states 
$\rhos(\e)$
and ``multifractal'' scaling of surface state 
wave functions. Predictions for a given class depend only on the bulk winding number $\nu$ \cite{WZWP2,WZWP4},
e.g.\  $\rhos(\e) \propto |\e|^{-1 / (2 \nu - 3)}$ for class DIII \cite{LeClair08,WZWP4}.

For the cubic-dispersing surface states of the hole-doped LSM with isotropic $p$-wave pairing,
we compare numerics to 
the corresponding $\nu = 3$ CFT
predictions \cite{YZCP1,WZWP4,GDF17}.
While our numerical results give disorder-independent multifractal
spectra for $\nu = 3$,
the agreement with the SO($n$)${}_3$ CFT \cite{WZWP4,GDF17} is rather poor. 
(Here $n \rightarrow 0$ counts replicas, used to define the disorder-averaged field theory \cite{Altland-Book}.)
We believe that this is due to the limited system size afforded by our numerics;
finite-size effects are expected to be worse for stronger multifractality \cite{YZCP1} 
and multifractality is maximized for \emph{lower} winding numbers \cite{WZWP4}. 
In addition, for $\nu = 3$ the power-law energy scaling of the average density of states predicted by the CFT 
\emph{accidentally} coincides with that due to the clean cubic dispersion, and is therefore not 
a useful indicator in this particular case.

Instead of performing a finite-size scaling analysis
(see Ref.~\cite{Ghorashi18}), we invent a 
\emph{generalized surface theory} that allows the investigation of
a Majorana surface fluid corresponding to a generic bulk winding number $\nu$. 
Computing both the scaling exponent for $\rhos(\e)$ and the zero-energy multifractal spectrum,
we find excellent agreement 
between the SO($n$)${}_\nu$ CFT \cite{WZWP4,GDF17}
and numerics 
for $\nu = 5,7$, predicted to exhibit much weaker multifractality.

Since the SO($n$)${}_\nu$ theory is known 
to be stable against the effects of interparticle interactions \cite{WZWP4},
our results imply that surface states enjoy robust
topological protection, with signatures such as 
the universal tunneling density of states $\rhos(\e)$ 
and the precisely quantized thermal conductivity \cite{WZWP3} 
that could be detected experimentally. 
That we find critical delocalization for any $\nu$ is surprising,
since the naive expectation would be a surface thermal metal phase. 
(The thermal metal would exhibit disorder-dependent spectra.)
Indeed, the CFT is
technically \emph{unstable} towards flowing into the thermal metal, see Fig.~\ref{Fig--CFTFlow}.
Our results 
for generic winding numbers
suggest that, in the presence of disorder, the topology fine-tunes the surface to the CFT.

We emphasize that the \emph{clean limit} for our generalized surface model exhibits a stronger
density of states van Hove singularity with larger $\nu$. This would suggest a stronger tendency 
at larger winding numbers for the disorder to induce a diffusive surface thermal metal, due to the high accumulation of
states in a narrow energy window that can be admixed by the disorder. 
It is all the more surprising that we recover universal, critical CFT results with \emph{better agreement} 
for increasing $\nu$.
We also expect that $\nu = 3$ (relevant to the LSM) would give results consistent 
with the SO($n$)${}_3$ CFT for bigger system sizes than we can access here, 
which could capture the highly rarified wave functions and predicted strong multifractality.
This extrapolation from results at larger winding numbers is in the same spirit 
as a large-$N$ expansion.

It is also interesting to note that the simple generalized surface theory
introduced here allows us to ``dial in'' any of the infinite class of Wess-Zumino-Novikov-Witten (WZNW)
SO($n$)${}_\nu$ conformal field theories (with $n \rightarrow 0$), 
simply by tuning one parameter $\nu \equiv 2 k + 1$, with $k \in \{1,2,3,\ldots\}$. 
By contrast, WZNW models with higher levels 
typically arise only by fine-tuning more and more parameters. This is because higher-level WZNW models
usually represent multicritical points in 1+1-quantum field theories \cite{Affleck-Spinchains,CFT-Book}. 
In the context of TSC surface states, (nonunitary) WZNW theories are robustly
realized \emph{without fine-tuning} \cite{WZWP4,GDF17,Ghorashi18}, 
an emerging novel aspect of ``topological protection'' for 3D topological superconductors.

\subsection{Outline}

This paper is organized as follows. 
The low-energy description of a LSM, 
possible pairings (both even- and odd-parity) and their classification are discussed in 
Sec.~\ref{pairing:luttinger}. 
The competition between even parity $s$- and $d$-wave superconductivity is discussed in Sec.~\ref{sec:s+id}. 
In Sec.~\ref{Weyl-SC} we focus on the competition amongst various $d$-wave pairings belonging to 
different representations, and the emergence of Weyl superconductivity at low temperatures. 
We also compute the nodal topology of Weyl pairings and its manifestation through anomalous 
pseudospin and thermal Hall conductivities. 
Sec.~\ref{externalstrain} is devoted to the effects of external strain, while the
effects of impurities on BdG-Weyl quasiparticles are addressed in Sec.~\ref{Weyl:disorder}. 
Connections with a recent experiment in YPtBi and possible future experiments to 
pin the pairing symmetry are presented in Sec.~\ref{experiment}. 
The bulk-boundary correspondence and the surface states of odd-parity $p$-wave pairing
are discussed in Sec.~\ref{sec: p-wave}. We conclude in Sec.~\ref{conclusions}. 
Appendix \ref{Luttinger:details} summarizes equivalent matrix formulations for the LSM Hamiltonian.
Additional technical details are relegated to appendices.


\pagebreak

\section{Pairing in the Luttinger semimetal}~\label{pairing:luttinger}

We review the low-energy description of a LSM, followed by even- and odd-parity Cooper pairing scenarios.
We enumerate the nodal-loop structure of all even-parity $d$-wave pairings.
Finally, we compute the free energy, gap equation and transition temperature within BCS theory.

\subsection{Luttinger Hamiltonian}

Quadratic touching of Kramers degenerate valence and conduction bands at an isolated point 
[taken to be the $\Gamma=(0,0,0)$ point] in the Brillouin zone in three spatial dimensions 
can be captured by the $k\cdot p$
Hamiltonian
\begin{equation}\label{HLDef}
	H_L 
	=
	\int \frac{d^3 {\mathbf k}}{(2 \pi)^3} \: \Psi^\dagger_{\mathbf k} \, \hat{h}_L(\mathbf k) \, \Psi_{\mathbf k},
\end{equation}   
where the four-component spinor $\Psi_{\mathbf k}$ is defined as 
\begin{equation}~\label{spinor-4}
	\Psi^\top_{\mathbf k}
	= 
	\left( c_{{\mathbf k}, +\frac{3}{2}}, c_{{\mathbf k}, +\frac{1}{2}}, c_{{\mathbf k}, -\frac{1}{2}}, c_{{\mathbf k}, -\frac{3}{2}}  \right).
\end{equation}
Here $c_{{\mathbf k}, m_s}$ is the band electron annihilation operator with spin projection 
$m_s \in \{3/2,1/2,-1/2,-3/2\}$. 
Such quadratic touching is protected by the cubic symmetry, which restricts the form of the 
Luttinger Hamiltonian~\cite{luttinger, murakami} operator to
\begin{eqnarray}~\label{luttinger}
	\hat{h}_L(\mathbf k) 
	&=& 
	\left( \frac{{\mathbf k}^2}{2 m_0} -\mu \right) \Gamma_0 
	- 
	\frac{1}{2 m_1} \sum^{3}_{a=1} d_a (\mathbf k) \Gamma_a 
\nonumber\\
	&-& 
	\frac{1}{2 m_2} \sum^{5}_{a=4} d_a (\mathbf k) \Gamma_a,  
\end{eqnarray} 
where $\mu$ is the chemical potential measured from the band touching point. 
The ${\mathbf d}$-vector appearing in the Luttinger Hamiltonian is given by ${\mathbf d} (\mathbf k)= k^2 \, \hat{\vex{d}}(\hat{\vex{k}})$, 
where $\hat{\vex{d}}(\hat{\vex{k}})$ is a five-dimensional unit vector that transforms in the $l = 2$ (``d-wave'') representation under orbital SO(3) rotations.
Its components can be constructed from the spherical harmonics $Y^m_{l=2}(\theta, \phi)$, 
see Appendix~\ref{Luttinger:details}. 
While $\Gamma_0$ is a four-dimensional unit matrix, the five mutually anti-commuting matrices appearing in the Luttinger Hamiltonian are given by 
\begin{equation}~\label{Gamma:definition}
\begin{gathered}
	\Gamma_1 =\kappa_3 \sigma_2,\quad 
	\Gamma_2=\kappa_3 \sigma_1,\quad 
	\Gamma_3=\kappa_2,
\\ 
	\Gamma_4=\kappa_1,\quad 
	\Gamma_5=\kappa_3 \sigma_3. 
\end{gathered}
\end{equation}
Two sets of Pauli matrices $\left\{ \kappa_\alpha \right\}$ and $\left\{ \sigma_\alpha \right\}$, with $\alpha=0,1,2,3$ operate 
respectively on the
sign [$\sgn(m_s)$] and the magnitude [$|m_s| \in \{1/2,3/2\}$] of the spin projection $m_s$. 
The $\Gamma$ matrices provide a basis for a symmetric traceless tensor operator formed
from bilinear products of $j = 3/2$ matrices
[Eqs.~(\ref{GammatoTmunu}) and (\ref{tracelesstensor:definition}) in Appendix~\ref{Luttinger:details}], 
and transform in the $j = 2$ representation of the spin SU(2). 
Consequently, the Hamiltonian in Eq.~(\ref{luttinger}), is an $A_{1g}$ quantity in a cubic environment. 
For $m_1 = m_2$, $\hat{h}_L(\mathbf k)$ exhibits continuous SO(3) rotational invariance.

Besides five mutually anticommuting $\Gamma$ matrices and the identity matrix ($\Gamma_0$), 
we can define ten commutators as $\Gamma_{ab}= \left[ \Gamma_a, \Gamma_b \right]/(2i) \equiv -i \Gamma_a \Gamma_b$ for $a,b=1, \cdots, 5$ with $a \neq b$ that 
together close the basis for all four dimensional matrices. The ten commutators are the generators of a (fictitious) SO(5) symmetry. 
Since ${\mathbf d}({\mathbf k}) = 0$ at the $\Gamma$ point of the Brillouin zone ${\mathbf k}=0$, the four degenerate 
bands possess an emergent SU(4) symmetry at this point. However, at finite momentum such symmetry gets reduced to SU(2) $\times$ SU(2),
stemming from the Kramers degeneracies of the valence and conduction bands. In addition, the Luttinger Hamiltonian is 
invariant under the time reversal transformation: 
${\mathbf k} \to -{\mathbf k}$ 
and 
$\Psi_{\mathbf k} \to \Gamma_{13} \Psi_{-\mathbf k}$. 
The anti-unitary time-reversal symmetry operator is given by ${\mathcal T}=\Gamma_1 \Gamma_3 K$, where $K$ is the complex conjugation
 and ${\mathcal T}^2=-1$. 
The Kramers degeneracy is protected by inversion symmetry ${\mathcal P}: {\mathbf k} \to -{\mathbf k}$.

Without any loss of generality, but for the sake of technical simplicity, we work with the isotropic 
Luttinger model for which $m_1 = m_2 \equiv m$. 
The Luttinger Hamiltonian then has the alternative representation, 
\begin{equation}~\label{luttinger-spin}
	\hat{h}_L(\mathbf k)
	=
	\left[ \left(\lambda_1 + 5 \lambda_2/2 \right) k^2 -\mu \right]\Gamma_0
	-
	2 \lambda_2 \left(\vex{J}\cdot\vex{k} \right)^2,
\end{equation}
with ${\mathbf J}=\left( J^x, J^y, J^z \right)$ and ${\mathbf k}=\left( k_x, k_y, k_z \right)$. 
Here $J^{x,y,z}$ are SU(2) generators in the 3/2 representation.
The correspondence between Eqs.~(\ref{luttinger}) and (\ref{luttinger-spin}) is 
$\lambda_1 = \left(2 m_0\right)^{-1}$, $\lambda_2=\left(4m\right)^{-1}$. 
The Luttinger Hamiltonian can be diagonalized as 
${\mathcal D}^\dagger \hat{H}_L({\mathbf k}) {\mathcal D}$, with the energy spectra 
\begin{equation}\label{lut_energy}
	\varepsilon_{\pm,\sigma}(\mathbf k)= \left(\frac{{\mathbf k}^2}{2 m_0} -\mu \right) \pm \frac{{\mathbf k}^2}{2 m}. 
\end{equation}
Here $+$ ($-$) corresponds to the $|m_s| = 1/2$ conduction ($|m_s| = 3/2$ valence) band.
We have assumed that $m_0 > m_1$, so that these two bands bend oppositely. 
The ``band pseudospin'' index $\sigma \in \pm 1$, and independence of $\varepsilon_{\pm,\sigma}(\mathbf k)$ on $\sigma$ specifies the Kramers degenerate states in each band.
For a given $\vex{k}$, one possible choice is $\sigma = \sgn(\vex{J}\cdot\vex{k})$ (i.e.\ pseudospin-momentum locking), 
but we will not need to fix this basis. 
The diagonalizing matrix ${\mathcal D}$ is given by~\cite{murakami} 
\begin{equation}~\label{diagonalizer}
	{\mathcal D}
	= 
	\left[ 2 \left( 1+ \hat{d}_5 \right) \right]^{-1/2}
	\left[ 
		\left( 1+\hat{d}_5\right) \Gamma_0 + i \sum^{4}_{a=1} \Gamma_{a5} 
	\right].
\end{equation}
The pseudospin locking in the valence and conduction bands becomes transparent with a specific choice of the momentum ${\bf k}=\left( 0,0,k_z \right)$ for which the Luttinger Hamiltonian from Eq.~(\ref{luttinger}) readily assumes a diagonal form
\begin{equation}
\hat{h}_L (k_z)= 	{\rm Diag.} \left[ -\frac{k^2_z}{2 m^{-}_\ast}, \frac{k^2_z}{2 m^{+}_\ast}, \frac{k^2_z}{2 m^{+}_\ast}, -\frac{k^2_z}{2 m^{-}_\ast}  \right]-\mu,
\end{equation} 
in the spinor basis defined in Eq.~(\ref{spinor-4}), where $m^{\pm}_\ast=m_0 m/|m_0 \pm m|$. 
Therefore, for $m_0>m$ the first and fourth (second and third) entries yield Kramers degenerate spectra 
for the valence (conduction) band. 
Hence, the pseudospin projection on the valence (conduction) band is $|m_s|=3/2$ $(1/2)$.\footnote{
Note that in the subsequent sections we use the same band-diagonalization procedure to investigate the form 
of even-parity local (or intra-unit cell) pairings as well as odd-parity non-local (or extended) pairings around the Fermi surface. 
When projected onto the valence (conduction) band, the pairings takes place among the spin-3/2 fermions
with spin projection $|m_s|=3/2$ $(|m_s|=1/2)$, respectively. 
It turns out that the form of the local pairings ($s$- and $d$-wave) around the Fermi surface \underline{do 
not depend} on the choice of the band, or equivalently, the spin projections (see Sec.~\ref{sec: evenp}), 
while the form of the non-local $p$-wave pairing crucially depends on whether the Fermi surface is 
realized in the valence or conduction band (see Sec.~\ref{sec: p-wave}).	
}

\subsection{Even-parity local pairings \label{sec: evenp}}

In this section we review even-parity local pairing operators
that give rise to pseudospin-singlet $s$- and $d$-wave channels
when projected to the Fermi surface~\cite{brydon,savary}.
We enumerate the nodal loop states that arise from individual $d$-wave pairing (see Table~\ref{table_projection}),
and which form the basis for nodal $d + i d$ Weyl superconductors in the 
sequel. Odd-parity pairings are considered in Sec.~\ref{sec: oddp}.

\begin{table*}[!ht]
\begin{tabular}{|c|c|c|c|}
\hline
{\bf 
$\stackrel{\displaystyle{\text{Pairing}}}{\displaystyle{\text{in LSM}}}$ 
} 
& 
{\bf Pairing near Fermi surface} 
&
{\bf 
$\stackrel{\displaystyle{\text{IREP}}}{\displaystyle{\text{(Nature)}}}$ 
}
& 
{\bf Quasiparticle spectrum}  \\
\hline \hline
$\Delta_0 \; \Gamma_0$ & $\Delta_0 \sigma_0$ & $A_{1g}$ ($s$-wave) & Fully gapped \\
\hline \hline
$\Delta_1 \; \Gamma_1$ & $\Delta_1 \hat{d}_1 \sigma_0 \equiv \sqrt{3} \Delta_1 \; \left( \hat{k}_y \hat{k}_z \right) \sigma_0$ & $T_{2g}$ ($d_{yz}$) 
& 
$
	\text{Gapless: 2 Nodal loops,}  
	\quad
	\displaystyle{    
	\left\{
	\begin{aligned}
	& k^2_x+k^2_z=k^2_F, \; k_y=0
	\\
	& k^2_x+k^2_y=k^2_F, \; k_z=0
	\end{aligned}
	\right\}
	}
$ 
\\
\hline
$\Delta_2 \; \Gamma_2$ & $\Delta_2 \hat{d}_2 \sigma_0 \equiv \sqrt{3} \Delta_2  \; \left( \hat{k}_x \hat{k}_z \right) \sigma_0$ & $T_{2g}$ ($d_{xz}$) 
& 
$
	\text{Gapless: 2 Nodal loops,}  
	\quad
	\displaystyle{    
	\left\{
	\begin{aligned}
	& k^2_y+k^2_z=k^2_F, \; k_x=0
	\\
	& k^2_x+k^2_y=k^2_F, \; k_z=0
	\end{aligned}
	\right\}
	}
$ 
\\
\hline
$\Delta_3 \; \Gamma_3$ & $\Delta_3 \hat{d}_3 \sigma_0 \equiv \sqrt{3} \Delta_3 \; \left( \hat{k}_x \hat{k}_y \right) \sigma_0$ & $T_{2g}$ ($d_{xy}$) 
& 
$
	\text{Gapless: 2 Nodal loops,}  
	\quad
	\displaystyle{    
	\left\{
	\begin{aligned}
	& k^2_y+k^2_z=k^2_F, \; k_x=0
	\\
	& k^2_x+k^2_z=k^2_F, \; k_y=0
	\end{aligned}
	\right\}
	}
$ 
\\
\hline \hline
$\Delta_4 \; \Gamma_4$ & $\Delta_4 \hat{d}_4 \sigma_0 \equiv \frac{\sqrt{3}}{2} \Delta_4 \; \left(\hat{k}^2_x- \hat{k}^2_y \right) \sigma_0$ & $E_{g}$ ($d_{x^2-y^2}$) 
& 
$
	\text{Gapless: 2 Nodal loops,}  
	\quad
	\displaystyle{    
	\left\{
	\begin{aligned}
	& k^2_\perp+k^2_z=k^2_F, \; k_x=+k_y
	\\
	& k^2_\perp+k^2_z=k^2_F, \; k_x=-k_y
	\end{aligned}
	\right\}
	}
$ 
\\
\hline
$\Delta_5 \; \Gamma_5$ & $\Delta_5 \hat{d}_5 \sigma_0 \equiv \frac{\Delta_5}{2} \; \left( 2\hat{k}^2_z-\hat{k}^2_x-\hat{k}^2_y\right) \sigma_0$ & $E_{g}$ ($d_{3z^2-r^2}$) 
& 
$
	\text{Gapless: 2 Nodal loops,}  
	\quad
	\displaystyle{    
	\left\{
	\begin{aligned}
	& k_\perp=k_F \sqrt{{\ts{\frac{2}{3}}}}, \; k_z=+ {\ts{\frac{1}{\sqrt{3}}}}k_F
	\\
	& k_\perp=k_F \sqrt{{\ts{\frac{2}{3}}}}, \; k_z=- {\ts{\frac{1}{\sqrt{3}}}}k_F
	\end{aligned}
	\right\}
	}
$ 
\\
\hline \hline
\end{tabular}
\caption{Classification of six local pairing operators in Eq.~(\ref{hBdG_Def}) for the Luttinger semimetal, and corresponding nodal-loop structures for five basis d-wave pairings. These pairing operators are time-reversal even (constituting a six dimensional basis for local pairings), and we shall consider the time-reversal odd $s+id$ (see Sec.~\ref{sec:s+id}) and $d+id$ (see Sec.~\ref{Weyl-SC} and Table~\ref{egt2g:competitiontable}) combinations in subsequent sections. First column: All six possible local pairings in a Luttinger semimetal. 
Second column: Representation of the corresponding pairing close to the Fermi surface in the conduction or valence band. 
Third column: Transformation of each pairing under specific irreducible representation of the octahedral group $O_h$ (nature of individual pairing). 
Fourth column: Quasiparticle spectrum inside each individual paired state. 
Here $k^{\pm}_F=\sqrt{2 m_\ast^\pm |\mu|}$ 
is the Fermi momentum in the conduction and valence band, respectively, and 
$k_\perp = \sqrt{k^2_x+k^2_y}$.
Since the form of the local pairing does not depend on the choice of band, we here take $k^{\pm} \to k_F$ for notational simplicity. For a general discussion on pairing in a cubic system, see Ref.~\cite{ueda-sigrist}. 
}~\label{table_projection}
\end{table*}

The effective single-particle pairing 
Hamiltonian in the presence of local or intra-unit cell superconductivity 
assumes the form 
\begin{equation}~\label{local-pairing-cond}
	H^{local}_{pp}
	=
	\Delta_M 
	\int d^3{\mathbf r} \, \Psi^\T M \Psi  + \text{H.c.}, 
\end{equation}  
where $M$ is a $4 \times 4$ matrix, $\Delta_M$ is the pairing amplitude, 
$\T$ is the matrix transpose, and H.c.\ denotes the Hermitian conjugate.  
The Pauli principle mandates that $M^\T = -M$, implying 
that there are only \emph{six} possible independent bilinears of the form
$\Psi^\T M \Psi$, since the allowed matrices correspond
to the generators of SO(4). 
Therefore, the effective single-particle Hamiltonian in the presence of \emph{all} 
possible local pairings reads as 
\begin{align}~\label{local-pairing-all}
	H^{local}_{pp} 
	=&\,
	\int d^3{\mathbf r} 
	\,
	\Psi^\T 
	\left[ 
	\begin{aligned}
	&\,
		\Delta_0 \Gamma_{13} +  \Delta_{1} \Gamma_3 + \Delta_2 \Gamma_{45}
	\\&\,
	+ 
		\Delta_3 \Gamma_1 + \Delta_4 \Gamma_{25} + \Delta_5 \Gamma_{24}
	\end{aligned}
	\right] 
	\Psi
\nonumber\\ 
	&\,
	+ \text{H.c.}, 
\end{align}  
where we have used the product basis for the Clifford algebra [Eq.~(\ref{Gamma:definition})]
to express the antisymmetric matrices. 
We stress that the existence of the above six local pairing operators 
does not depend on the character of the normal state; it relies only on the fact that the 
low-energy description of this state is captured by a four-component spinor. 
Identical pairing operators
arise for massive Dirac fermions describing either topological or normal 
insulators~\cite{ohsaku,fu-berg,goswami-roy}, 
Weyl semimetals 
(where the four-component representation accounts for a pair of Weyl nodes)~\cite{roy-goswami-juricic}, 
and ${\mathcal T}$-preserving nodal-loop semimetals~\cite{roy-NLSM}. 
However, the \emph{physical meaning} of these local pairings crucially depends 
on the band structure of the parent state.

To characterize local pairings, we now introduce an eight-component Nambu spinor      
\begin{eqnarray}~\label{nice-Nambu}
	\Psi_N= \left[ \begin{array}{c} 
	\Psi \\
	i \Gamma_{13} \left( \Psi^\dagger\right)^\top
\end{array} \right],
\end{eqnarray}
where $\Psi$ is the four-component spinor defined in Eq.~(\ref{spinor-4}). 
We have absorbed the unitary part of the time-reversal operator ${\mathcal T}$ in the lower block 
of the Nambu spinor. This ensures that $\Psi_N$ transforms the same way as $\Psi$ 
under spin SU(2) rotations, because the $j = 3/2$ generators $\{J^\mu\}$ satisfy the pseudoreality condition
\begin{align}\label{pseudoreal}
	- \Gamma_{13} \, \left(J^\mu\right)^\T \, \Gamma_{13} = J^\mu, \quad \mu \in \{x,y,z\}.
\end{align}

In this basis the single-particle Hamiltonian operator in the presence of six local pairings [introduced in Eq.~(\ref{local-pairing-all})] 
assumes a simple and instructive form 
\begin{align}
	H^{local}_{pp}
	=&\,
	\int d^3{\mathbf r} 
	\,
	\Psi_N^\dagger 
		\,
	\hat{h}^{local}_{pp}
		\,
	\Psi_N,
\nonumber\\
	\hat{h}^{local}_{pp}
	=&\,
	\left(\tau_1 \cos \phi + \tau_2 \sin \phi \right) 
\nonumber\\
	&\,\times
	\left[
	\begin{aligned}
		&\,
		\Delta_0 \Gamma_0 + \Delta_1 \Gamma_1 + \Delta_2 \Gamma_2 
		\\&\,+ 
		\Delta_3 \Gamma_3 + \Delta_4 \Gamma_4 + \Delta_5 \Gamma_5
	\end{aligned}
	\right],
		\label{hBdG_Def}
\end{align}    
where $\phi$ is the $U(1)$ superconducting phase. 
The Pauli matrices $\{\tau_\alpha\}$ act on the particle-hole (Nambu) space. 
The identity matrix $\Gamma_0$ represents $s$-wave pairing. 
By contrast, as the Clifford matrices transform irreducibly in the $j = 2$ 
representation of the spin SU(2), the corresponding pairing channels 
$\Delta_{1,\ldots,5}$ are all (effectively) $d$-wave.  
In terms of cubic symmetry, the pairing  proportional to $\Gamma_0$ belongs to the
trivial $A_{1g}$ representation. The three pairings proportional to $\Gamma_{1,2,3}$ 
transform as a triplet under the $T_{2g}$ representation. 
By contrast, those proportional to $\Gamma_4$ and $\Gamma_5$ transform
as a doublet under the $E_g$ representation in a cubic environment.

In the Nambu basis, a Bogoliubov-de Gennes Hamiltonian 
$\hat{h}(\vex{k})$ automatically satisfies the particle-hole symmetry 
\begin{align}\label{MajPH}
	- 
	\Mp \, \hat{h}^\T(-\vex{k}) \, \Mp = \hat{h}(\vex{k}), 
	\quad
	\Mp = \tau_2 \, \Gamma_{13},  
\end{align}
owing to the reality condition $\Psi_N^\dagger(\vex{k}) = \Psi_N^\T(-\vex{k}) \, \Mp$
and Pauli exclusion. For momentum-independent pairing operators,
Eqs.~(\ref{pseudoreal}) and (\ref{MajPH}) imply that only 
tensor operators composed from products with even numbers of 
spin generators (e.g. $\{J^\mu J^\nu\}$) are allowed. 
These are precisely the identity and the anticommuting Clifford matrices (see Appendix~\ref{Luttinger:details}). 
Therefore, all $d$-wave pairings can also be considered as \emph{quadrupolar pairings}. 
Since Eq.~(\ref{MajPH}) is automatic, we can combine it with the 
usual form of time-reversal symmetry to get the chiral condition
\begin{align}\label{ChiralTRI}
	- \Ms \, \hat{h}(\vex{k}) \, \Ms = \hat{h}(\vex{k}), \quad \Ms = \tau_2.
\end{align}
Thus pairings in Eq.~(\ref{hBdG_Def}) proportional to $\tau_1$ ($\tau_2$) are
even (odd) under time-reversal (for a fixed phase $\phi$).

We focus on superconductivity in an LSM doped to finite electron 
or hole density, away from charge neutrality. The nature of these 
pairings becomes transparent after projecting onto the valence or conduction band. 
With local pairings the result is the same for both bands,
see Appendix~\ref{bandprojection} for details. 
If we assume that pairing occurs only in close proximity to the Fermi surface 
and 
does not mix the bands, 
the projected pairing Hamiltonian takes the 
form
\begin{align}~\label{local-band}
	\hppbandloc
	= 
	\left( \tau_1 \cos \phi + \tau_2 \sin \phi \right) 
	\sigma_0 \bigg[ \Delta_0 + \sum^{5}_{j=1} \Delta_j \hat{d}_j \bigg],
\end{align}
as shown in Table~\ref{table_projection} (see also Appendix~\ref{Luttinger:details}). 
Here the Pauli matrices $\{\sigma_\alpha \}$ act on the pseudospin (Kramers) degenerate
states within the projected band. 
All even-parity pairing operators map to \emph{pseudospin singlets}.

The band-projected kinetic energy term arising from
Eq.~(\ref{luttinger})
assumes the simple form in the Nambu basis
\begin{equation}~\label{kinetic-band}
	\hat{h}^{band}_0 \left( {\mathbf k} \right)
	=
	\left[
		\pm
		\left( \frac{{\mathbf k}^2}{2 m_\ast^\pm}\right)
		- 
		\mu 
	\right] \tau_3 \sigma_0,	
\end{equation} 
where $m^\pm_\ast = m_0 m/|m_0 \pm m|$ (and $m_0 > m$).					
Here $+$ ($-$) denotes the $|m_s| = 1/2$ conduction ($|m_s| = 3/2$ valence) band.
While both the kinetic and pairing terms involve the Clifford matrices 
before the projection, only the kinetic term of the Luttinger Hamiltonian
depends on the five $d$-wave harmonics $\vex{d}(\vex{k})$. Post projection, the kinetic term is 
trivial and the five pairing operators become $\vex{d}(\vex{k})$ components.  
Therefore, the band structure in the normal state plays a paramount 
role in determining the projected form of the local pairing operators.
The Nambu-doubled four-component spinor describing quasiparticle excitations around the Fermi surface is defined as 
$
	\psi^\dagger_{\mathbf k}
	=
	\left[ 
		c^\dagger_{\uparrow, {\mathbf k}}, 
		c^\dagger_{\downarrow, {\mathbf k}}, 
		c_{\downarrow, -{\mathbf k}},
		-
		c_{\uparrow, -{\mathbf k}}  
	\right]$, 
and the time-reversal operator in the reduced space reads as 
${\mathcal T}=i \sigma_2 K$, so that ${\mathcal T}^2=-1$ as usual.

As a counterexample, we note that the same six local pairings projected 
into the valence/conduction bands in a massive Dirac semiconductor 
give rise to two even-parity $s$-wave and four pseudospin-triplet, odd-parity
 $p$-wave pairing, including analogous to the fully gapped B-phase and three 
polar pairings in $^3$He,
see Appendix~\ref{Dirac:bandprojection}, Table~\ref{Pairing:Dirac-projected}. Hence, realization of various superconductivity 
close to a Fermi surface crucially depends on the normal state band structure,
at least for low carrier densities.

The quasiparticle spectra inside each of the six local pairing states of the LSM
are as follows.
The $s$-wave pairing gives a fully gapped spectrum everywhere on the Fermi surface. 
On the other hand, each component of the five $d$-wave pairings supports two nodal loops, 
along which the Fermi surface remains gapless. 
The equations determining the nodal loop for each $d$-wave pairing 
are reported in the rightmost column of Table~\ref{table_projection}. 
As a result each $d$-wave pairing supports 
``topologically protected''
flatband surface states that span the images of the bulk nodal loops on each
surface \cite{brydon,brydon-2}. 
We note that mixing between the conduction and valence bands in paired states of the
LSM can also produce novel effects, such as bulk quasiparticle Fermi
surfaces (instead of nodal points or lines)~\cite{brydon-2,brydon-3}.

The density of states (DoS) in the presence of isolated nodal loops vanishes as 
$\varrho(E) \sim |E|$. 
In Sec.~\ref{Weyl-SC} we will show that underlying cubic symmetry causes 
specific phase locking among different components of the $d$-wave pairings 
belonging to either $T_{2g}$ or $E_g$ representation (see Table~\ref{table_projection}), at the cost of the 
time-reversal symmetry. Consequently, the paired state only supports simple 
Weyl nodes at a few isolated points on the Fermi surface, around which the 
DoS vanishes as $\varrho(E) \sim |E|^2$. Such reconstruction of the quasiparticle 
spectra thus causes a \emph{power-law} suppression of the DoS and that way 
optimizes the condensation energy gain. Thus we expect Weyl superconductors 
to be energetically favored over the nodal-loop pairings, at least within 
the framework of weak BCS superconductivity and for dominant $d$-wave pairing coupling strengths.

\subsection{Odd-parity, momentum-dependent pairings  \label{sec: oddp}}

In the isotropic case ($m_1=m_2=m$), odd-parity pairings can be classified via angular momentum addition \cite{savary}. 
A basis of 10 Hermitian, particle-hole--odd 
operators with well-defined SU(2) spin $j$ is given by 
[c.f.\ Eq.~(\ref{hBdG_Def})]
\begin{align}\label{OddPair}
	\tau_{1,2}
	\otimes
	\left\{
	\begin{array}{ll}
	J^\mu, \quad&\, j = 1 \\
	T^{\mu \nu \gamma}, \quad&\, j = 3
	\end{array}
	\right\}.
\end{align}
Here $T^{\mu \nu \gamma}$ is a completely symmetric, traceless tensor operator
formed from triple products of $J^\mu$ generators, see Eq.~(\ref{T3}).
Eq.~(\ref{MajPH}) implies that particle-hole allowed pairing operators 
obtain by multiplying any of the matrices in 
Eq.~(\ref{OddPair}) by odd powers of momentum. 
The resulting 
momentum-dependent
pairing operator with particle-hole matrix $\tau_1$ ($\tau_2$) 
is even (odd) under time-reversal [Eq.~(\ref{ChiralTRI})].

For orbital $p$-wave pairing ($l = 1$), angular momentum addition gives
$l \otimes j \equiv j_{\mathsf{tot}}= 0\,\oplus\,1\,\oplus\,2$ for $j = 1$ 
and 
$j_{\mathsf{tot}} = 2\,\oplus\,3\,\oplus\,4$ for $j = 3$.
We highlight a few combinations. 
The $j_{\mathsf{tot}}= 0$ corresponds to a fully gapped, isotropic $p$-wave superconductor.
For weak pairing this represents strong topological superconductivity \cite{SRFL2008,Fang2015,Yang16}, which we 
study in detail in Sec.~\ref{sec: p-wave}, below. 
The $j_{\mathsf{tot}}= 1$ state corresponds to gapless ``$p_x$-wave'' pairing.  
The $j_{\mathsf{tot}}= 2$ states arising from $j = 1$ and $j = 3$ spin can mix,
since only the total angular momentum is well-defined \cite{savary}.

The $p$-wave ``septet'' order considered in the context of YPtBi is also built from 
the $j = 3$ operator \cite{brydon,brydon-2,Exp:Paglione-2}. 
Finally, we note that isotropic $f$-wave pairing of the form
\[
	\tau_{1,2} \otimes T^{\mu \nu \gamma} k_\mu k_\nu k_\gamma, \quad  j_{\mathsf{tot}}= 0,
\]
turns out to give the same band-projected Bogoliubov-de Gennes Hamiltonians
as the isotropic $p$-wave case, Eq.~(\ref{hBdGProj}), except that the 
pairing potential is multiplied by an additional factor of $k^2$ in each case.


\subsection{Free energy and gap equation}\label{sec:d-wave}

We now discuss the free energy and resulting gap equations for the five $d$-wave pairings 
summarized in Table~\ref{table_projection}. 
At zero temperature the free energy of a superconductor is given by~\cite{Tinkham}
\be
		F_j 
		= 
		\frac{|\Delta_d|^2}{2g_d} 
		- 
		a^3 \!\!\!\!
		\int\limits_{|\xi_k|<\Omega_D}\!\!\!\!\!\! 
		\frac{\ud^3 \kk}{(2\pi)^3} 
		\sqrt{\xi_k^2 + |\Delta_d|^2 d_j^2(\kk) },
\label{eq:free0}
\ee
where $d_j(\kk),\, j\in\{1,\ldots,5\}$ are the $d$-wave harmonics 
(see the second column of Table~\ref{table_projection}), 
$\xi_k \equiv \frac{k^2}{2m^*}-\mu$, 
$a$ denotes the lattice spacing,
$g_d$ is the coupling strength,
and we set $\hbar=1$ throughout. 
The pairing is restricted to an energy window set by the (effective) Debye frequency $\Omega_D$. 
We introduce dimensionless variables defined via 
$f_j \equiv F_j/[\mu^2\varrho(\mu)]$, 
$x \equiv |\kk|/k_F$, 
$\lambda_d \equiv g_d \varrho(\mu)$, 
$\hat{\Delta}_d \equiv \Delta_d/\mu$, and
$\omega_D \equiv \Omega_D/\mu$, 
where 
$\varrho(\mu) = 2a^3m^\ast\sqrt{2m^\ast\mu}/(2\pi^2)$ is the DoS at the Fermi level.
In what follow we throughout use the above set of dimensionless variables.

In terms of these dimensionless variables the corresponding gap equation is of the form:
\be
	\frac{1}{\lambda_d} 
	= 
	\int\limits_{-\omega_D}^{\omega_D}  \ud y \frac{\sqrt{1+y}}{2} 
	\int \frac{\ud \Omega}{4\pi} \frac{d_j^2(y,\Omega)}{E_j(y,\Omega)} \tanh\left[\frac{E_j(y,\Omega)}{2k_B T}\right].
\label{eq:gapT}
\ee
where $y = x^2-1$,
$\Omega$ denotes the angular variables,  
and 
$E_j(y,\Omega) \equiv \sqrt{y^2 + {|\hat{\Delta}_d|^2} d_j^2(y,\Omega)}$ is the (dimensionless) bulk quasiparticle energy. 
At zero temperature, the above gap equation can be simplified 
provided the pairing takes place within a thin shell around 
the Fermi momentum
so that $\omega_D \ll 1$, yielding
\begin{equation}~\label{eq:gap}
	\frac{1}{\lambda_d} 
	= 
	\int\frac{\ud \Omega}{4\pi}\; 
	\hat{d}^2(\Omega)
	\ln\left[
		\frac{\omega_D+\sqrt{\omega_D^2+|\hat{\Delta}_d|^2 \cdot\hat{d}^2(\Omega)}
			}{
			\hat{\Delta}_d \cdot\hat{d}(\Omega)} 
	\right], 
\end{equation}
where $\hat{d}(\Omega)$ is a purely angle dependent form-factor.
This equation can be solved analytically in the standard BCS weak-coupling approximation, 
$|\Delta_d|\ll \omega_D$, yielding 
\bea
	\Delta_{x^2-y^2}(0) &=& \Delta_{xy}(0)= 3.355\, \omega_D \exp(-5/\lambda_d);
\nonumber \\
	\Delta_{3z^2-r^2}(0) &=& 3.501\, \omega_D \exp(-5/\lambda_d). \label{eq:gap0}
\eea 
The appearance of the factor of $5$ in the exponent can be traced to the angle-average 
of the $d$-wave form-factors $\int\ud \Omega\; \hat{d}^2(\Omega)/(4 \pi) = 1/5$,
resulting in an exponential suppression of the gap compared to the well known $s$-wave 
result: $\Delta_s(T=0) \approx 2\omega_D \exp(-1/\lambda_s)$~\cite{Tinkham}.

\begin{figure}[t!]
\includegraphics[width=0.46\textwidth]{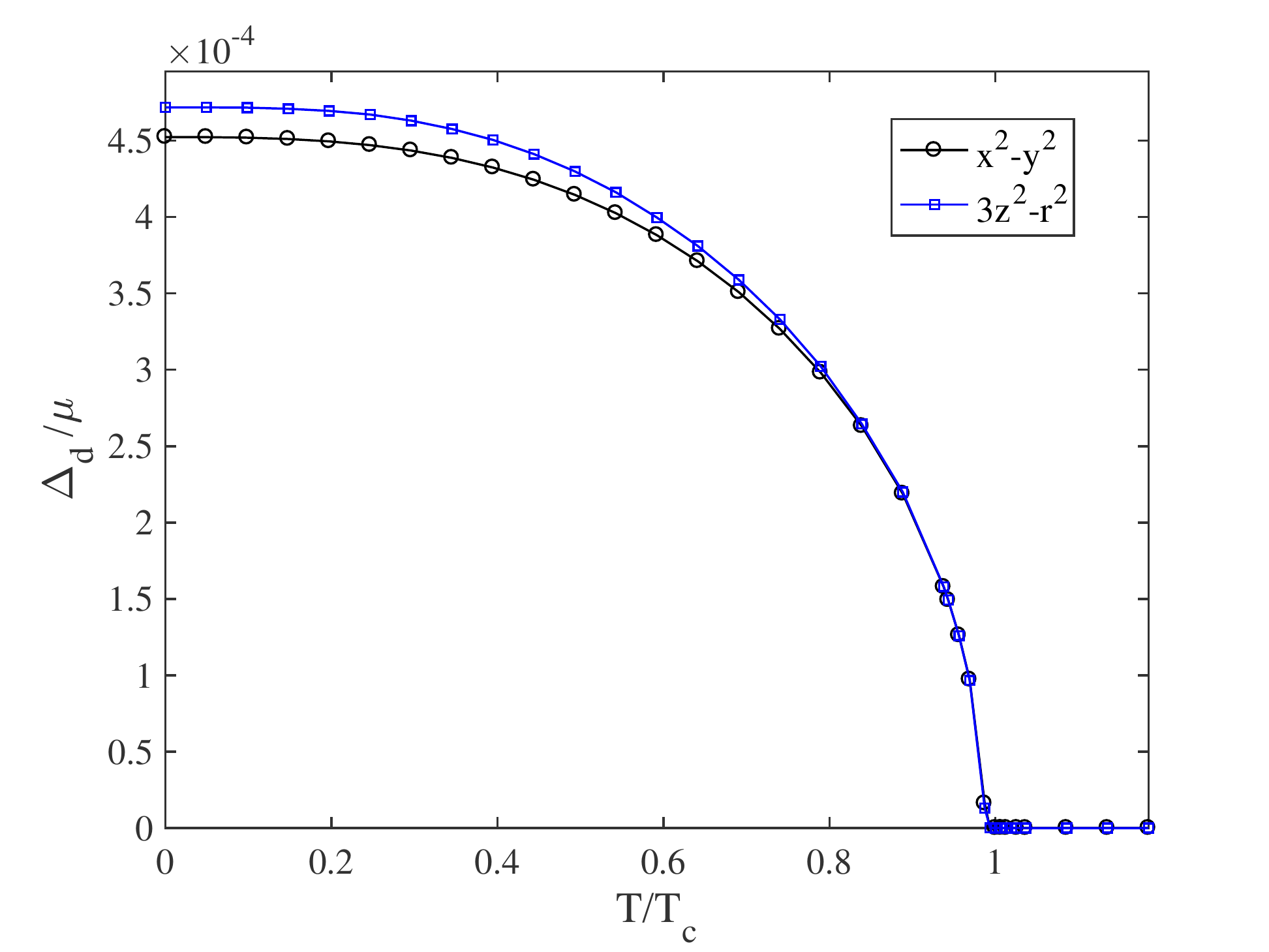}
\caption{The temperature dependence of the superconducting gap for 
(a) $d_{x^2-y^2}$ and (b) $d_{3z^2-r^2}$ pairings, 
calculated with $\lambda_d = 1$ and $\omega_D=0.02$. 
The two gaps have identical transition temperature ($T_c$) given by Eq.~(\ref{eq:Tc}), 
but are generally different at any lower temperature $T<T_c$, 
with the zero-temperature values approximately given by Eq.~(\ref{eq:gap0}).
\label{Fig--gapT}}
\end{figure} 

\begin{figure}[t!]
\includegraphics[width=0.44\textwidth]{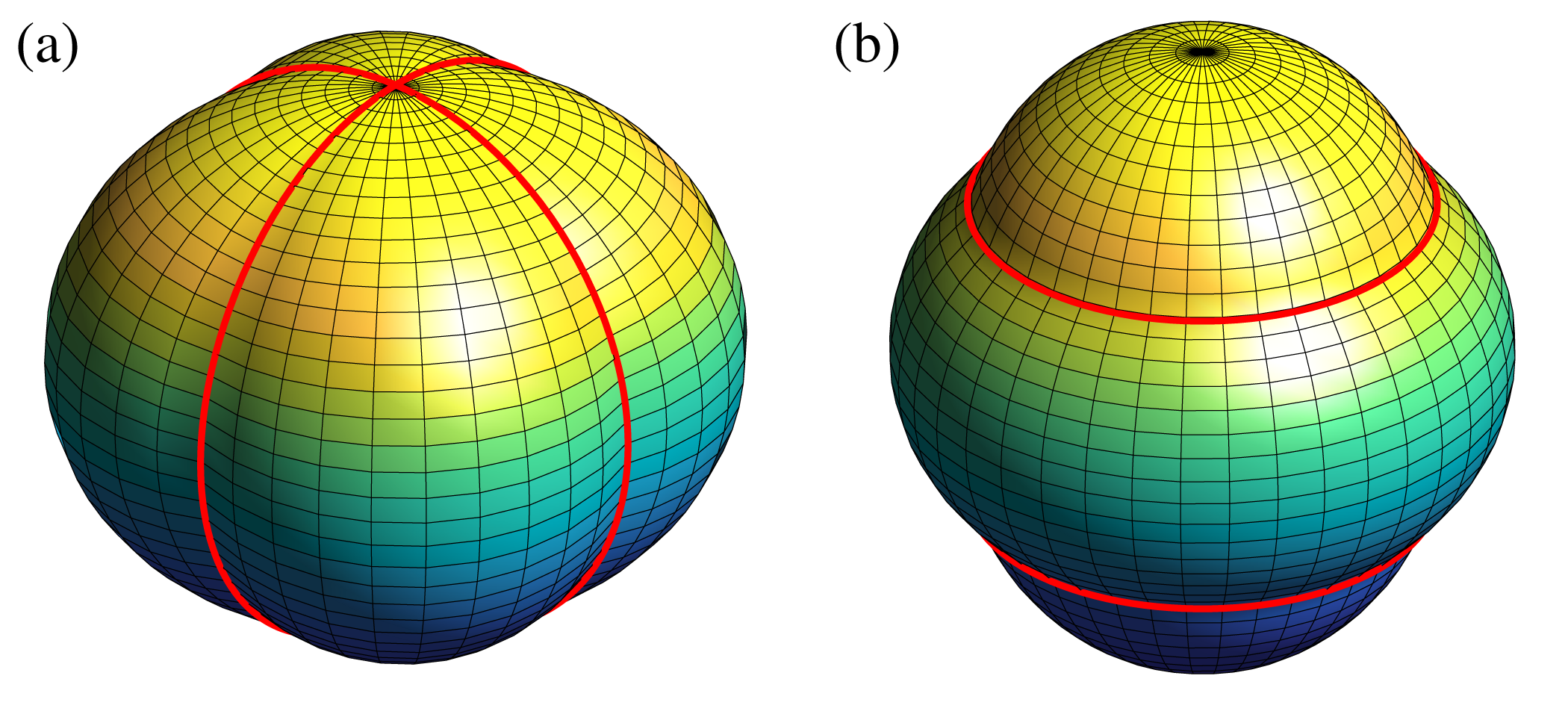}
\caption{Superconducting gap profiles for (a) $d_{x^2-y^2}$ and (b) $d_{3z^2-r^2}$ pairing on an isotropic 
Fermi sphere (realized for $m_1=m_2$), with the red curves representing the nodal lines of the gap (see Table~\ref{table_projection}). 
It is obvious that the two gap structures cannot be related by any SO(3) rotation. 
By contrast, any of the three $T_{2g}$ gaps can be obtained by an appropriate rotation of the $d_{x^2-y^2}$ gap, 
shown in panel (a).}
\label{Fig--Eg}
\end{figure}

It may appear surprising, at first glance, that the two $E_g$ $d$-wave harmonics have different values of the superconducting gap in Eq.~(\ref{eq:gap0}) at $T=0$. 
This is not a mistake, and the numerically exact solution of the gap equation Eq.~(\ref{eq:gapT}) 
leads to the identical conclusion, see Fig.~\ref{Fig--gapT}, with about $4\%$ difference between 
the zero-temperature values of the gap. 
This fact, identical to the case of $d$-wave pairing of spin-1/2 particles, is actually well documented~\cite{ueda-sigrist} but perhaps not always appreciated. What is true is that the two $E_g$ harmonics have the same $T_c$, by virtue of belonging to the same representation of the cubic point group, but the same cannot be said about the order parameters below $T_c$, as our Fig.~\ref{Fig--gapT} demonstrates.
Degeneracy lifting within the $E_g$ sector was previously discussed in Ref.~\cite{ueda-sigrist},
however necessitating consideration of the sixth order terms in the expansion of the Landau potential, valid near $T_c$. 
Here we demonstrate, perhaps more transparently, that the zero-temperature solutions of the gap equations 
(\ref{eq:gap0}) display splitting within the $E_g$ doublet. Our analysis is valid far away from $T_c$ (including at zero temperature) where the argument of Ref.~\onlinecite{ueda-sigrist} can no longer be applied. Simply put, the magnitudes of the superconducting order parameters within the $E_g$ doublet are not equal because  
of the different geometry of the nodal loops in $d_{x^2-y^2}$ 
and $d_{3z^2-r^2}$ paired states, as shown in Fig.~\ref{Fig--Eg}. Said differently, there is no way to rotate these two harmonics into each other by any SO(3) rotation (let alone by any operation of a cubic point group).
 The difference in the order parameter amplitudes 
becomes smaller on approaching $T_c$ (see Fig.~\ref{Fig--gapT}), consistent with the analysis in Ref.~\cite{ueda-sigrist}.~\footnote{This outcome can be contrasted with the scenario for a $p$-wave pairing. Let us assume that the system is spherically or SO(3) symmetric. Then each component of $p$-wave superconductor, namely $p_x$, $p_y$ and $p_z$ pairings, possesses identical transition temperature, free-energy and gap size at $T=0$, since the $p$-wave pairings transform as a ``vector" under SO(3) symmetry. By contrast, five $d$-wave pairings transform as components of a ``rank-2 tensor" under SO(3) rotation, leading to the mentioned degeneracy lifting in the free-energy and gap size at $T=0$.  }

We conclude that the two $E_g$ solutions have different gap values. 
These solutions however have the identical transition temperature $T_c$, 
protected by the  cubic symmetry. Indeed, the expression for $T_c$ follows 
from Eq.~(\ref{eq:gapT}), yielding
\be
	\frac{1}{\lambda_d} 
	= 
	\int_{-\omega_D}^{\omega_D}  \ud y \frac{(1+y)^{\frac{5}{2}}}{2y} 
	\tanh\left( \frac{y}{2k_B T_c}\right) \int \frac{\ud \Omega}{4\pi} \hat{d}_j^2(\Omega). 
\label{eq:gap-Tc}
\ee
Symmetry requires that \mbox{$\int \ud \Omega\, \hat{d}_j^2/(4\pi) = 1/5$} 
for all $d$-wave harmonics (belonging to $E_g$ and $T_{2g}$ representations), and thus all five 
$d$-wave pairings must have the identical $T_c$. 
Weak-coupling ($\omega_d \ll 1$) yields
\be
	k_B T_c = \frac{2e^\gamma}{\pi} \, \omega_D \, e^{-\frac{5}{\lambda}} \approx 1.134\, \omega_D \, e^{-\frac{5}{\lambda}}, \label{eq:Tc}
\ee
where $\gamma \approx 0.577$ is the Euler's number. It follows from the above and from Eq.~(\ref{eq:gap0}) that the ratio 
\be
\frac{\Delta_d(0)}{k_BT_c} = \left\{ 
\begin{array}{ll}
2.96 & \text{for } x^2-y^2,\, xy,\, xz,\, yz \\
3.09 & \text{for } 3z^2-r^2,
\end{array}
\right.
\ee
is non-universal, and should be contrasted with the well known result $\Delta_s(0)/k_BT_c = 1.76$ for $s$-wave 
pairing~\cite{Tinkham}.

Despite having the same transition temperature, $d_{3z^2-r^2}$ pairing will be realized as it has a higher 
(by modulus) condensation energy gain, which can be appreciated from the difference between the free energies 
in the normal (N) and superconducting (SC) states 
\be
	\left.\Delta f\right|_{T=0}
	= 
	\frac{\omega_D^2}{2} \int \frac{\ud \Omega}{4\pi} \left[\sqrt{1+\frac{\hat{\Delta}_d(0)^2}{\omega_D^2}\hat{d}_j^2(\Omega)} -1 \right]
\label{eq:free1}
\ee
in terms of the dimensionless parameters defined earlier (see Appendix~\ref{Append:free-energy} for the details of the derivation). 
Note that the above expression should \emph{not} be thought of as the Landau free energy---indeed, $\hat{\Delta}_d(0)$ 
here is not a variational parameter, but rather the self-consistent solution of the zero-temperature gap 
equation (\ref{eq:gap}). 	
Expanding the integrand in the powers of 
$(\hat{\Delta}_d/\omega_D)$, we obtain:
\bea
	f_{N} - f_{SC} 
	= \frac{|\hat{\Delta}_d|^2}{20} + \mathcal{O}(|\hat{\Delta}_d|^4).
	\label{eq:free1a}
\eea
As emphasized above, this equation expresses the well known fact that the Cooper pair condensation energy is proportional 
to the square of the superconducting order parameter. Consequently, the cubic harmonic with the largest value of $\Delta_d$, 
namely $3z^2-r^2$, will have the lowest energy, as verified by our numerical solution in Fig.~\ref{Fig--gapT}.

\subsection{${s+id}$ pairing}\label{sec:s+id}

Recall that in addition to $d$-wave pairings, the even-parity local pairing also contains an $s$-wave component that transforms under the $A_{1g}$ representation (first row in Table~\ref{table_projection}). Such solution is generically fully gapped, with the exception of  accidental nodes in an extended $s$-wave, which occur if the Fermi surface happens to cross the lines of nodes (for instance, $\Delta(\kk) = \cos(k_x)\cos(k_y)$ has nodes at $k_{x,y}=\pm \pi/2$). We exclude this latter possibility based on the fact that 
this would require a very large doping of the Luttinger semimetal in order to achieve the necessarily large $k_F$. 
For low carrier density ($k_F \to 0$) the amplitude of such an extended $s$-wave pairing also vanishes as 
we approach the band-touching points. Hence, the nucleation of extended $s$-wave pairing is energetically more expensive.

In this section, we instead investigate the possibility of a time-reversal symmetry breaking $s+id$ pairing. Such a solution necessarily involves a combination of two different irreducible representations, 
and one generically finds $\Delta_s \neq \Delta_d$ in the Bogoliubov quasiparticle dispersion:
\be
E_{s+id}(\kk)	=
	\bigg[ 
		 \xi^2_{{\bf k}}
		+ 
		|\Delta_s|^2 + |\Delta_d|^2 d^2(\kk)
	\bigg]^{1/2},
\ee
where the five 
$d$-wave form-factors $d(\kk)$ are listed in Appendix~\ref{Luttinger:details}.
It is intuitively clear that for such an $s+id$ solution to be realized, pairing strengths $\lambda_s$ and $\lambda_d$ need to be comparable: 
otherwise, a pure $s$-wave or a pure $d$-wave (more precisely, $d+id$) will dominate. 
One can therefore imagine that by tuning the ratio $r=\lambda_d/\lambda_s$, the $s+id$ solution might be realized in an intermediate parameter range.
To see whether this is indeed the case, we must solve a gap equation similar to Eq.~(\ref{eq:gap}) in Sec.~\ref{sec:d-wave} for each of the two gap components
\allowdisplaybreaks[4]
\bea
\!\!\!\!\!\frac{\hat{\Delta}_s}{\lambda_s} &\!\!=\!\!& \!\!
	\int\limits_{-\omega_D}^{\omega_D}  \!
	\frac{\ud y}{2}
	 \int \frac{\ud \Omega}{4\pi} \frac{\sqrt{1+y}\, \hat{\Delta}_s }{\sqrt{y^2 + {|\hat{\Delta}_s|^2 + |\hat{\Delta}_d|^2} d^2(y,\Omega)}},
\,\,\,\,\,\,\,\,\,
\label{eq:gap_s}\\
\!\!\!\!\!\frac{\hat{\Delta}_d}{\lambda_d} &\!\!=\!\!& \!\!
	 \int\limits_{-\omega_D}^{\omega_D}  \!
	\frac{\ud y}{2}  
	\int \frac{\ud \Omega}{4\pi} \frac{\sqrt{1+y}\, \hat{\Delta}_d\,d^2(y,\Omega)}{\sqrt{y^2 + |\hat{\Delta}_s|^2 + |\hat{\Delta}_d|^2 d_j^2(y,\Omega)}}, \label{eq:gap_d}
\eea
where $\hat{\Delta}_j = \Delta_j/\mu$ and $y=(k/k_F)^2-1$ are the dimensionless parameters, introduced in Sec.~\ref{sec:d-wave}.

\begin{figure}
\includegraphics[width=0.48\textwidth]{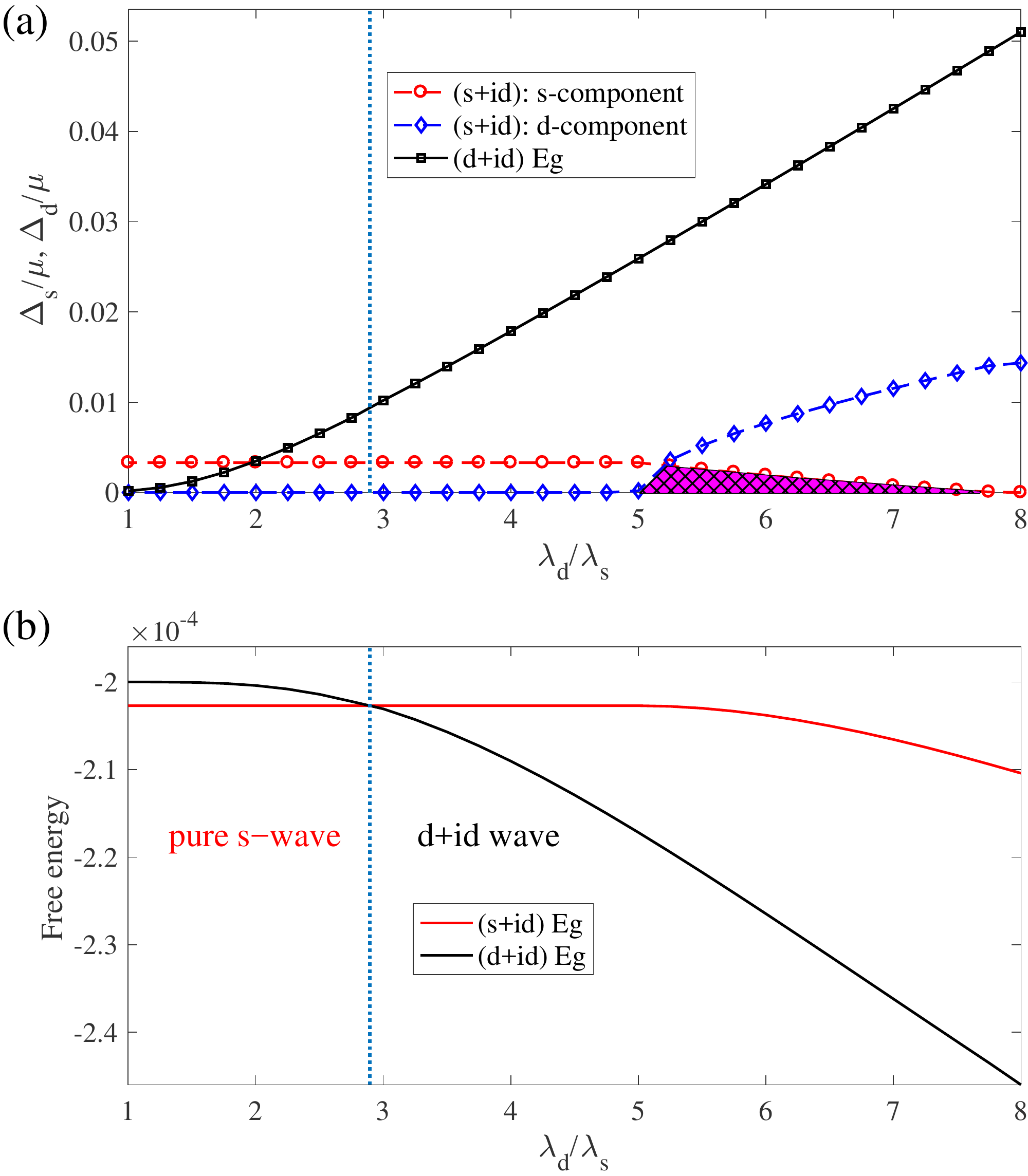}
\caption{Competition between $s+id_{x^2-y^2}$ and $d_{x^2-y^2}+id_{3z^2-r^2}$ pairings as a function of pairing strength ratio $\lambda_d/\lambda_s$, computed for a fixed $\lambda_s=0.4$. (a) Zero-temperature gap values in the $d+id$ channel (black squares), as well as the $s$-component (circles) and $d$-component (diamonds) of the $s+id$ solution. The shaded magenta region indicates the (unstable) $s+id$ phase with non-zero values of both $s$ and $d$-components. (b) The corresponding free energies in these two channels. The vertical dotted line indicates the position of a first-order phase transition from a pure $s$-wave pairing on the left, to the $(d+id)$ pairing on the right. 
} \label{Fig--s+id}
\end{figure}

The coupled system of equations (\ref{eq:gap_s})--(\ref{eq:gap_d}) does not lend itself to an analytical solution, nevertheless the solution can be obtained numerically, with the result shown in Fig.~\ref{Fig--s+id}(a). At first, for low values of $r=\lambda_d/\lambda_s$, the only solution to Eq.~(\ref{eq:gap_d}) is a trivial one: $\hat{\Delta}_d=0$, resulting in a pure $s$-wave solution. As the strength of $d$-wave pairing grows, a non-zero value of $\hat{\Delta}_d$ (blue diamonds) starts developing above a certain value $r_{c1}\approxeq 5$, and an $s+id$ solution appears in a finite region of the phase diagram $r_{c1}<r<r_{c2}$ (magenta shading in Fig.~\ref{Fig--s+id}a). Above the second critical point, $r>r_{c2}\approx 7.75$, only a trivial solution $\hat{\Delta}_s=0$ is possible, resulting in a pure $d$-wave for large coupling strength $\lambda_d$. See Appendix~\ref{Append:s+id} for details.

So far, it appears that the initial intuition was correct and that the $s+id$ solution exists in an intermediate regime of coupling strength $r_{c1}<r<r_{c2}$. However, one 
must carefully consider 
other competing orders: in particular, since we are entertaining the possibility of  time-reversal symmetry broken phases, we must also include $d+id$ order into the consideration. Allowing for the $d+id$ solution (specifically, $d_{x^2-y^2} + id_{3z^2-r^2}$ pairing, as it is energetically the most favorable state in the $E_g$ sector, see Sec.~\ref{phaselocking:eg}), we find that $\Delta_{d+id}$ rises precipitously with increasing $\lambda_d$ (black squares in Fig.~\ref{Fig--s+id}(a).
It is clear that the $d+id$ order parameter grows parametrically faster than that of the pure $d$-wave and that it should dominate for sufficiently large $\lambda_d$. This is intuitively clear since the $d+id$ solution only has point nodes, as discussed in Sec.~\ref{Weyl-SC}, and is therefore energetically more favorable than the pure $d$-wave with its line nodes. 
This argument can be made rigorous by comparing our result for $d_{x^2-y^2}$ from Eq.~(\ref{eq:gap0}) to that of $d_{x^2-y^2} + id_{3z^2-r^2}$ solution [see Sec.~\ref{phaselocking:eg} and Eq.~(\ref{eq:gap-d+id})]:
\allowdisplaybreaks[4]
\bea
\hat{\Delta}_{d} &=& 3.355 \;\omega_D \;   \exp\left(-\frac{5}{\lambda_d}\right), \nonumber \\
\hat{\Delta}_{d+id}&=& 2.705 \; \omega_D \; \exp\left(-\frac{5}{2\lambda_d}\right).
\eea 
Hence the $d+id$ order parameter is parametrically larger than the pure $d$-wave one because of the value in the exponent. 
Therefore the question is: can the $s+id$ phase survive the competition against its $d+id$ rival?

To answer this, we plot the energies of the two solutions in Fig.~\ref{Fig--s+id}(b), from which it becomes evident that $d+id$ has lower energy than pure $s$-wave or $s+id$, provided  $r>r_0\approx 2.9$ (to the right of the vertical dashed line in Fig.~\ref{Fig--s+id}). The entire region of existence of the putative $s+id$ phase lies at coupling strength $r>r_0$, and  we conclude that  the $s+id$ phase is therefore energetically unstable. Instead, there is a first-order phase transition (i.e. an energy level crossing) at $r=r_0$ from pure $s$-wave directly into the $d+id$ phase. 
The phase diagram is summarized in Figure~\ref{Fig--s+id}(b). 
Such outcome is rooted in the underlying cubic symmetry of the system, for which $E_g$ is a two-component representation, permitting a $d_{x^2-y^2}+i d_{3z^2-r^2}$ pairing to compete with (and finally win over) 
the $s+id_{x^2-y^2/3z^2-r^2}$ pairing. By contrast, in a tetragonal environment, 
{the $d_{x^2-y^2}$ pairing belongs to a single-component $B_{1g}$ representation,	} 
and consequently a $s+id_{x^2-y^2}$ 
pairing can easily be found for comparably strong $\lambda_s$ and $\lambda_d$ 
(for example, see the magenta shaded region in Fig.~\ref{Fig--s+id}). Therefore, our formalism is specifically tailored to address the competition among the pairings 
(including the local as well as the non-local ones), belonging to different \emph{multi-component} 
representations, in a cubic environment; even though we here explicitly study only the competition between 
the simplest $A_{1g}$ pairing and the $E_g$ pairings, it can be generalized to address 
the competition between $E_g$ and $T_{2g}$ pairings, as well as $A_{1g}$ and $T_{2g}$ pairings. We leave these 
exercises for future investigations.


\section{Weyl superconductors}~\label{Weyl-SC}

In this section we consider competition among the even-parity, $d$-wave pairings enumerated
above in Sec.~\ref{sec: evenp} and Table~\ref{table_projection}. The conclusions are identical for weak-pairing superconductivity
arising from a finite density Fermi surface in either the $|m_s| = 1/2$ conduction or
$|m_s| = 3/2$ valence bands, and thus, for notational simplicity, we take $m^\pm_\ast \to m_\ast$.

We explicitly demonstrate below that pairing energy minimization and the underlying cubic symmetry 
cause specific phase locking amongst various components of the $d$-wave pairings in both the $E_g$ and $T_{2g}$ sectors.
As a result, simple Weyl superconductors are expected to emerge at low temperature if the pairing strength in the $d$-wave
channel dominates.  
We first review the nodal topology of such Weyl superconductors, since we will be interested in its 
manifestation in various measurable quantities 
(such as the anomalous thermal and pseudospin Hall conductivities, discussed in Sec.~\ref{hall:thermal-spin}).

\subsection{Topology of Weyl superconductors}~\label{toplogy:definition}

Since all $d$-wave pairings are band pseudospin-singlets we can further simplify the reduced BCS Hamiltonian 
$\hat{h}_{pair}	= \hat{h}^{band}_0({\mathbf k}) + \hppbandloc$ 
[see Eqs.~(\ref{local-band}) and (\ref{kinetic-band})] as a direct sum of two $2 \times 2$ blocks 
(reflecting the pseudospin-degeneracy). 
To illustrate the nodal topology of such a system, it is now sufficient to consider one such block, which can schematically be 
written as 
\begin{equation}
	\hat{h}_{\mathbf k}= E_{\mathbf k} \; \left[ \hat{\mathbf n}_{\mathbf k} \cdot {\boldsymbol \tau} \right]. 
\end{equation}
For simple Weyl nodes $E_{\mathbf k}$ is a general linear function of all three momenta. Here $\hat{\mathbf n}_{\mathbf k}$ is a unit vector, a function of only polar ($\theta$) and
azimuthal ($\phi$) angles, and the ${\boldsymbol \tau}$s are three standard Pauli matrices 
operating on the particle-hole/Nambu index.
The monopole charge of a Weyl node ($W_n$) is then defined as 
\begin{equation}~\label{monopolecharge}
	W_n 
	= 
	\frac{1}{4 \pi} \; 
	\int^{2 \pi}_{0} d \phi_{\mathbf k} \; 
	\int^{\pi}_{0} d \theta_{\mathbf k} \;  
	\left[ 
		\hat{{\mathbf n}}_{\mathbf k} 
		\cdot 
		\left( 
			\frac{\partial \hat{{\mathbf n}}_{\mathbf k}}{\partial \theta_{\mathbf k}} 
			\times \frac{\partial \hat{{\mathbf n}}_{\mathbf k}}{\partial \phi_{\mathbf k}} 
		\right) 
	\right],
\end{equation}
which for simple Weyl nodes $W_n=\pm 1$ (see also discussion in Refs.~\cite{roy-goswami-juricic, goswami-balicas, pallab-andriy, sigrist-thomale, li-roy-dassarma, ezawa}). The Weyl node with monopole charge $+1$ ($-1$) corresponds to a
source (sink) of Abelian berry curvature of \emph{unit} strength. 

The topological nature of the BdG-Weyl quasiparticles can also be assessed from the gauge invariant Abelian 
Berry curvature (${\bf \Omega}_{n,{\bf k}}$), given by 
\begin{equation}
	\Omega_{n,{\bf k}, a}
	=
	\frac{(-1)^n}{4} 
	\; 
	\epsilon_{abc} \; 
	\hat{{\bf n}}_{\bf k}
	\cdot 
	\left[ 
		\frac{\partial \hat{{\bf n}}_{\bf k}}{\partial k_b} 
		\times  
		\frac{\partial \hat{{\bf n}}_{\bf k}}{\partial k_b} 
	\right],
\end{equation}
with $a,b,c=x,y,z$, and $n=1,2$ are the Bogoliubov band indices. 
The Berry curvature distribution in various Weyl superconducting phases 
will be displayed below.

Due to the bulk-boundary correspondence,
Weyl superconductors
(arising from time-reversal symmetry breaking $d+id$ pairings) 
support topologically protected pseudospin-degenerate Fermi arc surface states, which connect the 
projections of the Weyl nodes on the surface in the reciprocal space. 
By contrast, in the presence of a nodal-loop pairing the pseudospin degenerate surface states are 
completely flat and correspond to the images of the bulk loop~\cite{ryu-review-1}. 
A detailed analysis of these topologically protected surface Andreev bound states is left as a subject for a 
future investigation. In the absence of inversion symmetry such surface states lose the pseudospin degeneracy, 
which could be directly observed in scanning tunneling microscopy (STM) measurements.
Weyl nodes in the normal state can also be realized in a LSM via Floquet driving \cite{GhorashiFloquet}.

\subsection{$E_g$ pairing}~\label{phaselocking:eg}

\begin{figure}
\includegraphics[width=0.35\textwidth]{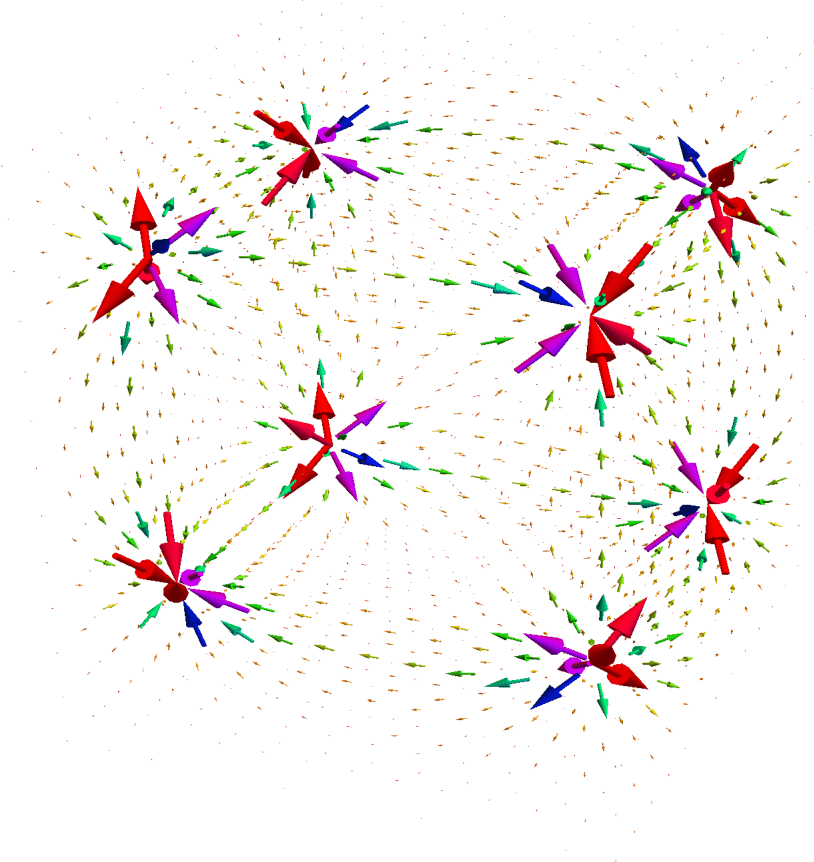}
\caption{Distribution of the Abelian Berry curvature for the nodal Weyl superconductor 
arising due to $d_{x^2-y^2}+i d_{3z^2-r^2}$ ($E_g$) pairing. 
The source (with outward arrows) and sink (with inward arrows) are symmetrically placed 
about the four possible body-diagonal directions of the spherical Fermi surface (an octupolar arrangement). 
The net Berry curvature through any high-symmetry plane therefore vanishes and the 
paired state does not support any anomalous pseudospin or thermal Hall conductivity. 
}~\label{Berry:Eg}
\end{figure}

We first investigate the effect of underlying cubic symmetry in the $E_g$ channel. 
Since $E_g$ is a two-component representation, encompassing $d_{x^2-y^2}$ and $d_{3z^2-r^2}$ pairings, 
optimal minimization of the condensation energy then enforces nucleation of $d_{x^2-y^2}+i d_{3z^2-r^2}$ pairing 
(within the framework of weak-coupling pairing). 
The matrix coefficients in the reduced BCS Hamiltonian 
\allowdisplaybreaks[4]
\begin{eqnarray}~\label{Eg:BCS_mother}
\hat{h}^{E_g}_{pair} &=& \left( \frac{{\mathbf k}^2}{2 m_\ast} - |\mu| \right) \tau_3 + \frac{\sqrt{3} |\Delta_4|}{2 k^2_F} \left( k^2_x-k^2_y \right) \tau_1 \nonumber \\
&+& \frac{|\Delta_5|}{2 k^2_F} \left( 2k^2_z -k^2_x-k^2_y \right) \tau_2,
\end{eqnarray} 
then appear as \emph{sum of the squares} in the expression for the Bogoliubov dispersion, 
where $k_F=\sqrt{2 m_\ast \mu}$ is the Fermi momentum. 
Here and in what follows, we assume that $\mu > 0$ for weak BCS superconductivity that arises from a spherical
Fermi surface in the conduction band.
The conclusions are identical for the hole-doped system. 
The time-reversal symmetry in such a paired state is spontaneously broken, and the quasiparticle spectra vanishes only at \emph{eight} isolated points on the Fermi surface $\pm k_x = \pm k_y = \pm k_z = k_F/\sqrt{3}$, 
precisely where the nodal loops for individual $d_{x^2-y^2}$ and $d_{3z^2-r^2}$ pairings cross each other 
(see Table~\ref{table_projection} and Fig.~\ref{Fig--Eg}). 
These isolated points are Weyl nodes and the phase can be considered a thermal Weyl semimetal,
since the BdG-Weyl quasiparticles carry well-defined energy (but not well-defined electric charge). 
At the cost of shedding the time-reversal symmetry, the $d_{x^2-y^2}+i d_{3z^2-r^2}$ 
paired state eliminates the line-nodes of its individual components 
(see the fifth and sixth rows of Table~\ref{table_projection}). The distribution of the Abelian Berry curvature for the $d_{x^2-y^2}+i d_{3z^2-r^2}$ Weyl superconductor is shown in Fig.~\ref{Berry:Eg}.
For possible $d_{x^2-y^2}+id_{3z^2-r^2}$ pairing in the close proximity to a Fermi surface of 
spin or pseudospin-1/2 
electrons
in heavy-fermion compounds see also Ref.~\cite{volovik-gorkov}.

\emph{Competition within $E_g$}: Given the discussion in Sec.~\ref{sec:d-wave}, we know that the two $E_g$ components 
have \emph{different} values of the superconducting gap below $T_c$. 
The basis of the $E_g$ representation obtains from two independent diagonal components of  
a symmetric, $3 \times 3$ traceless tensor [see Eq.~(\ref{tracelesstensor:definition})] and is therefore \emph{not unique}.
Indeed, dropping the normalization factors for brevity, one can choose the following basis functions 
either
\bea\label{BasisBasis}
	\text{Basis A: } 
	d_1(\kk) &\!=& \!k_x^2-k_y^2,
	\; 
	d_2(\kk)=2k_z^2-k_x^2 -k_y^2, \nonumber 
\\
	\text{Basis B: } 
	d_3(\kk)&\!=\!&k_z^2-k_x^2,\; 
	d_4(\kk)=k_z^2 -k_y^2,  
\\
	\text{Basis C: } 
	d_5(\kk)&\!=\!&2k_x^2 - k_y^2-k_z^2,\; 
	d_6(\kk)=2k_y^2 - k_x^2-k_z^2. 
	\nonumber
\eea
Note that in this subsection and in Appendix~\ref{Append:free-energy}, 
only
$d_{1,\ldots,6}$ are the above-defined $E_g$ sector harmonics.
This is a different notation than that employed everywhere else in this paper, as exemplified by Table~\ref{table_projection}. 

Notice that bases A,B,C are not independent of one another; 
for instance, $d_4 - d_3=d_1$, $d_3 + d_4=d_2$, $d_5+d_6=-d_2$ and so on. 
Nevertheless, these bases are distinct in the sense that no SO(3) rotation can convert one basis into another. 
As a result, the corresponding gaps will have different configurations of nodal loops that cannot be interconverted 
by rotations, and also different gap values! 
This raises a non-trivial question: which one of these three bases has the lowest energy, 
when we allow to form a time-reversal symmetry breaking $(d_m+id_n)$ order parameter? 
The details of the analysis 
are relegated to Appendix~\ref{Append:Eg}.
Here we quote only the final results.

The zero-temperature value of the superconducting gap
is
given by
\allowdisplaybreaks[4]
\be
 \Delta_{E_g}^{(X)}(T=0) 
	= 
	N_{X} \, \omega_d \exp\left(-\frac{5}{2\lambda_d}\right),
 	\label{eq:gap-d+id}
\ee
with $N_{A}=2.705$, $N_{B}=2.451$, $N_{C}=2.145$ 
non-universal numerical 
prefactors in the weak-coupling approximation, see
Appendix~\ref{Append:Eg}.
To the lowest order in $|\Delta|^2$, the condensation energy gain in the $(d_m+id_n)$ state is
\begin{align}
\Delta f
	\approx 
	-\frac{|\Delta_{E_g}|^2}{4} \int\frac{\ud \Omega}{4\pi}\; \sum^{n}_{j=m} \hat{d}_j ^2  
	=	-\frac{|\Delta_{E_g}|^2}{10} + \mathcal{O}(|\Delta_d|^4),
\label{eq:free-d+id}
\end{align}
and thus the solution with the largest value of the zero-temperature gap, namely 
the $(d_{x^2-y^2}+id_{3z^2-r^2})$ paired state, has the lowest energy. However, the location 
of the nodal points in any $d+id$ paired state is insensitive to the choice of basis.  

Finally, we note that the preferred pairing channel is selected by the minimum of the free energy,
independent of the chosen basis. We introduce different bases [$L \equiv A,B,C$ in Eq.~(\ref{BasisBasis})]
only because it is conventional to express (e.g.) $L_1 + i L_2$ paired states in terms of components $L_{1,2}$ 
that are each themselves basis elements, not linear combinations thereof.

\emph{Nodal topology}: We now investigate the nodal topology of the eight isolated Weyl nodes inside the $d_{x^2-y^2}+i d_{3z^2-r^2}$ paired state. 
Since the Weyl nodes are placed along eight possible $[1,1,1]$ directions, 
we introduce a rotated co-ordinate frame 
\begin{equation}~\label{momentum_rotation}
	q_x=\frac{k_x + k_y -2 k_z}{\sqrt{6}}, \;
	q_y=\frac{k_x-k_y}{\sqrt{2}}, \;
	q_z=\frac{k_x + k_y + k_z}{\sqrt{3}},
\end{equation}
keeping our focus around ${\mathbf k}=\pm (1,1,1)k_F/\sqrt{3}$. 
In this rotated co-ordinate system the Weyl nodes are located at 
${\bf q}=(0,0,\pm 1) k_F$. The reduced BCS Hamiltonian [see Eq.~(\ref{Eg:BCS_mother})] for the $d_{x^2-y^2}+i d_{3z^2-r^2}$ state then becomes
\begin{equation}
	\hat{h}^{E_g}_{pair}=\pm \left[ v_x \tau_1 q_x + v_x \tau_2 q_y + v_z \tau_3 \delta q_z \right] + {\mathcal O}(k^{-2}_F),  
\end{equation}
where $v_x = v_y = \sqrt{2} |\Delta_{E_g}| /k_F$, $v_z = k_F/m_\ast$ and $\delta q_z=q_z \pm k_F$.
Eq.~(\ref{monopolecharge}) then implies that the Weyl nodes located at 
${\mathbf k}=\pm (1,1,1)k_F/\sqrt{3}$ are characterized by monopole charge $W_n =\pm 1$ [see Eq.~(\ref{monopolecharge})]. 
We also find that the Weyl nodes located at 
${\mathbf k}=\left( -1, -1,  1 \right)k_F/\sqrt{3}$, 
$\left( -1, 1,  -1 \right)k_F/\sqrt{3}$ 
and 
$\left( 1, -1,  -1 \right)k_F/\sqrt{3}$ 
are characterized by monopole charge $W_n=+1$. 
On the other hand, the Weyl nodes located at 
${\mathbf k}=\left( 1, 1,  -1 \right)k_F/\sqrt{3}$, 
$\left( -1, 1,  1 \right)k_F/\sqrt{3}$ 
and 
$\left( 1, -1,  1 \right)k_F/\sqrt{3}$ 
have monopole charge $W_n=-1$. See also Fig.~\ref{Berry:Eg}.  
For an illustration of nodal topology of $d_{x^2-y^2}+id_{3z^2-r^2}$ paired state, also consult
Fig.~1 of Ref.~\cite{volovik-gorkov}.

\emph{DoS}: 
The nodal topology determines
the scaling of the DoS inside the 
$d_{x^2-y^2}+i d_{3z^2-r^2}$ paired state. Since the Weyl nodes bear monopole charge $W_n=\pm 1$, 
the DoS at low enough energy vanishes as $\varrho(E) \sim |E|^{2/|W_n|} \sim |E|^2$. 
Recall the DoS in the presence of a nodal line also scales as $\varrho(E) \sim |E|$. 
Therefore, by sacrificing the time-reversal symmetry the system gains condensation energy 
through power-law suppression of the DoS at low energies.

\subsection{$T_{2g}$ pairing}~\label{Phaselock:T2g}

\begin{figure}[t!]
\includegraphics[width=0.36\textwidth]{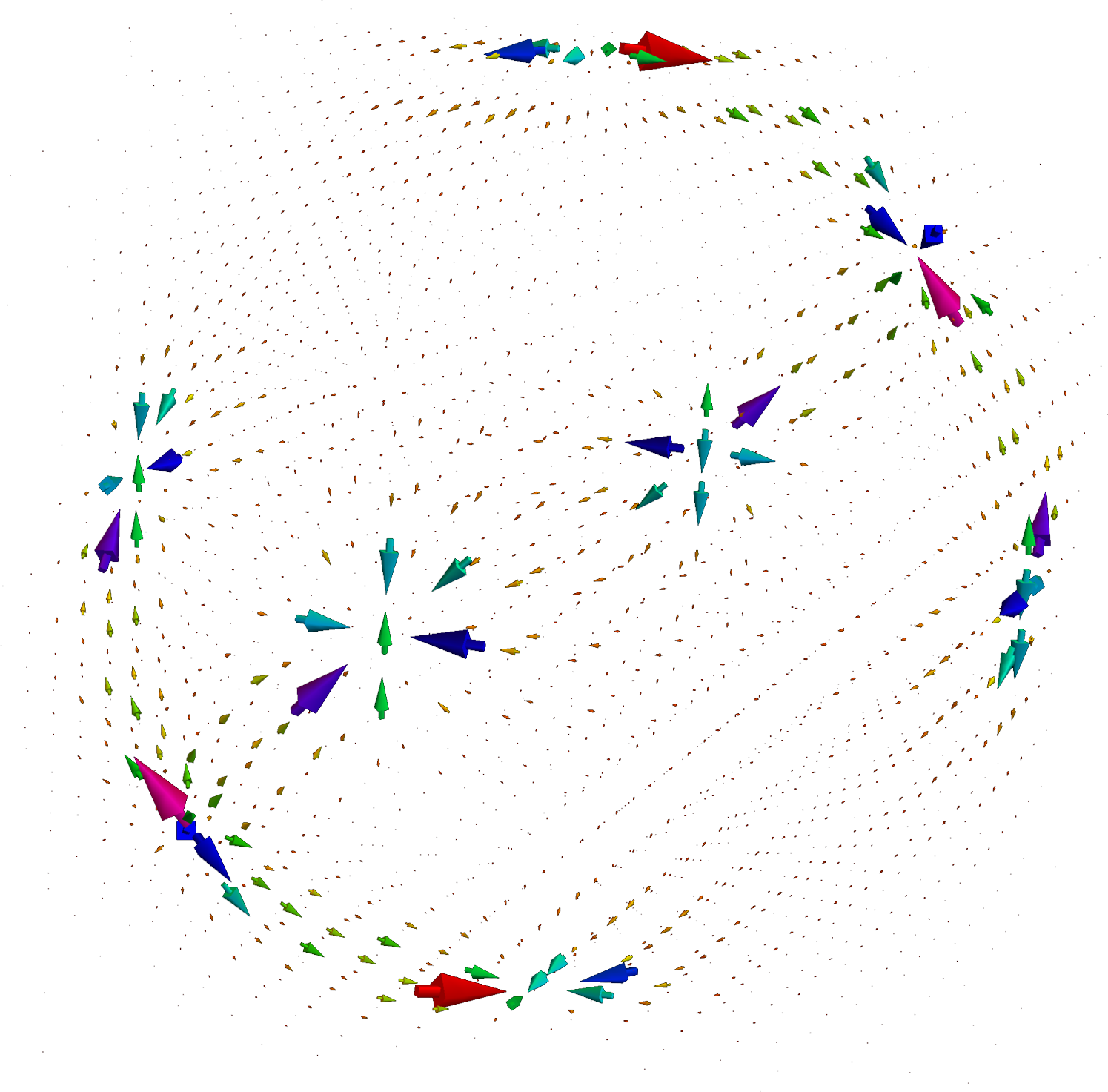}
\caption{Distribution of the Abelian Berry curvature for the nodal Weyl superconductor arising from 
a specific phase locking amongst $d_{xy}$, $d_{xz}$ and $d_{yz}$ ($T_{2g}$) pairings, given by Eq.~(\ref{root3locking}). 
Weyl node pairs are enumerated in Eq.~(\ref{gapless_t2g}).
In the figure, 
the top and bottom pair of nodes represent set $(a)$, 
left and right pair set $(b)$, 
front and back pair set $(c)$, 
and the two diagonally opposite ones 
set $(d)$. 
The net Abelian Berry curvature of such a paired state does not vanish 
through any high-symmetry plane, and concomitantly the paired state exhibits 
nontrivial anomalous pseudospin and thermal Hall effects.}~\label{Berry:T2g}
\end{figure}

Since $T_{2g}$ is a three-component representation, we denote the phases of the complex superconducting 
pairing amplitudes associated with the $d_{xy}$, $d_{xz}$ and $d_{yz}$ pairings as $\phi_{xy}$, $\phi_{xz}$ and $\phi_{yz}$, respectively. 
The nodal loops associated to each of the three pairing channels in isolation 
can only be eliminated by the choice~\footnote{
This paired state is characterized by an eight-fold degeneracy, which can be appreciated in the following way.  
Four degenerate states are realized with $\left( \phi_{xy}, \phi_{xz}, \phi_{yz} \right)=\left(0, 2\pi/3, 4\pi/3 \right)$, 
$\left(0, 2\pi/3+\pi, 4\pi/3 \right)$, $\left(0, 2\pi/3, 4\pi/3 + \pi \right)$, $\left(0, 2\pi/3 +\pi, 4\pi/3 + \pi \right)$. 
The remaining 4-fold degeneracy is achieved by $\phi_{xz} \leftrightarrow \phi_{yz}$~\cite{brydon, volovik-gorkov}, 
leaving $\phi_{xy}$ unchanged (set by the global $U(1)$ phase locking).		}	
\begin{equation}\label{root3locking}
	\left( \phi_{xy}, \phi_{xz}, \phi_{yz} \right)= \left(0, \frac{2\pi}{3}, \frac{4\pi}{3} \right),	
\end{equation}    
The resulting quasiparticle spectrum exhibits eight isolated gapless points on the Fermi surface. 
In particular, for the specific choice $\left( \phi_{xy}, \phi_{xz}, \phi_{yz} \right)= \left(0, 2\pi/3, 4\pi/3 \right)$
the reduced BCS Hamiltonian reads as 
\begin{eqnarray}
\hat{h}^{T_{2g}}_{pair} &=& \left( \frac{{\mathbf k}^2}{2 m_\ast} - |\mu| \right) \tau_3 + \frac{\Delta_{T_{2g}}}{k^2_F} \bigg[ \frac{\sqrt{3}}{4} k_z\left( k_x -k_y \right) \tau_2    \nonumber \\
&+& \frac{\sqrt{3}}{2} \left( k_x k_y -\frac{1}{2} k_z k_x -\frac{1}{2} k_z k_y \right) \tau_1   \bigg],
\end{eqnarray}         
and the energy spectrum
vanishes at
\begin{align}~\label{gapless_t2g}
	(a)= \left( 0,0, \pm 1 \right) k_F,\;\; 
	(b)=&\, \left( 0, \pm 1, 0 \right)k_F, \; 
\nonumber \\ 
	(c)= \left( \pm 1, 0, 0 \right)k_F, \;\;
	(d)=&\,\pm \left(1, 1, 1\right) k_F / \sqrt{3}.
\end{align}
Note that the pairs of Weyl nodes denoted by 
$(a)$, $(b)$ and $(c)$ are located on the three $C_{4v}$ axes, 
while the Weyl nodes $(d)$ are located on one of the four $C_{3v}$ axes. 
As we discuss below, any other phase locking amongst the three components of the 
$d$-wave pairing produces at least one nodal loop in the quasiparticle spectrum. 
Thus within the framework of a weak-coupling pairing mechanism the above phase locking is
energetically most favored. 
The distribution of the Abelian Berry curvature in the presence of this pairing is shown in Fig.~\ref{Berry:T2g}. 
Next we discuss the nodal topology of the Weyl nodes 
reported in Eq.~(\ref{gapless_t2g}).

\emph{Nodal topology}: The reduced BCS Hamiltonian in the close proximity to the Weyl nodes $(a)$ assumes the form 
\begin{equation}
	\hat{h}^{T_{2g}}_{pair,(a)}=\pm \left[ \tau_3 v_z \delta p_z -\tau_1 v_x p_x + \tau_2 v_y p_y \right], 
\end{equation}
where $\delta p_z =k_z \pm k_F$, $p_x=\frac{k_x+k_y}{\sqrt{2}}$, $p_y=\frac{k_x-k_y}{\sqrt{2}}$, $v_z = k_F / m_\ast$, 
$v_x = v_y = \sqrt{6} |\Delta_{T_{2g}}| / 4 k_F$.
Therefore, the Weyl nodes located at ${\bf k}=(0,0, \pm k_F)$ have monopole charge $W_n=\mp 1$. Following a similar analysis 
we find that the Weyl nodes located at ${\mathbf k}=(0,\pm k_F,0)$ are accompanied by monopole charge $W_n= \pm 1$, 
and those residing at ${\mathbf k}=(\pm k_F,0,0)$ have monopole charge $W_n=\mp 1$.

Following the discussion presented in Sec.~\ref{phaselocking:eg}, we can immediately come to the 
conclusion that the Weyl nodes $(d)$ [see Eq.~(\ref{gapless_t2g})] are also simple, and 
the members ${\mathbf k}=\pm (1,1,1)k_F/\sqrt{3}$ have monopole charge $W_n=\pm 1$. 
Therefore, the DoS around all eight simple Weyl nodes vanishes as $\varrho(E) \sim |E|^2$.
All eight Weyl nodes arising due to the pairing in the $T_{2g}$ sector are simple Weyl nodes with unit monopole charge, 
similar to the situation for $E_g$ pairing. However, the arrangement of these Weyl nodes on the Fermi surface are 
completely different in these two sectors 
(compare Figs.~\ref{Berry:Eg} and ~\ref{Berry:T2g}), which bears  
important consequences for 
the anomalous thermal and pseudospin Hall conductivities, see Sec.~\ref{hall:thermal-spin}.

\begin{figure}[t!]
\includegraphics[width=0.4\textwidth]{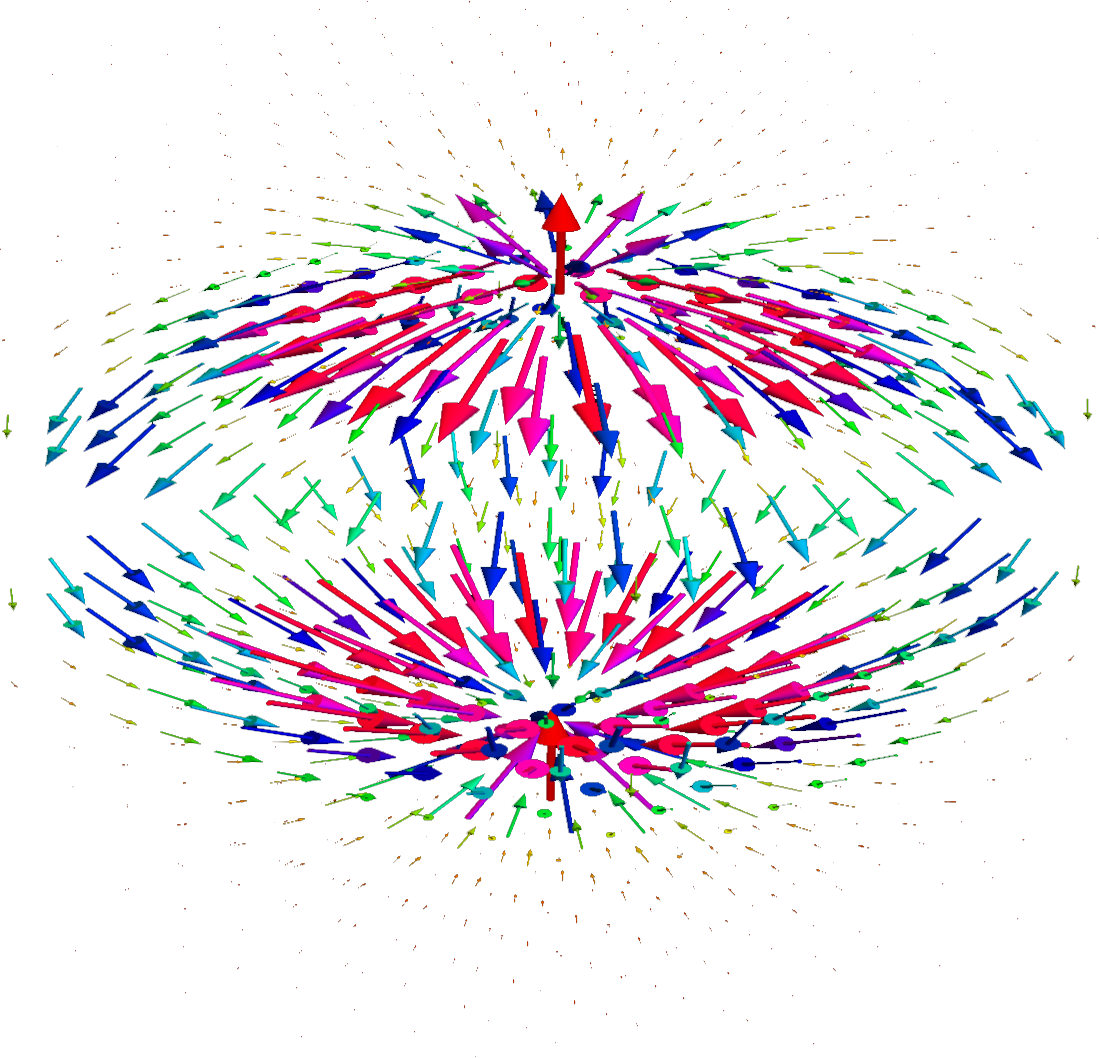}
\caption{The distribution of the Abelian Berry curvature in the presence of 
$k_z(k_x+ik_y)$ pairing on the spherical Fermi surface. 
Notice that there is no flux line along the equator of the Fermi surface where the BdG quasiparticles 
support a nodal loop. The Berry flux through the $k_x-k_y$ plane is finite and consequently 
the paired state supports non-zero anomalous pseudospin and thermal Hall conductivities in the $xy$ plane, given by Eq.~(\ref{Hall:kzkxky}). 
}~\label{Berry:T2g_kxkykz}
\end{figure}

\emph{Alternative phase locking}: We now briefly discuss a few other possible phase lockings among 
three components of $T_{2g}$pairings: 
(i) 	$\Delta_1 = \Delta_2=0$, 
(ii)  $\Delta_3 = 0$,	$\left( \phi_{xy}, \phi_{xz} \right) = (0,\pi/2)$, 
(iii) 	$\left( \phi_{xy}, \phi_{xz}, \phi_{yz} \right)=(0,0,0)$. 
The single-component paired state (i) supports two nodal loops.
The equations of these two nodal loops, along which the gap on the Fermi surface vanishes, 
are given in the fourth row of Table~\ref{table_projection}.

The reduced BCS Hamiltonian with relative phase locking (ii) in the above list reads as 
\begin{equation}
	\hat{h}^{T_{2g}}_{pair, {\rm (ii)}}
	=
	\left( \frac{{\mathbf k}^2}{2 m_\ast}-\mu \right) \tau_3 
	+ 
	\frac{\sqrt{3} \; |{\Delta_{T_{2g}}|}}{k^2_F} 
	k_z \left(k_x \tau_1 + k_y \tau_2\right).
\end{equation}
The quasiparticle spectrum in the ordered phase supports 
(a) a pair of simple Weyl nodes at the north and south poles of the Fermi surface, 
i.e.\ at ${\bf k}=(0,0, \pm k_F)$, and 
(b) a nodal loop along the equator of the Fermi surface ($k_z=0$) with radius $k_F$. 
The reduced BCS Hamiltonian around the isolated nodal points are given by 
\begin{equation}
\hat{h}^{T_{2g}, poles}_{pair, {\rm (ii)}}=\pm \left[v_z \delta k_z \tau_3 + v_x k_x \tau_1 +v_y k_y \tau_2 \right],
\end{equation}
where $\delta k_z=k_z \pm k_F$, $v_z=k_F/m_\ast$ and $v_x=v_y=\sqrt{3}|\Delta_{T_{2g}}|/k_F$. Therefore, the Weyl nodes residing at the opposite poles of the Fermi surface are characterized by the monopole charge $W_n=\pm 1$.
While the DoS due to isolated nodal points vanishes as $\varrho(E) \sim |E|^2$, 
that arising from the nodal loop scales as $\varrho(E) \sim |E|$. 
Therefore, the low-energy thermodynamic responses are dominated by the nodal loop. 
This is also commonly referred as $k_z(k_x+ik_y)$ pairing. 
The distribution of the Abelian Berry curvature on the Fermi surface in the presence of $k_z(k_x+ik_y)$ pairing is displayed in 
Fig.~\ref{Berry:T2g_kxkykz}. 
In a cubic environment this paired state is degenerate with $k_x(k_z+i k_y)$ and $k_y(k_z+i k_x)$ pairings. 
The nodal topology of these two states are same as the former one.

Finally, in the presence of phase locking appearing as (iii), 
the quasiparticle spectrum supports two isolated nodal loops, determined by 
\begin{eqnarray}
	3 {\bf k}^2-\sum_{i,j=x,y,z} k_i k_j=2k^2_F, 
\sum_{j=x,y,z}	k_j =\pm k_F.
\end{eqnarray} 
These two nodal loops are symmetrically placed around one of the body-diagonal $[1,1,1]$ directions.

Note that while (i) and (iii) produce \emph{two} nodal loops in the spectrum of the BdG quasiparticles, 
(ii) yields only one nodal loop and a pair of simple Weyl points. 
Therefore, at least within the weak coupling scenario for pairing 
(ii) appears to be energetically more favored among these three possibilities. 
Recently, a similar pairing [$k_z(k_x+ik_y)$] has also been discussed in the context of 
URu$_2$Si$_2$~\cite{goswami-balicas}, possessing tetragonal symmetry. However, in a cubic environment the 
$k_z(k_x+ik_y)$ pairing can be energetically inferior to the one discussed in Sec.~\ref{Phaselock:T2g}, 
with 
$\left( \phi_{xy}, \phi_{xz}, \phi_{yz} \right)= \left(0, 2\pi/3, 4\pi/3 \right)$ for example, since this 
pairing only produces eight isolated simple Weyl nodes on the Fermi surface, yielding $\varrho(E) \sim |E|^2$ 
and thereby causing power-law suppression of the DoS at low energies.

\begin{table*}[t!]
\begin{tabular}{|c|c|c|c|c|c|c|}
\hline
{\bf Paired State} & {\bf Nodes} & {\bf Locations} & {\bf Nodal topology} & {\bf DoS} & {\bf ASHC} & {\bf ATHC} \\
\hline \hline
$d_{xy}+i d_{x^2-y^2}$ & 2 & $\left( 0,0, \pm 1\right) k_F$ & Double Weyl ($W_n= \pm 2$) & $\varrho(E) \sim |E|$ 
	& 
	$\sigma^{s,0}_{xy}=\frac{2 \hbar}{8\pi} \frac{k_F}{\pi}$ 
	& 
	$\kappa^0_{xy}=\frac{2 \pi^2 k^2_B T}{3 h}\frac{k_F}{\pi}$ \\
\hline
$d_{xy}+i d_{3 z^2-r^2}$ & 8 
& 
	$\displaystyle{   \stackrel[]{\textstyle{\left( 0,\pm \frac{\sqrt{2}}{\sqrt{3}}, \pm \frac{1}{\sqrt{3}} \right) k_F,}}{\textstyle{\left( \pm \frac{\sqrt{2}}{\sqrt{3}}, 0, \pm \frac{1}{\sqrt{3}} \right) k_F}  }}$ 
& 
Single Weyl ($W_n= \pm 1$) & $\varrho(E) \sim |E|^2$ & $0$ & $0$ \\
\hline
	$d_{xz}+i d_{x^2-y^2}$ 
& 
	
	$\displaystyle{
		\stackrel[]{
			\phantom{\bigg[}\textstyle{2}\phantom{\bigg[}
			}{
			\textstyle{4}  
			}
		}$
&  

		$\displaystyle{
		\stackrel[]{
			\phantom{\Big[}\textstyle{\left( 0,0, \pm 1\right) k_F,}\phantom{\Big[}
			}{
			\textstyle{\left( \pm \frac{1}{\sqrt{2}},\pm \frac{1}{\sqrt{2}}, 0 \right) k_F}  
			}
		}$
	
& 
  $\displaystyle{
		\stackrel[]{
			\phantom{\bigg[}\textstyle{ \mbox{Accidental nodes}}\phantom{\bigg[}
			}{
			\textstyle{ \mbox{Single Weyl} (W_n= \pm 1)}  
			}
		}$
& 
	$\displaystyle{
		\stackrel[]{
			\phantom{\bigg[}\textstyle{\varrho(E) \sim |E|^{3/2}}\phantom{\bigg[}
			}{
			\textstyle{\varrho(E) \sim |E|^2}  
			}
		}$

& 
  	$\displaystyle{
		\stackrel[]{
			\phantom{\bigg[}\textstyle{0}\phantom{\bigg[}
			}{
			\textstyle{0}  
			}
		}$

& 
  	$\displaystyle{
		\stackrel[]{
			\phantom{\bigg[}\textstyle{0}\phantom{\bigg[}
			}{
			\textstyle{0}  
			}
		}$  \\
\hline
$d_{yz}+i d_{x^2-y^2}$
& 
	
	$\displaystyle{
		\stackrel[]{
			\phantom{\bigg[}\textstyle{2}\phantom{\bigg[}
			}{
			\textstyle{4}  
			}
		}$
&  

		$\displaystyle{
		\stackrel[]{
			\phantom{\Big[}\textstyle{\left( 0,0, \pm 1\right) k_F,}\phantom{\Big[}
			}{
			\textstyle{\left( \pm \frac{1}{\sqrt{2}},\pm \frac{1}{\sqrt{2}}, 0 \right) k_F}  
			}
		}$
	
& 
  $\displaystyle{
		\stackrel[]{
			\phantom{\bigg[}\textstyle{ \mbox{Accidental nodes}}\phantom{\bigg[}
			}{
			\textstyle{ \mbox{Single Weyl} (W_n= \pm 1)}  
			}
		}$
& 
	$\displaystyle{
		\stackrel[]{
			\phantom{\bigg[}\textstyle{\varrho(E) \sim |E|^{3/2}}\phantom{\bigg[}
			}{
			\textstyle{\varrho(E) \sim |E|^2}  
			}
		}$

& 
  	$\displaystyle{
		\stackrel[]{
			\phantom{\bigg[}\textstyle{0}\phantom{\bigg[}
			}{
			\textstyle{0}  
			}
		}$

& 
  	$\displaystyle{
		\stackrel[]{
			\phantom{\bigg[}\textstyle{0}\phantom{\bigg[}
			}{
			\textstyle{0}  
			}
		}$  \\
\hline
$d_{xz}+i d_{3z^2-r^2}$ & 4 &  $\left( 0,\pm \frac{\sqrt{2}}{\sqrt{3}}, \pm \frac{1}{\sqrt{3}} \right) k_F$ & Single Weyl ($W_n= \pm 1$) & $\varrho(E) \sim |E|^2$ & $0$ & $0$ \\
\hline 
$d_{yz}+i d_{3z^2-r^2}$ & 4 &  $\left(\pm \frac{\sqrt{2}}{\sqrt{3}}, 0, \pm \frac{1}{\sqrt{3}} \right) k_F$ & Single Weyl ($W_n= \pm 1$) & $\varrho(E) \sim |E|^2$ & $0$ & $0$ \\
\hline \hline
\end{tabular}
\caption{Weyl superconductors that obtain via $d+id$ combinations of $T_{2g}$ and $E_g$ local pairings. 
First column: Various possible (all together 6) broken time-reversal paired states resulting from the competition between $T_{2g}$ and $E_g$ pairings. 
Second column: The number of nodes in the spectrum of BdG quasiparticles inside corresponding paired state. 
Third column: The location of the gapless points on the Fermi surface. 
Fourth column: The nodal topology of the BdG-Weyl fermions. 
Fifth column: Scaling of the density of states (DoS) of the BdG quasiparticles around the nodes. 
Sixth and seventh columns: 
Anomalous (pseudo)spin Hall conductivity (ASHC) 
and 
anomalous thermal Hall conductivity (ATHC) 
in each such time-reversal symmetry breaking paired state. Here, $k_F=\sqrt{2 m_\ast \mu}$ is the Fermi momentum. 
How the various Weyl nodes arise from nodal-loop intersections is shown in 
Fig.~\ref{Fig--egt2g}.
Note that $d_{xz/yz} + i d_{x^2-y^2}$ also support a pairs of nodes at the opposite poles of 
the Fermi surface. These nodes are, however, accidental and do not bear any topological charge nor do they contribute to ASHC or ATHC. 
}~\label{egt2g:competitiontable}
\end{table*}

\subsection{Competition between $E_g$ and $T_{2g}$ pairings}~\label{EgT2g:Competition}

We now briefly discuss the competition among various $d$-wave pairings when the pairing interaction in the 
$E_g$ and $T_{2g}$ channels, respectively denoted by $g_{E_g}$ and $g_{T_{2g}}$ (say), are of comparable strength. 
Under this circumstance, two distinct possibilities can arise: 
(a) These two paired states are separated by a first-order transition with the pairings discussed in 
Sec.~\ref{phaselocking:eg} and Sec.~\ref{Phaselock:T2g}, residing on opposite sides of the discontinuous transition, 
respectively for $g_{E_g}>g_{T_{2g}}$ and $g_{T_{2g}}>g_{E_g}$, or 
(b) there can be a region, roughly when $g_{E_g} \sim g_{T_{2g}}$, where pairings belonging to these two distinct 
representation can coexist. 
Leaving aside the possibility (a), we here further elaborate on the second scenario, by restricting ourselves to a weak coupling pairing picture.

When pairing from these two channels coexists, at the cost of the time-reversal symmetry, one can minimize the number of 
gapless points on the Fermi surface (thereby causing gain in the condensation energy). Since $T_{2g}$ and $E_g$ channels 
are respectively three- and two-component representations, all together we can find \emph{six} possible time-reversal 
symmetry breaking paired phases (note these are simplest possibilities), 
shown in the first column of Table~\ref{egt2g:competitiontable}.  

Following the discussion and methodology presented earlier in this section, we realize that only 
the $d_{x^2-y^2}+id_{xy}$ paired state gives rise to double-Weyl points, with $W_n=\pm 2$, on two poles of the Fermi surface. 
The DoS of low-energy BdG quasiparticles in the presence of double-Weyl nodes goes as $\varrho(E) \sim |E|$. 
More detailed discussion on the nodal topology of the $d_{x^2-y^2}+id_{xy}$ paired state is presented in the next section. 
The rest of the pairings only support simple Weyl nodes with $W_n=\pm 1$ [see Appendix~\ref{Append:t2geg}], 
and result in $\varrho(E) \sim |E|^2$ at low energies. 

We also note that in the $d_{xz}+id_{x^2-y^2}$ and $d_{yz}+id_{x^2-y^2}$ paired states, 
besides the simple Weyl nodes in the $k_x-k_y$ plane there also exist a pair of nodes at two 
opposite poles of the Fermi surface. With the former pairing the reduced BCS Hamiltonian around the poles reads as 
\begin{eqnarray}
	H^{pole}_{d+id}= \pm v_z \delta k_z \tau_3 \pm v_x k_x - \frac{|\Delta_{T_{2g}}|}{k^2_F} k^2_y \tau_2, 
\end{eqnarray} 
where $v_z=k_F/m_\ast$, $v_x=|\Delta_{T_{2g}}|/k_F$, and $\delta k_z=k_z \pm k_F$. 
For such isolated nodes $W_n=0$. Therefore, this pair of nodes are non-topological in nature and their existence 
is purely \emph{accidental}. However, if such a node exists the DoS near the pole vanishes as $\varrho(E) \sim |E|^{3/2}$, 
and the low-energy thermodynamic responses of the $d_{xz/yz}+id_{x^2-y^2}$ states will be dominated by these accidental nodes. 
We postpone any further discussion on the competition among all six time-reversal symmetry breaking paired states 
and the nature of the ultimate ground state for a future work.

\subsection{Anomalous thermal and spin Hall conductivities}~\label{hall:thermal-spin}

One hallmark signature of spin-singlet pairing is the separation of the spin and charge degrees of freedom. 
Electric charge is carried by the superconducting condensate, a macroscopic collection of charge $2e$ spinless bosonic Cooper pairs, 
while spin is fully carried by the fermionic excitations (BdG quasiparticles) that do not carry definite electric charge. 
In particular, such spin-charge separation bears important consequences for non-$s$-wave (such as $d$-wave) singlet pairing. 
For example, in a spin-singlet $d$-wave superconductor with broken time-reversal symmetry, the BdG quasiparticles
can give rise to anomalous spin and thermal Hall conductivities.

One well-studied example is the 
$d_{x^2 - y^2} + i d_{xy}$ state, which could be germane to cuprate high-T$_c$ 
superconductors~\cite{hightc-1, hightc-2, hightc-3, hightc-4, hightc-5, hightc-6}.
A state with this symmetry is also possible in the LSM (see Table~\ref{egt2g:competitiontable}).
Recently this pairing has also been discussed in the context of URu$_2$Si$_2$~\cite{goswami-balicas} and SrPtAs~\cite{sigrist-thomale}. 
Such a paired state bears close resemblance to the integer quantum Hall effect. 
In two dimensions (where it is fully gapped), the $d_{x^2 - y^2} + i d_{xy}$ state
supports quantized spin (since spin is a conserved quantity) and thermal (since energy is conserved) Hall 
conductivities~\cite{spinthermal-0A,spinthermal-0B,spinthermal-1,spinthermal-2,spinthermal-3}.
We here do not discuss the experimental setup for the measurement of the anomalous spin or thermal Hall conductivities, which are readily available in 
the literature~\cite{spinthermal-1, spinthermal-2,spinthermal-3}. 
Instead we emphasize these two responses inside various Weyl superconductors 
that can directly probe the net Berry flux enclosed by the paired phase,
while the lack of the time-reversal symmetry can directly be probed by Faraday and Kerr rotations~\cite{kapitulnik}. 
We also note that in the absence of inversion symmetry 
(which is the situation in half-Heusler compounds) 
the notion of (pseudo)spin Hall conductivity becomes moot, 
while thermal Hall conductivity remains well-defined. 
 
Let us first pick a specific example of a Weyl superconductor, $d_{x^2-y^2}+i d_{xy}$ pairing, 
accommodating the Weyl nodes with monopole charge $W_n=\pm 2$ at ${\bf k}=(0,0, \pm k_F)$. 
The reduced BCS Hamiltonian for such a pairing in the $k_z=0$ plane is 
\allowdisplaybreaks[4]
\begin{eqnarray}~\label{did:inplane}
	\hat{h}_{d+i d}({\bf k}, k_z=0) 
	&=& 
	\bigg\{ \left( \frac{{\mathbf{k}^2_\perp}}{2 m_\ast} -\mu \right) \tau_3 + \frac{\Delta_{T_{2g}}}{k^2_F} \left(2 k_x k_y \right) \tau_1      
\nonumber \\
	&+& \frac{\Delta_{E_{g}}}{k^2_F} \left(k^2_x -k^2_y \right) \tau_2 \bigg\} \otimes \sigma_0,
\end{eqnarray}  
where ${\mathbf k}^2_\perp=k^2_x+k^2_y$, which describes a quantum anomalous thermal/spin Hall insulator, characterized by the Chern-Number $C_n=2$ in the $\left(k_x,k_y \right)$ plane. Appearance of the Pauli matrix $\sigma_0$ reflects that the band pseudospin is a good quantum number inside the paired state. Note that the pseudospin texture in the $\left( k_x,k_y \right)$ plane associated with the reduced BCS Hamiltonian in Eq.~(\ref{did:inplane}) assumes the form of a \emph{skyrmion}, and the skyrmion number is the Chern number ($C_n$). If we express the above Hamiltonian as 
$\hat{h}_{d+i d}({\bf k}, k_z=0)=E_{{\bf k}_\perp} \left[\hat{{\bf n}}_{{\bf k}_\perp} \cdot {\boldsymbol \tau}\right]$, the in-plane skyrmion number is given by 
\begin{equation}
C_n = \int \frac{d^2{\bf k}_\perp}{4 \pi} \;  \left[ \hat{{\bf n}}_{{\bf k}_\perp} \cdot \left( \frac{\partial \hat{{\bf n}}_{{\bf k}_\perp}}{\partial k_x} \times \frac{\partial \hat{{\bf n}}_{{\bf k}_\perp}}{\partial k_y} \right) \right].
\end{equation}
At $T=0$, such time-reversal symmetry breaking thermal insulator yields a \emph{quantized} spin Hall conductivity 
\begin{equation}~\label{QSH}
\sigma^{xy,0}_s=\sigma^{xy}_s(T=0)= \frac{\hbar}{8\pi} \times C_n \equiv \frac{\hbar}{4 \pi}, 
\end{equation}
in the $xy$-plane, where $\hbar/2$ is the \emph{spin-charge} and $\left( \hbar/2 \right)^2/h \equiv \hbar/(8\pi)$ is the \emph{quantum of spin Hall conductance}. The above thermal insulator also supports nonzero thermal Hall conductivity, which as $T \to 0$ is given by 
\allowdisplaybreaks[4]
\begin{equation}~\label{ThH}
\kappa^{0}_{xy}=\lim_{T \to 0} \kappa_{xy} (T) = 2 \times \frac{\pi^2 k^2_B T}{6 h} C_n = \frac{2 \pi^2 k^2_B T}{3 h}.
\end{equation}  
In the above expression addition factor of $2$ comes from the spin-degeneracy as two components of the spin projection carry heat-current in the same direction. In two dimensions the unit of anomalous spin and thermal Hall conductivities are respectively ${\rm J} {\rm s}$ and ${\rm W} {\rm K}^{-1}$. Between the spin and thermal Hall conductivity as $T \to 0$ there exists a modified Wiedemann-Franz relation, given by 
\allowdisplaybreaks[4]
\begin{equation}~\label{wiedemenfranz}
\lim_{T \to 0} \frac{\kappa_{xy} (T)/T}{\sigma^{xy,0}_s} = \frac{4 \pi^2 }{3} \left( \frac{k_B}{\hbar} \right)^2 = L_m,
\end{equation} 
where $L_m\approx 2.2731 \times 10^{23} \; {\rm K}^{-2} {\rm s}^{-2}$ is the modified Lorentz number.

Note that the three-dimensional $d_{x^2-y^2}+i d_{xy}$ Weyl superconductor can be envisioned as stacking (in the momentum space) of corresponding two-dimensional 
class C spin quantum Hall Chern insulators 
[described by Eq.~(\ref{did:inplane})] along the $k_z$ direction within the range $ -k_F \leq k_z \leq k_F$. The interlayer tunneling is captured by $(k^2_z/2m)\tau_3 \sigma_0$. Concomitantly, the contribution to the anomalous spin and thermal Hall conductivity from each such layer is respectively given by Eq.~(\ref{QSH}) and Eq.~(\ref{ThH}). Therefore, the anomalous spin and thermal Hall conductivities as $T \to 0$ of a three dimensional $d_{x^2-y^2}+i d_{xy}$ paired state are respectively given by 
\allowdisplaybreaks[4]
\begin{eqnarray}
\sigma^{xy,0}_{s,3D} &=& \sigma^{xy,0}_s \; \int^{k_F}_{-k_F} \; \frac{dk_z}{2 \pi} = \frac{\hbar}{4 \pi} \times \left( \frac{k_F}{\pi} \right),~\label{QSH:3D} \\
\kappa^{0}_{xy,3D} &=& \kappa^{0}_{xy} \int^{k_F}_{-k_F} \; \frac{dk_z}{2 \pi}= \frac{2 \pi^2 k^2_B T}{3 h} \times \left( \frac{k_F}{\pi} \right).~\label{ThH:3D}
\end{eqnarray}
In three dimensions the unit of anomalous spin and thermal Hall conductivities are respectively ${\rm J} {\rm s} {\rm m}^{-1}$ and ${\rm W} {\rm K}^{-1} {\rm m}^{-1}$. Also note that the two double-Weyl nodes located at ${\bf k}=(0,0,\pm k_F)$ acts as source and sink of Abelian Berry curvature in the reciprocal space, and the $(k_x,k_y)$ plane encloses quantized Berry flux. The anomalous spin Hall conductivity [and thus also the anomalous thermal Hall conductivity, tied with the spin Hall conductivity via the modified Wiedemann-Franz relation, see Eq.~(\ref{wiedemenfranz})] is directly proportional to the enclosed Berry flux. 
Upon unveiling the topological source of anomalous spin and thermal Hall conductivities in a Weyl superconductor, we can now proceed with the estimation of these two quantities in the $E_{g}$ and $T_{2g}$ paired states.

\subsubsection{ \bf Anomalous responses for $E_g$ pairing}

We first focus on the $E_g$ channel. Recall that the $d_{x^2-y^2}+id_{3z^2-r^2}$ paired state supports eight simple Weyl nodes with $W_n=\pm 1$. From the arrangement of the source and sink of the Abelian Berry curvature discussed in Sec.~\ref{phaselocking:eg}, we immediately come to the conclusion that the net Berry flux passing through any high symmetry plane is precisely \emph{zero} (see Fig.~\ref{Berry:Eg}). Therefore, the $d_{x^2-y^2}+id_{3z^2-r^2}$ paired state, despite possessing Weyl nodes, gives rise to net \emph{zero} anomalous spin or thermal Hall conductivity. Qualitatively, this situation is similar to the all-in all-out ordered phase in the presence of sufficiently strong repulsive electronic interactions~\cite{goswami-roy-dassarma}.

\subsubsection{ \bf Anomalous responses for $T_{2g}$ pairing}

In the $T_{2g}$ paired state, with a specific phase locking $\left( \phi_{xy}, \phi_{xz}, \phi_{yz} \right)=\left(0, 2\pi/3, 4\pi/3 \right)$, shown in Sec.~\ref{Phaselock:T2g}, the low-temperature phase also supports eight simple Weyl nodes [see Eq.~(\ref{gapless_t2g})] with monopole charge $W_n=\pm 1$. The topology of each such nodal point has been discussed in details in Sec.~\ref{Phaselock:T2g} and the distribution of the Abelian Berry curvature is depicted in Fig.~\ref{Berry:T2g}. Notice that even through time-reversal symmetry breaking $E_g$ and $T_{2g}$ pairings supports eight simple Weyl nodes, their location and distribution of the Berry flux in various high symmetry planes are completely different (compare Figs.~\ref{Berry:Eg} and \ref{Berry:T2g}). 
Consequently, the anomalous spin and thermal Hall conductivity in the $T_{2g}$ paired state are distinct from its counterpart in the $E_g$ channel. For concreteness, we here focus on these two responses in the $xy$ plane and a plane perpendicular to a $[1,1,1]$ direction.

For anomalous spin and thermal Hall conductivity in the $xy$ plane the Weyl nodes denoted as $(b)$ and $(c)$ do not contribute and contributions come only from the two pairs of Weyl nodes identified as $(a)$ and $(d)$ in Eq.~(\ref{gapless_t2g}). 
After carefully accounting for the enclosed Berry flux we find the anomalous spin and thermal Hall conductivities in the $xy$ plane are respectively given by 
\begin{equation}
\sigma^{xy,0}_{s,3D} = \frac{\hbar k_F}{8 \pi^2} N_{xy}, \:
\kappa^{0}_{xy,3D} = \frac{\pi^2 k^2_B T k_F}{3 h \pi} N_{xy},
\end{equation}
where $N_{xy}=1-1/\sqrt{3}$. Following the same spirit, we find that these two quantities in the $yz$ plane are identical to the above expressions, while those in the $xz$ plane is obtained by replacing $N_{xy} \to N_{yz}=1+1/\sqrt{3}$.

By contrast, in a plane perpendicular to the $[1,1,1]$ direction all four pairs of Weyl nodes contribute to anomalous spin and thermal Hall conductivities, yielding 
\begin{eqnarray}
	\sigma^{[111],0}_{s,3D} = \frac{2}{3} \times \frac{\hbar k_F}{8 \pi^2}, \:
	\kappa^{0}_{[111],3D} = \frac{2}{3} \times \frac{\pi^2 k^2_B T K_F}{3 h \pi}.\,\, 
	\end{eqnarray}
Note that anomalous spin and thermal Hall conductivities are also finite along three other body diagonals.

Recall that the $k_z(k_x+ik_y)$ paired state supports only a single pair of simple Weyl nodes at two poles of the Fermi surface. 
Consequently, a net non-zero quantized Berry flux is enclosed by the $k_x-k_y$ plane, yielding 
\begin{equation}~\label{Hall:kzkxky}
	\sigma^{xy,0}_{s,3D}=\frac{\hbar}{8 \pi} \times \left( \frac{k_F}{\pi} \right), \:
	\kappa^0_{xy,3D}=\frac{\pi^2 k^2_B T}{3 h} \times \left( \frac{k_F}{\pi} \right).
\end{equation}
Similarly, the two other degenerate paired states $k_x(k_z+ik_y)$ and $k_y(k_z+ik_x)$ support anomalous spin and thermal Hall 
conductivities of equal magnitude, but respectively in the $k_z-k_y$ and $k_z-k_x$ planes.

Following the same set of arguments we find that all five time-reversal odd paired states, resulting from the competition 
between $T_{2g}$ and $E_g$, yield net zero anomalous spin and thermal Hall conductivities, apart from the $d_{x^2-y^2}+id_{xy}$ phase, as shown in Table~\ref{egt2g:competitiontable}.


\section{External strain and $s+d$ pairing \label{externalstrain}}

\begin{figure}[t!]
\subfigure[]{
\includegraphics[width=4cm,height=3.0cm]{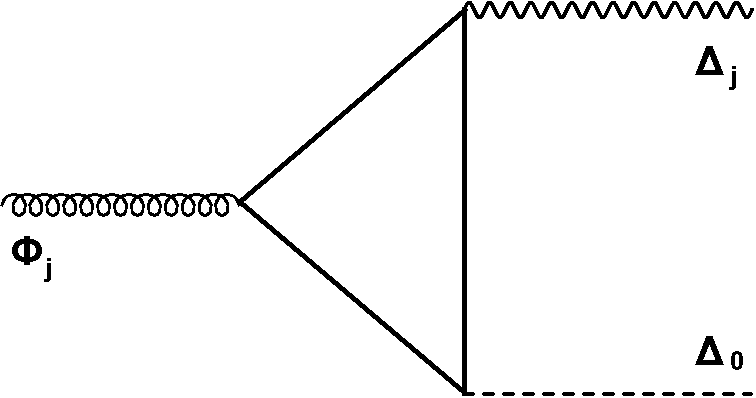}
\label{triangle_Landau}
}
\subfigure[]{
\includegraphics[width=4cm,height=4cm]{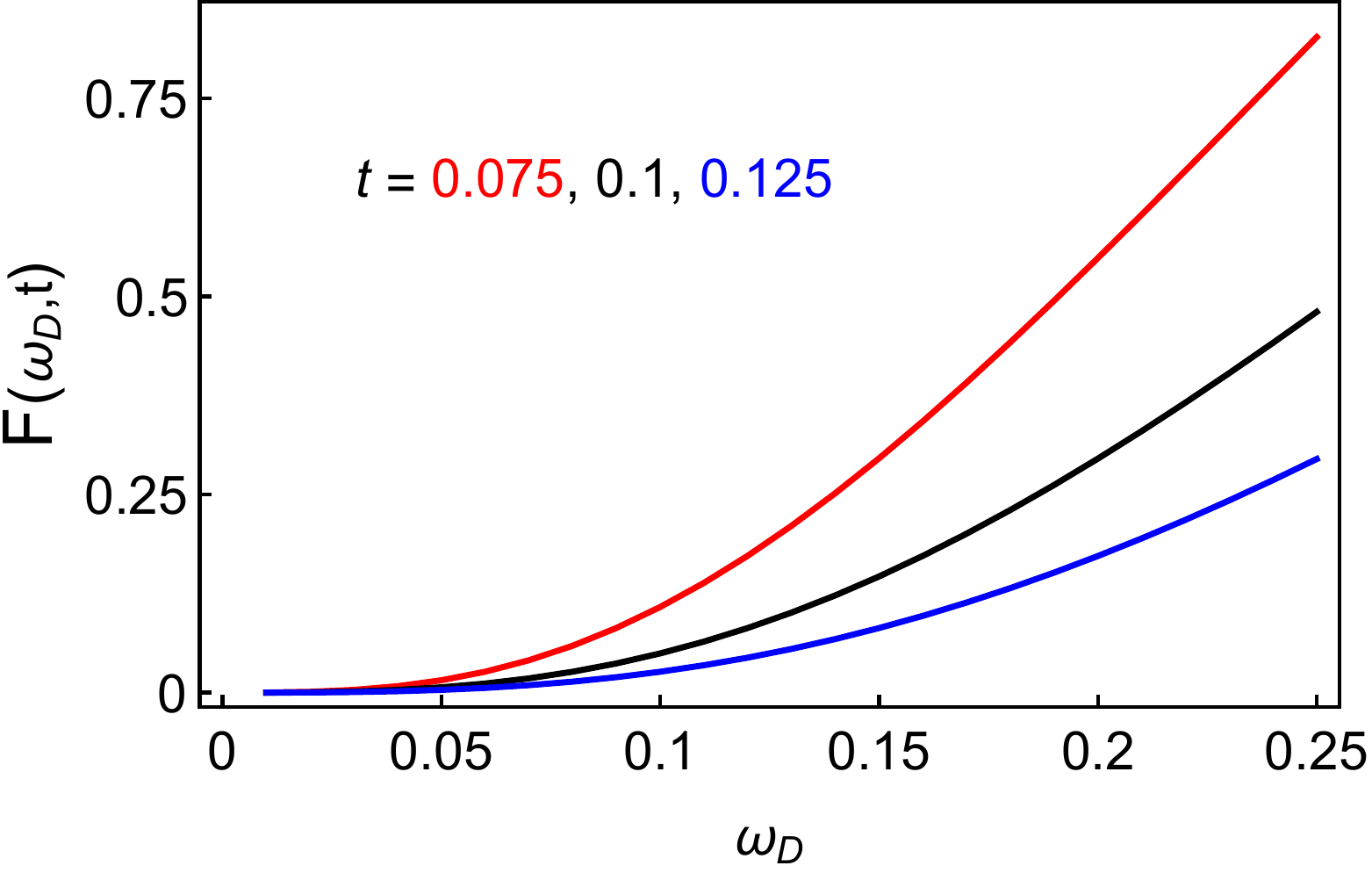}
\label{strain_Function}
}
\caption{(a) The Feynman diagram contributing to the Landau free energy 
$f_{str} \sim \Phi_j \Delta_{j} \Delta_0$, capturing the non-trivial coupling amongst 
the $s$-wave and $d$-wave pairings, and an external strain. 
Here solid lines represent fermions, wavy (dashed) lines $d$ ($s$)-wave pairing, 
and the spiral line the strain field. 
(b) Scaling of the universal function 
$F\left( \omega_D, t \right)$ on $\omega_D$ for various fixed values of $t$ (quoted in the figure), 
appearing in Eq.~(\ref{strain_function}), where $\omega_D=\Omega_D/E_F$, $t=k_B T/E_F$, 
$\Omega_D$ is the Debye frequency, $E_F$ is the Fermi energy, and $T$ is the temperature. 
}~\label{Fig:strain-together}
\end{figure}

We now discuss the effects of external strain on the paired states. 
Generic external strain in a Luttinger semimetal can be captured by the Hamiltonian 
\begin{equation}
	\hat{h}_{str}= \Phi_1 \; \Gamma_1 +  \Phi_2 \; \Gamma_2 +  \Phi_3 \; \Gamma_3 +  \Phi_4 \; \Gamma_4 +  \Phi_5 \; \Gamma_5,
\end{equation}
where $\Phi_j$ (for $j=1,\ldots,5$) represents the strength of the strain. 
\emph{Since we are interested in the effects of external strain on the paired state that only exists in the close proximity 
to the Fermi surface, we also project the above five strain operators onto the Fermi surface. 
We assume that the external strain is too weak to significantly mix the valence and conduction bands.} 
In the proximity to the Fermi surface the effects of generic strain are then encoded in 
\begin{eqnarray}
	\hat{h}^{FS}_{str} \equiv \sum^{5}_{j=1} \Phi_j \; \hat{d}_j \; \left( \tau_3 \sigma_0 \right),		
\end{eqnarray} 
where $\tau_3$ is the diagonal particle-hole (Nambu space) matrix and the $\hat{d}_j$s are 
defined in Appendix~\ref{Luttinger:details} [Eq.~(\ref{d-hats})]. 
Note that external strain does not couple with the spin degrees of freedom and preserves 
time-reversal and inversion symmetries, but breaks the cubic symmetry.

Since each component of $d$-wave pairing breaks the cubic symmetry, 
nucleation of any such pairing causes a small lattice distortion or electronic nematicity. 
In experiment, the onset of such nematicity can be probed from the measurement of 
the divergent nematic susceptibility around the transition temperature ($T_c$) 
(see for example Refs.~\cite{ian-fisher, kivelson-review}). 
Externally applied strain can directly couple with the appropriate 
$d$-wave pairing (depending on the direction of the applied strain), and in that way can be 
conducive for the nucleation of a specific component of this pairing. 
In other words, strain couples with $d$-wave pairing as an external field. 
In particular, an externally applied strain induces nontrivial coupling 
between $s$-wave and $d$-wave pairings and such coupling enters the 
expression for Landau free energy as 
$f_{str} \sim \Phi_j \Delta_{j} \Delta_0$. 
Here the index $j$ corresponds to a particular 
component of 
external strain/$d$-wave pairing, bearing the same symmetry,
and $\Delta_0$ is the order parameter for $s$-wave pairing. 
To gain quantitative estimation of such non-trivial coupling 
we compute the \emph{triangle diagram} shown in Fig.~\ref{triangle_Landau}.{\footnote{For discussion on the 
coupling amongst various magnetic, namely the all-in all-out and itinerant spin-ice, 
orders with an external strain, see Ref.~\cite{goswami-roy-dassarma}.}
\footnote{Notice that coupling between $d$-wave and $s$-wave pairings with electronic nematicity 
or external strain relies solely on the symmetry of the LSM of spin-3/2 quasiparticles. Such coupling is non-trivial if we
compute the term $f_{str}$ from the full band structure of the doped LSM, as shown in Appendix~\ref{strain:LSMfull} (see Fig.~\ref{strain_fullband}).}

The contribution of the triangle diagram to the Landau free energy is 
\begin{eqnarray}
	&& 
	f_{str} 
	= 
	-
	\Phi_j \Delta^{\mu}_l \Delta^{\nu}_0  
	\: 
	\frac{1}{\beta} \sum^{n=\infty}_{n=-\infty} 
	\; 
	\int 
	\; 
	\frac{d^3 {\mathbf k}}{(2 \pi)^3} 
	\:\: 
	{\mathbf {Tr}} \bigg[ \left(\tau_\mu \hat{d}_l \right) 
\nonumber \\
	&\times& 
	G\left(i\omega_n, {\mathbf k} \right) 
	\,
	\left(\tau_3 \hat{d}_j \right) 
	G\left(i\omega_n, {\mathbf k} \right) 
	\, 
	\left(\tau_\nu \right) 
	\, 
	G\left(i\omega_n, {\mathbf k} \right) \bigg], 
\end{eqnarray}     
where $\beta=\left( k_B T\right)^{-1}$ is the inverse temperature 
and $\omega_n=\beta^{-1}(2 n+1)\pi$ is the fermionic Matsubara frequency. 
In the above expression $\mu, \nu=1,2$. To test whether the coexistence of $s$ and $d$ wave pairing breaks time-reversal symmetry or not we have introduced the superscript $\mu, \nu$ to the pairing amplitudes 
$\Delta_l$ and $\Delta_0$, 
 respectively for these two channels.	 
Specifically, non-zero ${\mathbf {Tr}}$ for (i) $\mu=\nu$ corresponds to time-reversal symmetry preserving $s+d$ pairing, (ii) $\mu \neq \nu$ implies onset of time-reversal symmetry breaking $s+id$ pairing, due to an external strain. The minus ($-$) sign in the above expression comes from the fermion bubble. We here assume that all bosonic fields [see Fig.~~\ref{triangle_Landau}] are carrying zero external momentum and frequency, yielding the leading order contribution to the Landau potential. The fermionic Greens function is  
\begin{equation}
	G\left(i\omega_n, {\mathbf k} \right)
	=
	-
	\frac{i \omega_n + \tau_3 \xi_{\mathbf k}}{\omega^2_n + \xi^2_{\mathbf k}}, 
	\quad \xi_{\mathbf k} = \frac{|{\mathbf k}|^2}{2 m}-\mu.
\end{equation} 
We find that $f_{str} \propto {\mathbf {Tr}} \left[ \tau_\mu \tau_\nu \right] = 2 \delta_{\mu \nu}$. 
Hence, \emph{external strain supports a time-reversal-symmetry preserving combination of $s$-wave and $d$-wave pairings}. 
For strain assisted time-reversal symmetry breaking superconductivity, see Ref.~\cite{roy-juricic}. 
After the ${\mathbf {Tr}}$ algebra we arrive at the following expression 
\begin{eqnarray}
	f_{str} 
	&=& 
	2 \delta_{\mu \nu} \, 
	\frac{\Phi_j \Delta^{\mu}_l \Delta^{\nu}_0}{\beta} 
	\sum^{n=\infty}_{n=-\infty} 
	\int \, \frac{d^3 {\mathbf k}}{(2 \pi)^3} 
	\,
	\frac{\xi_{\mathbf k} \left(\hat{d}_l \; \hat{d}_j \right)}{\left[ \omega^2_n + \xi^2_{\mathbf k} \right]^2} 
\nonumber \\
	&=& 
	\frac{1}{2} \delta_{\mu \nu} \, \delta_{j,l} \, 
	\frac{\Phi_j \Delta^{\mu,l}_d \Delta^{\nu}_s}{40 \pi^2} 
	\, 
	\int dk 
	\, 
	\frac{k^2}{\xi^2_{\mathbf k}} 
\nonumber \\
	&\times& 
	\sech^{2}\left( \frac{\xi_{\mathbf k}}{2 k_B T}\right) 
	\, 
	\left[ 
		\sinh \left( \frac{\xi_{\mathbf k}}{ k_B T} \right) - \frac{\xi_{\mathbf k}}{ k_B T}
	\right]. 
\end{eqnarray}
In the final expression the Kronecker delta $\delta_{j,l}$ arises from the integral over the solid angle in three dimensions. 
This delta function indicates that the external strain and the $d$-wave pairing must break the cubic symmetry in the exact same way, 
such that $f_{str}$ is ultimately an $A_{1g}$ quantity. 
The final integral over momentum will be performed in the close proximity to the Fermi surface. 
We then arrive at the final expression 
\begin{eqnarray}~\label{strain_function}
	&&
	f_{str} 
	= 
	\left[ \delta_{\mu \nu} \; \delta_{j,l} \; \Phi_j \Delta^{\mu}_l \Delta^{\nu}_0 \right] 
	\; 
	\frac{\varrho\left(E_F\right)}{40 \pi^2 E_F} 
	\int^{\sqrt{1+\frac{\Omega_D}{E_F}}}_{\sqrt{1-\frac{\Omega_D}{E_F}}} \frac{x^2}{\left( x^2-1\right)^2} 
\nonumber \\
	&\times& 
	\sech^{2} \left( \frac{x^2-1}{2 t}\right) 
	\, 
	\left[ \sinh \left( \frac{x^2-1}{t} \right) - \frac{x^2-1}{t}\right] dx 
\nonumber \\ 
	&=& \left[ \delta_{\mu \nu} \; \delta_{j,l} \; \Phi_j \Delta^{\mu}_l \Delta^{\nu}_0 \right] \, 
	\frac{\varrho\left(E_F\right)}{40 \pi^2 E_F} \, 
	F\left( \omega_D, t \right),
\end{eqnarray}
where $\varrho\left(E_F \right)$ is the DoS at Fermi energy $E_F$, 
$\Omega_D$ is the Debye frequency, $\omega_D=\Omega_D/E_F$, and $t=k_B T/E_F$. 
The functional dependence of $F(x,y)$ is displayed in Fig.~\ref{strain_Function}. 
Next we discuss some specific examples when external strain is applied along certain 
high symmetry directions.

\emph{Strain along $[0,0,1]$}: First, we consider a situation when the external strain is applied along one of the $C_{4v}$ axes. 
For the sake of simplicity we consider the external strain to be applied along the $\hat{z}$ direction. 
Such strain can only couple with $d_{3z^2 - r^2}$ pairing. Thus, a strain along $\hat{z}$ direction results 
in an $s+d_{3z^2-r^2}$ paired state, a time-reversal symmetry preserving combination of $s$-wave and $E_g$ pairings.

\emph{Strain along $[1,1,1]$}: Next we consider a situation when the external strain is applied along one of the body diagonal or $[1,1,1]$ directions (one of the $C_{3v}$ axes). The coupling between such strain and the $d$-wave pairings can be appreciated most conveniently 
if we rotate the reference coordinate according to Eq.~(\ref{momentum_rotation}). In the rotated basis strain is applied along the $\hat{z}$ direction (now aligned along the body diagonal). After performing the same transformation for all $d$-wave pairings, 
augmented by the argument we presented above, we realize that only $d_{xy} + d_{yz} + d_{xz}$ 
pairing directly couples with the $[1,1,1]$ strain. Thus, strain along $[1,1,1]$ direction results in 
an $s+d_{xy}+d_{yz}+d_{xz}$ paired state, a time-reversal symmetry preserving combination of $s$-wave and $T_{2g}$ pairings.

\emph{Strain along $[1,1,0]$}: Finally, we discuss the effect of an in-plane external strain, applied along $[1,1,0]$ direction. 
Following the same set of arguments we conclude that when the strain is applied along the $[1,1,0]$, 
it directly couples with $d_{xy} + d_{3z^2-r^2}$ pairing. 
Thus, an external strain along $[1,1,0]$ direction is conducive to the formation of 
an $s + d_{xy} + d_{3z^2-r^2}$ paired state, a time-reversal symmetry preserving combination of $s$-wave, $T_{2g}$ and $E_g$ pairings.

We conclude that by applied strain along different directions one can engineer various time-reversal symmetry preserving 
combinations of $s$- and $d$-wave pairings. This mechanism can in particular be useful to induce exotic paired states 
in weakly correlated materials, such as HgTe and gray tin, which possibly can only accommodate phonon-driven $s$-wave pairing 
in the absence of strain.
The above outcome can also be stated in a slightly different words as follows. Anytime a $d$-wave pairing
nucleates in a Luttinger metal, it immediately causes a lattice distortion or nematicity. 
Consequently, any $d$-wave pairing will always be accompanied by an \emph{induced} $s$-wave component, which, as discussed in 
Sec.~\ref{experiment}, may bear important consequences in experiments.
In Appendix~\ref{strain:LSMfull} we show that induced s-wave component due to lattice distortion is indeed \emph{finite}, by minimizing a phenomenological Landau potential, where symmetry allowed terms up to the quartic order are taken into account. 
Notice existence of a small $s$-wave component does not break any additional symmetry deep inside the $d$-wave (or $d+id$-type Weyl) paired state. Thus, a non-trivial coupling between $d$-wave and $s$-wave pairings and the lattice distortion does not affect flat-band (for pure $d$-wave pairing) or Fermi arc (for $d+id$-type pairing) surface states as long as the pairing interaction in the $d$-wave channel dominates. 
Appearance of induced s-wave component plays an important role in the interpretation of the penetration depth data in YPtBi, discussed in Sec.~\ref{experiment}.


\section{Effects of impurities on B$\mathbf{d}$G-Weyl quasiparticles \label{Weyl:disorder}}

\begin{figure}[t!]
\includegraphics[width=0.35\textwidth]{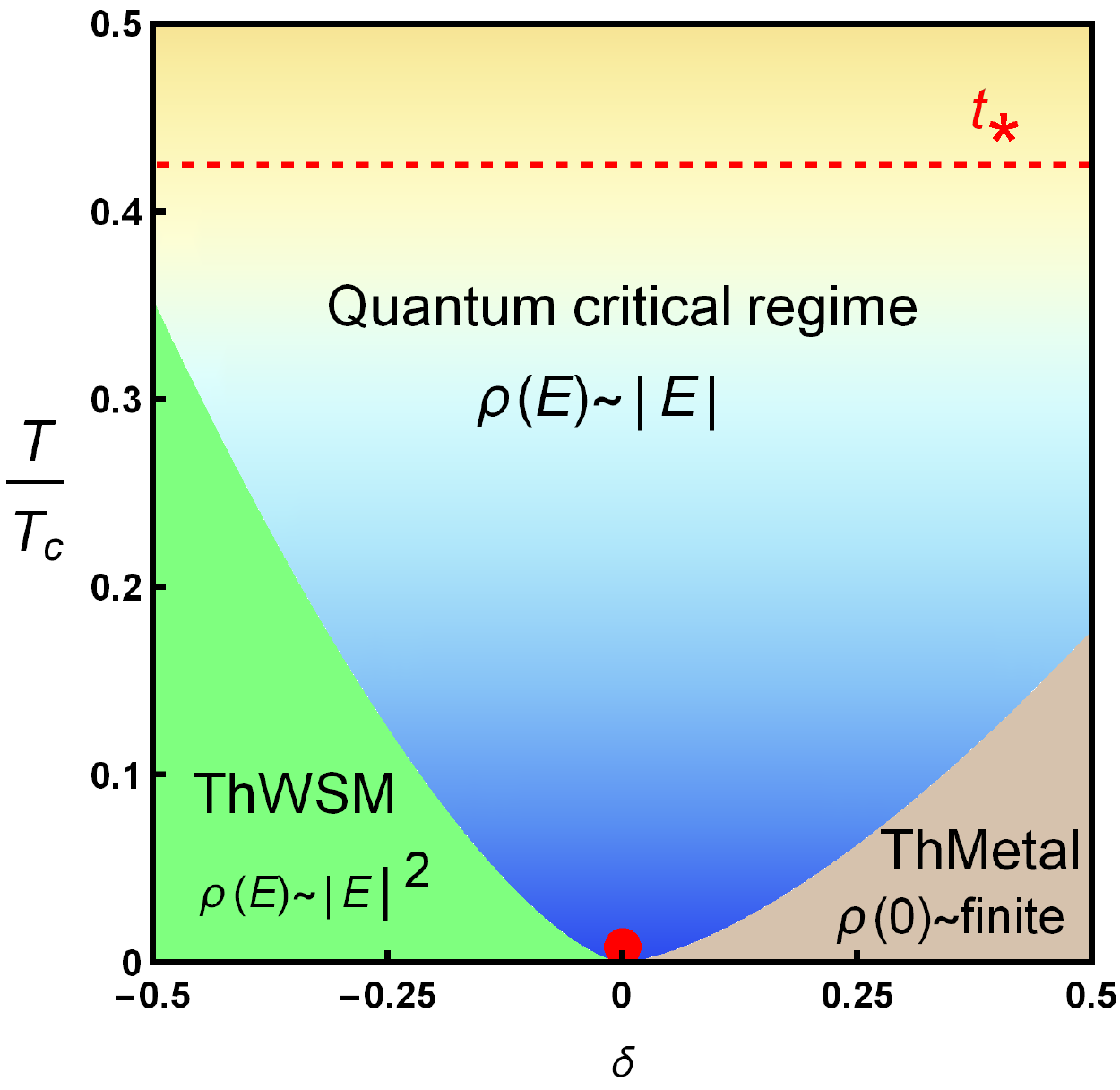}
\caption{A qualitative phase diagram of a dirty thermal Weyl semimetal (ThWSM) at finite temperature ($T$). 
Here $T_c$ is the superconducting transition temperature and $t_\ast(=T_\ast/T_C)$ is a crossover temperature 
above which the BdG-Weyl quasiparticles are not sharp. 
The red dot at $\delta=0$ represents the disorder-controlled ThWSM-thermal metal (ThMetal) 
quantum critical point, and the shaded region represents the associated quantum critical regime 
($\delta$ is the reduced disorder strength, see text). 
The shape of the crossover boundaries at finite temperature are roughly determined by 
$|\delta|^{\nu z} \sim |\delta|^{1.5}$, leading to a wide quantum critical regime. 
For a more quantitative estimation of a similar phase diagram at finite temperature or energy, 
see Refs.~\cite{pixley-goswami-dassarma, bera-sau-roy}. 
Scaling of the average density of states [$\varrho(E)$] in various regimes of the phase diagram 
is displayed in the figure. Here the labeling of the temperature and disorder axes are qualitative.    
}~\label{simpleweyl_dirty}
\end{figure}

We now discuss the effects of quenched disorder (static impurities) on BdG-Weyl quasiparticles. 
Understanding the effects of impurities on regular Weyl and Dirac fermions has attracted ample attention in recent 
times~\cite{fradkin, pallab-sudip2011, herbut-disorder, brouwer, roy-dassarma, radzihovsky, pixley-goswami-dassarma, bera-sau-roy, carpentier, pallab-sudip2016, radzihovsky-2, roy-juricic2016, roy-juricic-dassarma, carpentier-2, wilson-pixley, roy-alavirad, slager-fermiarc, nandkishore-rareregion, pixley-huse-dassarma, brouwer-2}. However, the role of randomness on BdG-Weyl/Dirac quasiparticles is still at an early stage of exploration 
(see however Refs.~\cite{roy-alavirad, wilson-pixley}).

In the context of superconductivity in the Luttinger semimetal (LSM),
the band projection (Sec.~\ref{sec: evenp} and Appendix~\ref{bandprojection})
modifies the form of the particle-hole and chiral time-reversal symmetries in Eqs.~(\ref{MajPH}) and (\ref{ChiralTRI}), respectively. 
For the conduction band (say), we can express particle-hole ($\mathsf{P}$), time-reversal ($\mathsf{T}$), and chiral time-reversal 
($\mathsf{S} \equiv \mathsf{P} \otimes \mathsf{T}$) symmetry conditions 
in terms of the 
$4 \times 4$ band-projected Bogoliubov-de Gennes (BdG) Hamiltonian $\hat{h}(\vex{k})$ as follows,
\begin{align}\label{BandSym}
\begin{aligned}
\!\!	- \Mpb \, \hat{h}^\T(-\vex{k}) \, \Mpb = \hat{h}(\vex{k}), 	&&& \Mpb = \tau_2 \,\sigma_2\!\! 	&& \mathsf{:P},	 	\\
\!\!	 \Mtb \, \hat{h}^*(-\vex{k}) \, \Mtb = \hat{h}(\vex{k}), 	&&& \Mtb = \sigma_2\!\! 		&& \mathsf{:T}, 		\\
\!\!	- \Msb \, \hat{h}(\vex{k}) \, \Msb = \hat{h}(\vex{k}), 		&&& \Msb = \tau_2\!\!	 		&& \mathsf{:S}.
\end{aligned}
\end{align}
The band Hamiltonian has indices in Nambu ($\tau$) and band pseudospin ($\sigma$) spaces. 
Eq.~(\ref{BandSym}) obtains from the corresponding conditions in the $8 \times 8$ LSM-BdG Hamiltonian
by replacing $\Gamma_{13} \rightarrow \sigma_2$, which is the band projection of the (unitary part) of the
time-reversal operator, see Eq.~(\ref{MajPH}).

Even though all candidates for Weyl superconductors have multiple Weyl nodes ($>2$), 
for the sake of the simplicity of the discussion, we consider its simplest realization with only two Weyl 
nodes, with opposite chiralities (left and right), located at $\pm {\mathbf K}$. 
Linearizing the band Hamiltonian $\hat{h}(\vex{k})$ in the vicinity of the pair, we get
\begin{align}\label{hWeyl}
	\hat{h}^\pup{W}_0 = -i \eta_3 \, \sigma_0 \, \sum_{j = 1}^3 v_j \, \tau_j \, \partial_j.
\end{align}
Here $\eta_3 = +1$ ($-1$) for the Weyl node with $W_n = +1$ ($-1$) [Eq.~(\ref{monopolecharge})].  
The Weyl Hamiltonian is $8 \times 8$, with Pauli matrices $\{\eta_\mu\}$ acting on 
the chirality.

The symmetry conditions in Eq.~(\ref{BandSym}) become  
\begin{align}\label{WeylSym}
\begin{aligned}
\!\!	- \MpW \, \left(\hat{h}^\pup{W}\right)^\T(-\vex{k}) \, \MpW 	= \hat{h}^\pup{W}(\vex{k}),\!\!\! 	&&& \MpW = \eta_1 \, \tau_2 \,\sigma_2,\!\! 	\\
\!\!	 \MtW \, \left(\hat{h}^\pup{W}\right)^*(-\vex{k}) \, \MtW 	= \hat{h}^\pup{W}(\vex{k}),\!\!\! 	&&& \MtW = \eta_1 \, \sigma_2,\!\! 		\\
\!\!	- \MsW \, \hat{h}^\pup{W}(\vex{k}) \, \MsW 			= \hat{h}^\pup{W}(\vex{k}),\!\!\! 	&&& \MsW = \tau_2.\!\!	 			
\end{aligned}
\end{align}
The particle-hole and time-reversal matrices $\MpW$ and $\MtW$ both flip the node chirality.
By contrast, the ``chiral'' version of time-reversal is the same in all cases [$\Ms = \tau_2$
in Eqs.~(\ref{ChiralTRI}), (\ref{BandSym}), and (\ref{WeylSym})]. 
Note also that in all cases, the time-reversal matrix is antisymmetric [e.g., $\MtW = -(\MtW)^\T$],
while the particle-hole matrix is symmetric [e.g., $\MpW = (\MpW)^\T$]. 
These conditions imply that $\mathsf{T}^2 = -1$ and $\mathsf{P}^2 = +1$ \cite{SRFL2008}.

It is easy to check that Eq.~(\ref{hWeyl}) satisfies particle-hole symmetry with $\MpW$, as defined above.  
In the LSM, such a Weyl pair arises from 
$d + i d$ pairing. It means that one component of the Hamiltonian $\hat{h}^\pup{W}_0$
is time-reversal-odd. Here the component proportional to $\tau_2$ breaks time-reversal invariance. 
The symmetry 
can be restored by 
setting $v_2 = 0$, although this flattens the band along the parent nodal loop.

Since Weyl superconductors in our analysis arise from singlet pairings, 
we will assume that the disorder preserves band pseudospin ($\sigma$) SU(2) symmetry. 
Band pseudospin is \emph{not} identical to the $m_s = \pm 1/2$ (conduction band)
or $m_s = \pm 3/2$ (valence band) physical spin-3/2 index, since the band projection
of the spin generator $\left[\equiv J_{|m_s|}^\mu(\vex{k})\right]$ 
is momentum-dependent, and moreover dependent upon the explicit gauge choice of the band diagonalizer.
Nevertheless, one can check that 
\[
	- \sigma_2 \, \left[J_{|m_s|}^\mu(-\vex{k})\right]^\T \, \sigma_2 = J_{|m_s|}^\mu(\vex{k}),
\]
so that the projected spin operators are odd under time-reversal, just like the band pseudospin operators
$\{\sigma_{1,2,3}\}$. Magnetic impurities would therefore effectively couple to the band pseudospin. 
Although time-reversal is broken in any of the $d + i d$ scenarios outlined in Sec.~\ref{Weyl-SC} that give rise
to isolated Weyl nodes, we assume that there is no magnetic impurity that breaks 
pseudospin symmetry.

Under this assumption, the BdG-Weyl Hamiltonian $\hat{h}^\pup{W}$ commutes
with all $\{\sigma_\nu\}$. Then we can replace the $\mathsf{P}^2 = +1$ 
physical particle-hole condition in Eq.~(\ref{WeylSym}) with an \emph{effective} 
$\mathsf{P}^2 = -1$ one,
\begin{align}\label{ClassC--PH}
\!\!	
	- \MpWe \, \left(\hat{h}^\pup{W}\right)^\T \, \MpWe 	
	= 
	\hat{h}^\pup{W},
	\;\;	
	\MpWe = \eta_1 \, \tau_2.
\end{align}
Since time-reversal symmetry is already broken in the absence of 
disorder, Eq.~(\ref{ClassC--PH}) is the only effective symmetry
expected to hold even in the presence of disorder. 
Since $\mathsf{P}^2 = -1$, the system belongs to class C. 
By contrast, the $\mathsf{P}^2 =+1$ condition in Eq.~(\ref{WeylSym})
would give class D. As often occurs, a continuous symmetry 
[here band pseudospin SU(2)]
changes the random matrix classification of a Hamiltonian with a given
``microscopic'' specification of $\mathsf{P}$, $\mathsf{T}$, and $\mathsf{S}$ \cite{SRFL2008}.

Incorporating generic quenched disorder in class C, we get the 
BdG-Weyl Hamiltonian
\begin{align}\label{hWeylDirty}
	\hat{h}^\pup{W}
	=
	\hat{h}^\pup{W}_0
	+
	&\,
	\vex{A}_0(\vex{r}) \cdot \vex{\tau}
	+
	\vex{B}_1(\vex{r}) \cdot \vex{\tau} \, \eta_1
	\nonumber\\	
	+
	&\,
	\vex{B}_2(\vex{r}) \cdot \vex{\tau} \, \eta_2
	+	
	v_3(\vex{r}) \, \eta_3. 
\end{align}
In this equation, $\vex{\tau} = \tau_1 \hat{x} + \tau_2 \hat{y} + \tau_3 \hat{z}$ 
is the vector of Nambu matrices. 
There are 10 allowed perturbations, which take the form 
of 
(i) an axial [since it is missing the $\eta_3$, c.f.\ Eq.~(\ref{hWeyl})] vector potential $\vex{A}_0(\vex{r})$,
(ii) two vector components $\vex{B}_{1,2}(\vex{r})$ of a tensor disorder potential,\footnote{
In relativistic notation, the 6 independent components of 
$\vex{B}_{1}$ and $\vex{B}_2$ 
couple to the independent elements of $\sigma^{\mu \nu} \equiv i \gamma^\mu \gamma^\nu$,
where $\{\gamma^\mu\}$ are the four $4 \times 4$ $\gamma$-matrices acting 
on the spinor field formed from the sum of left- and right-handed Weyl components. See for example~\cite{Peskin}.}  
and 
(iii) a Weyl node-graded (axial) scalar potential $v_3(\vex{r})$.

We will further simplify our treatment by neglecting quenched random
fluctuations of the Weyl ($d + i d$) pairing amplitudes, so that we drop disorder
terms in $\vex{B}_{1,2}$ that couple to $\tau_{1,2}$. 
Note that $\vex{B}_{1,2}$ disorder can qualitatively account 
for pair breaking effects. Since we are interested in sufficiently low energies
or temperatures ($T \ll T_c$), the amplitude of the $d$-wave pairings can be assumed
to be frozen. Under this circumstance, we are allowed to neglect disorder proportional
to $\vex{B}_{1,2}$. However, we note that close to $T_c$ one should account for all symmetry allowed 
disorder, appearing in Eq.~(\ref{hWeylDirty}), which goes beyond the scope of the present analysis.
In our renormalization group (RG) scheme explained below,
the remaining subset still closes under the one-loop RG. 
Thus we retain six random potentials, $\{A_0^{1,2,3},B_1^3,B_2^3,v_3\}$. 
Physically, $A_0^3(\vex{r})$ corresponds to the electric charge density,
i.e.\ encodes scattering off of Coulomb impurities  
(despite the fact that it appears as a vector potential component to the BdG-Weyl quasiparticles). 
Real and imaginary quenched fluctuations of the $s$-wave pairing 
are encoded in $\{A_0^{1,2}\}$. 
The potential $v_3(\vex{r})$ is a node-staggered chemical potential or ``random Doppler'' shift. 
$\vex{A}_0$ and $v_3$ scatter only within a given node.
The remaining potentials $\{B_1^3,B_2^3\}$ describe internode backscattering
due to short-ranged impurities.\footnote{ 
Note that random charge impurities couple as the third component 
of the axial vector potential, while the two planar components  
stem from the real and imaginary components of the random singlet $s$-wave pairing. 
Therefore, the strength of these two types of disorder coupling 
(respectively described by $\Delta_0$ and $\Delta_{1}$ below) 
at the microscopic level are different. 
In the presence of generic disorder the Fermi velocities along the 
$z$ direction (along which the quasiparticle spectrum supports Weyl nodes in clean system), 
denoted by $v_3$ and in the $x-y$ plane, denoted by $v_1=v_2=v_\perp$ receive different 
renormalizations from the disorder. 
Even if we impose isotropy at the bare level 
(assuming $v_1=v_2=v_3=v$), 
such symmetry is no longer respected at intermediate scale as 
we coarse-grain the theory. 
Nonetheless, we are allowed to perform the perturbative RG analysis 
with one Fermi velocity, but we need to treat the anisotropy parameter, 
defined here as $\alpha=v_\perp/v_3$, as a running coupling. 
But, as we demonstrate below that at the clean Weyl fixed point as well as the 
thermal Weyl semimetal-thermal metal quantum critical point, $\alpha$ is a 
marginal variable and does not affect the disorder-driven quantum critical behavior 
in a dirty thermal Weyl semimetal (at least to the one-loop order).}

In Eqs.~(\ref{DirtCorr-r}) and (\ref{DirtCorr-p}) above, the disorder potentials $\varphi_j \in \{A_0^{1,2,3},B_1^3,B_2^3,v_3\}$. 
However, for these six we assign only four variances, 
\begin{align}
\!\!\!
\begin{aligned}
	A_0^{3}:&\;\;\,		\Delta_0, \;\; \text{(Electric potential),} \\
	A_0^{1,2}:&\;\;\,	\Delta_1, \;\; \text{(Real and imaginary $s$-wave pairing),}\\
	B_{1,2}^3:&\;\;\,	\Delta_2, \;\; \text{(Internode backscattering),}\\
	v_3:&\;\;\,		\Delta_3, \;\; \text{(Random Doppler [axial potential]).} 
\end{aligned}
\end{align}
We here control the renormalization group (RG) calculation in the following way. Each 
disorder field is assumed to obey the following distribution 
\begin{equation}\label{DirtCorr-r}
	\langle 
		\varphi_j \left( \vex{x} \right) 
		\varphi_k \left( \vex{y} \right) 
	\rangle
	= 
	\delta_{jk} 
	\; 
	\frac{\Delta_j}{|\vec{x}-\vec{y}|^{d-m}}  
\end{equation}
in position space or
\begin{equation}\label{DirtCorr-p}
	\langle 
		\varphi_j \left( \vec{q} \right) 
		\varphi_k \left(0 \right) 
	\rangle
	= 
	\delta_{jk} \; \frac{\Delta_j}{|{\mathbf q}|^{m}},
\end{equation}
in momentum space and the limit $m\rightarrow0$ corresponds to the Gaussian white noise distribution, which we are ultimately interested in. 
This form of the white noise distribution stems from the following representation of the $d$-dimensional $\delta$-function
\begin{equation}
	\delta^{(d)}({\bf x}-{\bf y})
	=
	\lim_{m\rightarrow 0} 
	\: 
	\frac{\Gamma\left(\frac{d-m}{2}\right)}{2^m\pi^{d/2}\Gamma(m/2)} 
	\: 
	\frac{1}{|{\bf x}-{\bf y}|^{d-m}}.
\end{equation}
For additional details of this methodology readers should consult Refs.~\cite{pallab-sudip2016, roy-juricic2016}.
An $\epsilon$-expansion can be performed with the construction 
$m = 1 - \epsilon$, and ultimately for Gaussian white noise disorder we set $\epsilon=1$ at the end of the calculation.  

The RG flow equations to the leading order in the $\epsilon$-expansion read as 
\allowdisplaybreaks[4]
\begin{eqnarray}
  \beta_{v_3} &=& -\frac{4 v_3}{3} \left[\Delta_0+\Delta_1 +\Delta_2 +\Delta_3 \right] =v_3 (1-z), \nonumber \\
	\beta_{\alpha} &=& \alpha \; \frac{2}{3} \left[ \Delta_0 -\Delta_{1} \right], \: \:
	\beta_{\Delta_0} = \Delta_0 \; 
				\left[ -\epsilon- \frac{8}{3} \Delta_{2} \right], \nonumber \\
	\beta_{\Delta_{1}}&=& \Delta_{1} \; 
				\left[ -\epsilon + \frac{4}{3} \left(\Delta_0 - \Delta_{1}  \right) \right], \nonumber \\
	\beta_{\Delta_{2}} &=& \Delta_{2} \; 
				\left[ -\epsilon+ \frac{4}{3} \left(\Delta_0 - 2 \Delta_{1} - \Delta_{2} \right) \right], \nonumber \\
	\beta_{\Delta_3} &=& \Delta_3 \; 
				\left[ -\epsilon + \frac{8}{3} \left( \Delta_0 + \Delta_{1} - 2\Delta_{2} + \Delta_3\right) \right], 			
\end{eqnarray}
in terms of dimensionless disorder couplings $\hat{\Delta}_j=\Delta_{j} \Lambda^{\epsilon}/(2 \pi^2 v^2_3)$. 
For brevity we drop the `hat' notation in the above flow equations. Here, $\Lambda$ is the ultraviolet momentum up to which BdG-Weyl quasiparticles possess linear dispersion. The flow equation of $v_3$ then yields a scale-dependent dynamic scaling exponent
\begin{equation}~\label{Z:RG-QCP}
	z
	=
	1
	+
	\frac{4}{3} \; \left[ \Delta_0 + \Delta_{1} +\Delta_{2} + \Delta_3 \right].
\end{equation}
Notice that the bare dimension for all disorder couplings $\left[ \Delta_j \right] = -\epsilon$. 
Therefore sufficiently weak disorder is irrelevant at the BdG-Weyl fixed point. 
We here tacitly bypass the possibility of coexisting rare regions in this system~\cite{wilson-pixley, nandkishore-rareregion, pixley-huse-dassarma},
which suggest that Weyl fermions may become unstable for infinitesimal strength of disorder and enter 
into a metallic phase, where the DoS at zero-energy is finite. 
However, the DoS at zero energy due to the rare regions is extremely small and over almost 
the entire energy window the outcomes from the perturbative analysis hold.

\begin{figure}[t!]
\includegraphics[width=0.35\textwidth]{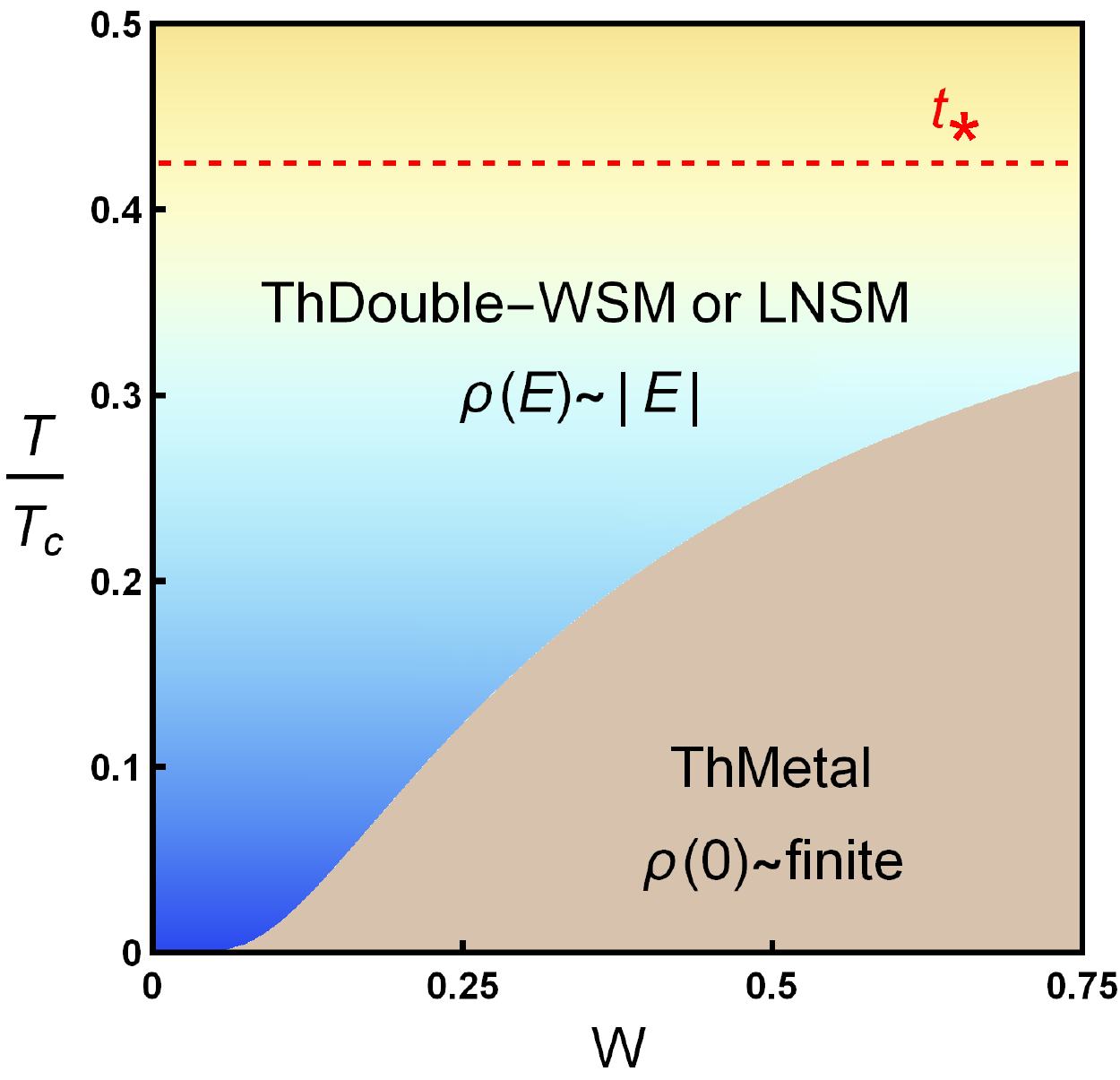}
\caption{A qualitative phase diagram of a dirty thermal double Weyl semimetal (WSM) 
or thermal line-node semimetal (LNSM) at finite temperature. 
Above a crossover temperature $t_\ast=T_\ast/T_C$, BdG quasiparticles are not sharp. 
The crossover boundary between the pseudo-ballistic semimetallic phase and the diffusive thermal metallic phase 
scales as $\sim \exp[-A/W]$, where $W$ denote the strength of disorder and $A(=0.35 \; \mbox{here})$ 
is, however, a non-universal (material dependent) constant. 
Scaling of the density of states in various regimes of the phase diagram is quoted in the figure. 
For quantitative estimation of this phase diagram see Ref.~\cite{bera-sau-roy}. 
Here the labeling of the temperature and disorder axes are qualitative.  
}~\label{doubleweyl_dirty}
\end{figure}

The above coupled flow equations support only two fixed points in the
{$\left( \Delta_0, \Delta_1, \Delta_2, \Delta_3\right)$	} 
plane: 

1. The stable fixed point located at {$\left( \Delta_0, \Delta_1, \Delta_2, \Delta_3 \right) =\left( 0,0,0,0 \right)$	} 
	is the BdG-Weyl ``thermal semimetal'' 
	phase.
	
2. On the other hand, the fixed point located at 
	{$\left( \Delta_0, \Delta_1, \Delta_2, \Delta_3 \right) =\left( 0,0,0,3/8 \right) \epsilon$	} 
	has only 
	one unstable direction. It represents a quantum critical point (QCP), describing a quantum phase transition (QPT) from thermal Weyl semimetal 
	to diffusive thermal metal. In the thermal metallic phase the DoS at zero energy is finite, and BdG quasiparticles possess finite elastic
	impurity lifetime and mean-free path (in the plane-wave basis).

It is worth pointing out that even though we can tune the strength of any one of the four disorder couplings, 
the thermal Weyl semimetal-thermal metal QPT is ultimately driven by the random Doppler shift, which couples to 
the Weyl fermions as the axial potential. Also note that the anisotropy parameter $\alpha$ is a marginal parameter at both fixed points. 
Therefore both fixed points are multicritical in the five-dimensional space $\{\alpha,\Delta_{0,1,2,3}\}$. 
For now we neglect the effect of $\alpha$ and focus only on the four-dimensional subspace spanned by the disorder.

At the thermal Weyl semimetal-metal QCP the dynamic scaling exponent [see Eq.~(\ref{Z:RG-QCP})] is given by 
\begin{equation}\label{DirtyDynCrit}
	z = 1 +  \epsilon/2  \quad \Rightarrow \quad z = 3/2,
\end{equation}
for the Gaussian white noise distribution ($\epsilon=1$). 
The correlation length exponent at the disorder controlled QCP is given by 
\begin{equation}
	\nu^{-1} = \epsilon \quad \Rightarrow \quad \nu=1, 
\end{equation} 
for $\epsilon=1$. Eq.~(\ref{DirtyDynCrit}) implies that 
the average density of states at the QCP scales as $\varrho(E) \sim |E|^{(d-z)/z} = |E|$. 
The product $\nu z = 3/2$. 
Consequently, the crossover boundaries at finite temperature or energy are determined by 
$T^{\ast} \; {\rm or} \; E^{\ast} \sim |\delta|^{\nu z}$ 
and are \emph{concave upward}, where $\delta=\left( \Delta-\Delta_\ast \right)/\Delta_\ast$ 
is the reduced distance from the disorder-controlled QCP located at $\Delta=\Delta_\ast$. 
As a result a wide quantum critical regime occupies the largest portion of the phase diagram 
of a dirty thermal Weyl semimetal at finite temperature, as shown in Fig.~\ref{simpleweyl_dirty}.

On the other hand, the thermal double-Weyl semimetal as well as the thermal nodal-loop semimetal become unstable towards 
the formation of a thermal metal for arbitrary weak strength of disorder due to the clean linearly vanishing density of states 
$\varrho(E) \sim |E|$~\cite{pallab-andriy, bera-sau-roy, brouwer-2}. This can be substantiated from the computation of the 
scattering lifetime ($\tau$) from a self-consistent Born approximation, leading to 
\begin{equation}
	W \int^{E_\Lambda}_0 dE \; \frac{\varrho(E)}{\hbar \tau^{-2}+E^2}=1 \;
	\Rightarrow 
	\frac{\hbar}{\tau}= E_\Lambda \exp \left( -\frac{A}{W}\right),
\end{equation}  
where $W$ is the strength of disorder, $E_\Lambda$ is the ultraviolet energy cut-off, and
$A$ is a non-universal (material dependent) constant. Thus, the self-consistent solution of 
$\tau$ indicates that BdG-Weyl fermions in thermal double Weyl and nodal-loop semimetals 
acquire finite lifetime and the system immediately becomes a diffusive thermal metal. 
Note that the DoS at zero energy also follows the profile of $1/\tau$, and the phase boundary 
in Fig.~\ref{doubleweyl_dirty} follows the functional form in the above equation.
This outcome is in agreement with the scaling analysis, which suggests that disorder is 
a marginally relevant perturbation in the presence of double Weyl nodes or line nodes~\cite{bera-sau-roy}. 
\footnote{
We note that the accidental nodes found at two opposite poles in the presence of $d_{xz/yz}+id_{x^2-y^2}$ pairing 
(see Table~\ref{egt2g:competitiontable}) are, however, stable against sufficiently weak randomness~\cite{roy-juricic2016}.}

Strong pairing that causes intermixing between the conduction and valence bands could induce small 
``inflated node'' Bogoliubov Fermi surfaces~\cite{brydon-2,brydon-3}. Arbitrarily weak disorder would smear these, leading
to diffusive thermal metallic behavior with a nonzero density of states at sufficiently small energy. 
The considerations of this section would still apply for energy scales larger than that of the band intermixing.


\section{Connection with experiments: Penetration depth in $\mathbf{YPtBi}$}~\label{experiment}

\begin{figure}
\includegraphics[width=0.48\textwidth]{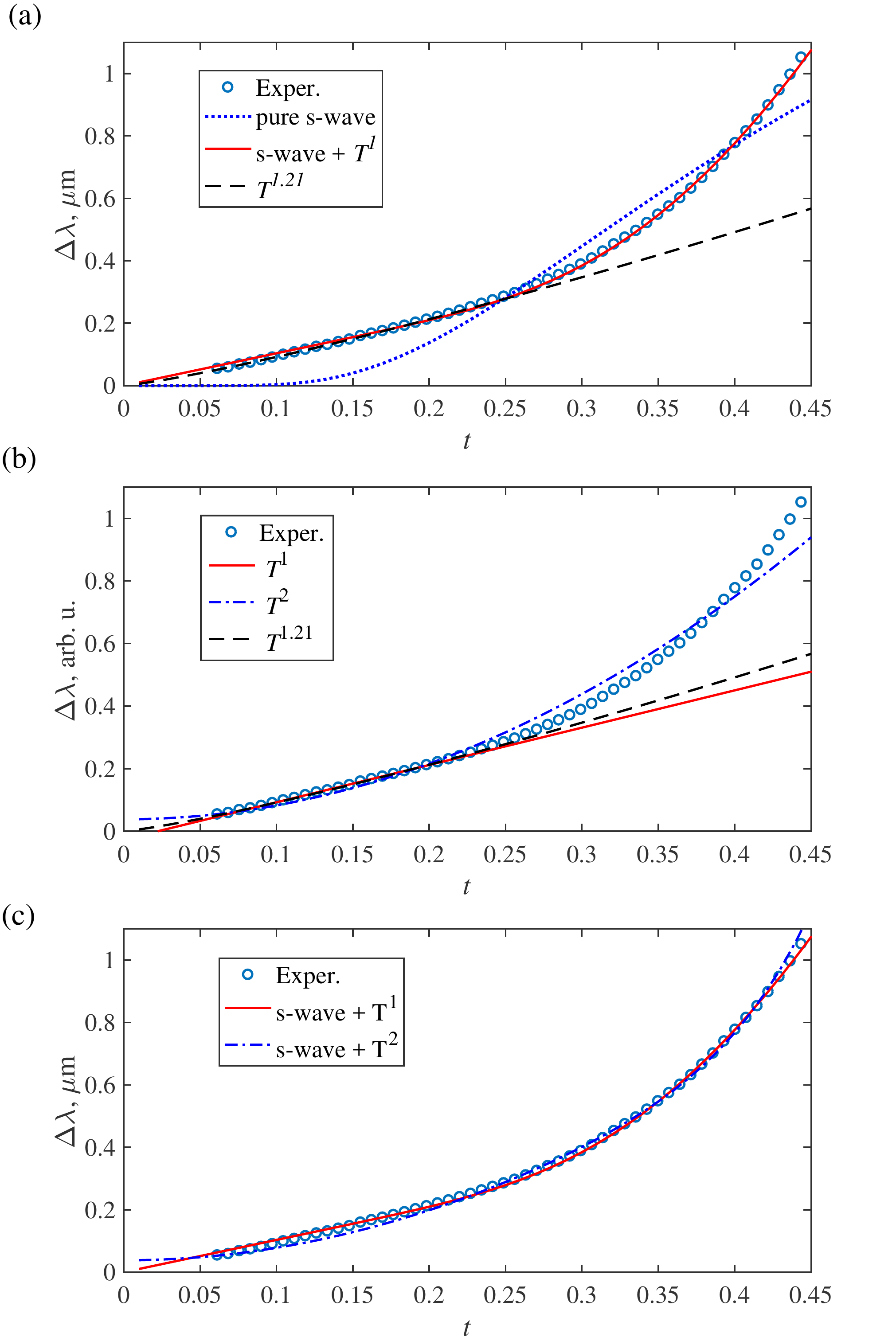}
\caption{ Various possible fits for the experimentally measured data of the penetration depth ($\Delta \lambda$) in YPtBi as a function of the reduced temperature $t=T/T_c$ (replotted from Ref.~\onlinecite{Exp:Paglione-2} with permission). Here $T_c=0.78$K is the superconducting transition temperature in YPtBi. For details of these fittings and the corresponding physical picture, see Sec.~\ref{experiment}.  
}~\label{fitlambda:expthr}
\end{figure}

Even though there exists experimental evidence suggestive of superconductivity in some half-Heusler materials, 
the actual nature of the pairing in these compounds is not very clear at this stage~\cite{Exp:Takabatake, Exp:Paglione-1a, Exp:Bay2012, Exp:visser-1, Exp:Taillefer, Exp:Zhang, Exp:visser-2, Exp:Paglione-1, Exp:Paglione-2}. In this respect, a recent experiment revealed a very interesting feature through the measurement of the penetration depth~\cite{Exp:Paglione-2}. This experiment~\cite{Exp:Paglione-2} suggests that the change in the penetration depth ($\Delta \lambda$) vanishes in a \emph{power-law fashion with temperature} $\Delta \lambda \sim T^{n}$,
which is suggestive of the existence of gapless BdG quasiparticles inside the paired state. However, the precise value of $n$ remains a subject of debate. 
In Ref.~\cite{Exp:Paglione-2} a reasonably good fit was found with $n=1.20 \pm 0.02$, but only over a limited window of temperature $0.1 \leq T/T_c \leq 0.2$ (approximately), where $T_c \approx 0.78$K is the superconducting transition temperature in YPtBi. Such power-law dependence was then interpreted as the signature of a paired state with \emph{nodal loops}, for which the DoS vanishes as $\varrho(E) \sim |E|$ in a clean system. Since $\Delta \lambda$ follows the power-law of the DoS, the deviation from a pure $T$-linear dependence was attributed to impurities.

We here make an independent attempt to understand the dependence of the penetration depth ($\Delta \lambda$) on the reduced temperature $t=T/T_c$ and fit the available data with the following functional form 
\begin{equation}~\label{fitfunction}
\Delta \lambda(t) = \lambda(0) \sqrt{\frac{\pi \Delta_0}{2 k_B T_c}\frac{1}{t}} \: \exp\left(-\frac{\Delta_0}{k_BT_c}\frac{1}{t}\right) + c_n \, t^n,
\end{equation} 
and provide alternative explanation for the experimental observation in Ref.~\cite{Exp:Paglione-2}. The first term on the right-hand side [denoted by $\Delta \lambda_s(t)$] is the canonical penetration depth dependence in an $s$-wave superconductor, whereas the power law terms correspond to the presence of gapless BdG quasiparticles for which the DoS vanishes as  $\varrho(E) \sim |E|^n$. The resulting fits are displayed in Fig.~\ref{fitlambda:expthr}. 
With the above functional dependence of the penetration depth $\Delta \lambda(t)$ we search for the fitting parameters, such as $\lambda(0)$ (zero temperature penetration depth of the $s$-wave component), $\Delta_0/(k_B T_c)$ and $n$ (takes only integer values), to obtain the best possible fit within the temperature window $0.05 \leq t \leq 0.33$, such that gapless BdG quasiparticles are sharp. For $t>0.33$ agreement between the fit function and experimental data is \emph{fortuitous}. 
A possible microscopic justification for the scaling of $\Delta \lambda(t)$, see Eq.~(\ref{fitfunction}), is presented later in this section.

The solid line in Fig.~\ref{fitlambda:expthr}(a) shows the best fit obtained by keeping only the linear in temperature term in Eq.~(\ref{fitfunction}) in addition to the $s$-wave contribution. We find the best fit with the zero-temperature value of the $s$-wave gap $\Delta_0 = 2.08 k_BT_c$, which is close to the BCS value $\Delta_{BCS} = 1.76 k_BT_c$, with the amplitude of the linear term $c_1 = 1.04 \mu$m. Interestingly, the fit results in the zero-temperature value of the penetration depth $\lambda(0) \approx 23 \mu$m, which is three orders of magnitude larger than found in conventional superconductors such as aluminum, as remarked by the authors of Ref.~\onlinecite{Exp:Bay2012} who also measured a high value of $\lambda(0)_\text{exp} = 2 \mu$m in YPtBi. As Fig.~\ref{fitlambda:expthr}(a) illustrates, a pure $s$-wave dependence (dotted line) is not a good fit to the data, even though it would result is a value of $\lambda(0) \approx 3\mu$m that is closer to the experiment~\cite{Exp:Bay2012}. 
{ We obtain a good agreement with the experimentally observed scaling of the penetration depth over a large temperature window by adding an $s$-wave component for which the zero temperature absolute value of penetration depth 
$\lambda(0) \approx 23\mu$m $\gg \lambda(0)_\text{exp}(=2\mu\mbox{m})$~\cite{Exp:Bay2012}. 
This observation indicates that the requisite strength of the $s$-wave component is rather small in 
comparison to the dominant $d$-wave pairing, as the superfluid density contribution scales with $1/\lambda(T)^2$. 
}

On the other hand, including the $s$-wave component is crucial to accurately fit the data. Indeed, attempts to fit to a pure power-law, see Fig.~\ref{fitlambda:expthr}(b), are unsatisfactory as the fits only work in the low-temperature regime $t \lesssim 0.2$. Moreover, the extracted power-law exponent depends sensitively on the width of the fitted temperature region and given the narrow temperature range, it is difficult to distinguish between a $T$-linear fit [solid red line in Fig.~\ref{fitlambda:expthr}(b)], the $T^2$ fit [dash-dotted blue line in Fig.~\ref{fitlambda:expthr}(b)] or, say, the $T^{1.2}$ fit adopted by the authors in Ref.~\onlinecite{Exp:Paglione-2} [dashed line in panels (a) and (b) of Fig.~\ref{fitlambda:expthr}]. Of course the three curves deviate from each other at higher temperatures $t>0.25$, but by that time neither one of them fits the experimental data even remotely.
In principle, one can try to fit the experimental data by varying the contribution from the $s$-wave component, which, however, immediately affects the temperature window over which the fitting function from Eq.~(\ref{fitfunction}) yields agreement with the experiment. Such semi-quantitative variation of our procedure does not change the fact that an $s$-wave contribution is necessary to produce sensible agreement within reasonable temperature window.    
Experimentally it is conceivable to find the contribution of the $s$-wave component by 
comparing the gap size at the nodal and anti-nodal points (since only the $s$-wave component is uniform
over the entire Fermi surface), as it has been done in YBCO~\cite{YBCO-s-wave}.

Based on the above analysis and the comparison between panels (a) and (b) in Fig.~\ref{fitlambda:expthr}, we conclude that it is necessary to include the $s$-wave component to properly fit the penetration depth data. The pure $s$-wave fit is unsatisfactory, as remarked earlier, and a power-law contribution must be considered. We now turn to the comparison between different such power-law contributions. The red and blue lines in  Fig.~\ref{fitlambda:expthr}(c) show the $\Delta\lambda_s(t)+ c_1 t$ and $\Delta\lambda_s(t)+ c_2 t^2$ fits to the data, respectively. The red curve incorporating the $T$-linear component fits the data better [the same fit as the solid line in panel (a)]. By contrast, attempts to fit the data with $\Delta\lambda_s(t)+c_2 t^2$ form not only fall below the target in the range $0.05 < t < 0.25$, but also requires in an unphysically large fitting parameter $\lambda(0) \sim 10^3\mu$m. We cannot however exclude possible presence of a small $T^2$ component in addition to the dominant $s$-wave and $T$-linear terms.

Based on the above analysis, we conclude that superconductivity in YPtBi is best described by a combination of a fully gapped $s$-wave component and a gapless BdG quasiparticles with linear in energy density of states. The latter is commonly attributed to the nodal lines in the gap, such as a $d$-wave or $p$-wave. However we stress that this is \emph{not a unique explanation} and we list several possible sources of $T$-linear dependence below.

\emph{Source of $T^2$ dependence}: Within a simple picture for pairing in the Luttinger system, there is only one possible source for $T^2$ dependence of the penetration depth: namely, the existence of gapless quasiparticles at isolated points on the Fermi surface where the DoS vanishes as $\varrho(E) \sim |E|^2$ at low energy. Only a \emph{Weyl superconductor}, constituted by Weyl nodes with monopole charge $W_n=\pm 1$ yields such DoS, see Secs.~\ref{phaselocking:eg} and \ref{Phaselock:T2g}. In this work, we have presented several examples of Weyl superconductors; any such candidate is capable of producing a $T^2$ dependence of the penetration depth at low temperatures.

\emph{Sources of $T$-linear dependence}: Unlike the aforementioned case of $T^2$ dependence, the origin of a $T$ linear contribution to $\Delta \lambda$ is not unique, with several possible sources resulting in the gapless BdG fermions displaying the linear scaling $\varrho(E) \sim |E|$ of the DoS. Once again within a simple picture of pairing in the Luttinger system, we can identify \emph{three} possible origins of such $E$-linear scaling of the DoS. They are  the following.

$1.$ A double Weyl superconductor, with isolated Weyl nodes characterized by the monopole charge $W_n =\pm 2$. Such Weyl nodes are also referred to as \emph{double-Weyl nodes}, and yield $\varrho(E) \sim |E|$ at low energies (below the superconducting gap). But, as we have seen above [see Table~\ref{egt2g:competitiontable} and Sec.~\ref{EgT2g:Competition}], examples of such a double-Weyl superconductor are sparse and so far we can only identify one candidate, namely, the $d_{x^2-y^2}+id_{xy}$ paired state, that can support a linear scaling of the DoS. Thus, based on various examples of Weyl superconductors discussed in this work, one may conclude that this possibility is the least likely one. Nucleation of such a phase will, however, be associated with a \emph{two-stage} transition. 

$2.$ A tempting source of $\varrho(E) \sim |E|$ is existence of at least one nodal loop in the spectrum of BdG quasiparticles. So far we have found ample examples in the presence of local or intra-unit cell pairings that support nodal loops in the ordered phase 
(see Table~\ref{table_projection}, 
Sec.~\ref{Phaselock:T2g}). 
However, for each such possibility there is always a competing ordered phase of the $d+id$ type that accommodates simple and isolated Weyl nodes (with $W_n=\pm 1$) on the Fermi surface. Since $\varrho(E) \sim |E|^2$ inside the simple Weyl superconductors, nucleation of such paired state will cause power-law suppression of the DoS, thus optimizing the gain in the condensation energy. We therefore conclude that Weyl nodal-point superconductors are energetically superior to the ones with nodal loops, at least within the framework of the weak coupling BCS pairing.  

$3.$ The last, but most likely, possibility is the following. The underlying paired state supports simple BdG-Weyl quasiparticles with the Weyl nodes that are characterized by the monopole charge $W_n=\pm 1$ in Eq.~(\ref{monopolecharge}). But, due to the presence of impurities in the system, the thermodynamic responses measuring the DoS (for instance, the penetration depth $\Delta \lambda$) are determined by the disorder-driven \emph{quantum critical regime} associated with the Weyl superconductor-thermal metal QPT, see Fig.~\ref{simpleweyl_dirty}. At such a transition, the dynamical critical exponent $z=3/2$ 
(possibly exact in light of recent field-theoretic and numerical works, see Sec.~\ref{Weyl:disorder}) yields the density of states $\varrho(E) \sim |E|^{-1+d/z} \sim |E|$, linear in energy. Such an origin of a $T$-linear contribution to the penetration depth is quite natural in YPtBi, since the carrier density is extremely low, making the system 
susceptible
to impurities.     

We  believe that the last possibility is the most likely scenario in YPtBi and other half-Heusler compounds for the following reasons. Nucleation of simple-Weyl nodes in the gap is possibly energetically the best option among various available candidates for \textit{nodal}
 pairings as it provides the optimal power-law suppression of DoS at low-energy (leaving aside fully gapped paired states that require non-local pairing, see Sec.~\ref{sec: oddp}). But, due to the presence of randomness/impurities, the thermodynamic responses are dominated by the wide quantum critical regime associated with disorder-controlled Weyl-to-thermal metal QPT. Note that  this critical regime occupies the largest portion of the phase diagram of the disordered Weyl superconductor at finite temperature, as shown in Fig.~\ref{simpleweyl_dirty}.

\emph{Sources of $s$-wave component}: Perhaps the most enigmatic aspect of our analysis of the recent experimental data on the penetration depth~\cite{Exp:Paglione-2} is the unambiguous presence of an $s$-wave component, which is however quite natural in light of the discussion presented in Sec.~\ref{externalstrain}. Recall that nucleation of any $d$-wave pairing (or any combination of multiple $d$-wave pairing) breaks the cubic symmetry, which naturally introduces a lattice distortion or electronic nematicity in the system. In turn, the cooperative effect of the $d$-wave pairing and such lattice distortion introduces a non-trivial $s$-wave pairing in the system. Therefore, incorporating the contribution of  
$s$-wave pairing is fully consistent with the symmetry of the problem: indeed, the $s$-wave $A_{1g}$ pairing appears on equal footing with the nodal $d$-wave channels in the vicinity of the Fermi surface, see Table~\ref{table_projection}. Additionally, one could of course imagine a more trivial origin of the $s$-wave pairing, such as due to the electron-phonon coupling, considered in a recent theoretical work~\cite{savary}.

Since the chemical potential in superconducting half-Heusler compounds, such as YPtBi, lies in close proximity to the quadratic band touching points, it is quite natural to anticipate that intra-unit cell pairings or local pairings ($s$-wave and five $d$-waves listed in Table~\ref{table_projection}) stand as prominent candidates. This is the reason why so far we have focused on these pairings, leaving aside non-local or longer-range pairings, which will be the subject of discussion in Sec.~\ref{sec: p-wave}. 
Since some of the half-Heusler compounds display magnetic order, we also believe that non-$s$-wave pairing is possibly the dominant intra-unit cell pairing, which has received some support from simple microscopic calculations~\cite{herbut-2,savary}. In light of the above discussion, the $s$-wave component, when it is manifest, is not interaction driven but rather an \emph{induced} one. 
Our whole discussion on experimental aspects of pairing in Luttinger system thus evolves around various $d$-wave pairings that can ultimately lead to simple Weyl superconductors via the formation of the $d+id$ state (see section~\ref{Weyl-SC}).

The proposed scaling of the penetration depth in Eq.~(\ref{fitfunction}) may require the existence two Fermi surfaces, 
as a small (induced) $s$-wave component in the presence of a dominant $d+id$ type pairing only shifts the location of the Weyl nodes 
(discussed in Appendix~\ref{Append:sdNodes}). The presence of two Fermi surfaces is quite natural in YPtBi, since the inversion 
symmetry is broken in half-Heuslers. Consequently, the Kramers (or pseudo-spin) degeneracy of the Fermi surface is lost. 
Under that circumstance, it is conceivable that only one of the Fermi surfaces hosts Weyl nodes, while the other one becomes 
fully gapped, in the presence of an $s+d+id$ pairing, justifying the proposed scaling of penetration depth in Eq.~(\ref{fitfunction}). 
The details of the analysis is presented in Appendix~\ref{Append:Inversionassymetry}. 
{On the other hand, from the fits shown in Fig.~\ref{fitlambda:expthr}, 
we realize that inclusion of the $s$-wave component is important to obtain a good fit 
at slightly higher temperatures, where the DoS can deviate from pure $|E|$-linear dependence 
due to the induced $s$-wave component. Hence, the existence of two Fermi surfaces (one being gapless 
while the other one being fully gapped) may not be necessary for the applicability 
of Eq.~(\ref{fitfunction}), as the component is operative at higher temperature. 
Only future experiments can resolve these two competing scenarios.}
We also note that even in the absence of inversion symmetry 
both Fermi surfaces remain gapless for a pure $d+id$ pairing. It should, however, be noted that $s$- and $d$-wave pairings are mixed 
solely due to the spin-3/2 nature of quasiparticles (see Sec.~\ref{externalstrain}) not the inversion asymmetry, as both pairings 
are even under the spatial inversion. We should mention that once $s$-wave pairing is induced by a (dominant) $d$-wave pairing, 
it can further be amplified by electron-phonon coupling, which is always present in any real material.

As a final remark of this section, we should point out another possible source of an $s$-wave component in YPtBi or any half-Heusler compound. Note that half-Heusler compounds break the inversion symmetry, which mandates that even (such as $s$- and $d$-wave) and odd (such as $p$- and $f$-wave) parity pairings always coexist~\cite{rashba-gorkov}. Presently the strength of inversion symmetry breaking is not clear in this class of materials (likely to be weak as no experiment has found a clear signature of inversion asymmetry). {Nevertheless, when the interaction is conducive for a \emph{unixial} $p$-wave pairing, which can produce line-nodes in the ordered phase and thus yield $\varrho(E) \sim |E|$, the ordered phase is always accompanied by an $s$-wave component~\cite{Exp:Paglione-2,brydon}. The $s+p$ paired state can be immune to pair-breaking effects when 
inversion symmetry breaking is sufficiently strong~\cite{Agterberg-Sigrist}. 
However, such a uniaxial $p$-wave pairing can be energetically inferior to an isotropic, but fully 
gapped $p$-wave pairing, discussed in Sec.~\ref{sec: p-wave}. On the other hand, for low carrier density it is  likely that local $d$-wave pairings are energetically favored over the non-local pairings (such as $p$- and $f$-wave), and naturally accompanied by an $s$-wave component. Given the uncertainty in our knowledge of various quintessential parameters in these materials, it is of absolute necessity to perform complimentary thermodynamic and transport measurement to unambiguously determine the symmetry of the paired state. Finally, we briefly comment on such possible future experiments that can help pin down the pairing symmetry in these compounds.

\subsection{Future experiments and pairing symmetry}

The existence of simple Weyl superconductors can be pinned down, at least in principle, by systematically controlling the impurity concentration in the system, for example. Note that with the decreasing strength of impurity scattering (as the material gets cleaner), the $T$-linear dependence of the penetration depth is expected to get suppressed and a $T^2$ dependence should become dominant. 
This feature, for example, can be probed by the following measurements: 

$\bullet \;$ Specific heat ($C_v$) measurements inside the superconducting phase, although difficult given the low critical temperatures ($0.78$ K in YPtBi) in half-Heuslers, can be instrumental in unveiling the pairing symmetry. With an underlying simple Weyl superconductor, $C_v$ should display a gradual onset of $T^3$ dependence at low temperature ($T \ll T_c$) and disappearance of $T^2$ scaling as the concentration of impurities is reduced (the system then falls on the Weyl side of the phase diagram, escaping the critical regime), see Fig.~\ref{simpleweyl_dirty}. By contrast, with increasing  impurity scattering, the $T^2$ dependence of specific heat is expected to gradually get replaced by the Fermi-liquid-like  $T$-linear dependence as the system moves into the thermal metallic side of the transition. 

$\bullet \;$ Measurements of the anomalous thermal Hall conductivity [see Sec.~\ref{hall:thermal-spin}]~\footnote{With the current estimation of various parameters in YPtBi~\cite{brydon, savary} we find $\kappa_{xy} \sim C \; 1.5 \times 10^{-4}$ Wk$^{-1}$m$^{-1}$, where the parameter $C \sim 1$ depends on the pairing symmetry [see Sec.~\ref{hall:thermal-spin}], which can, in principle, be measured~\cite{ong}.}, as well as probing the surface Andreev bound Fermi arc states with the STM quasi-particle interference (QPI) techniques are expected to distinguish among various candidates of Weyl pairing listed in Table~\ref{egt2g:competitiontable}.  

$\bullet \;$ The nuclear magnetic resonance (NMR) relaxation time ($T_1$) can also be a good probe to elucidate the scaling of the low-energy DoS since 
$1/T_1 \sim T [\varrho(E \to T)]^2$ (Korringa's relation). Therefore, as the impurity concentration is gradually increased, the inverse of the NMR relaxation time should display a crossover from $T_1^{-1} \sim T^5$ (dominated by BdG-Weyl quasiparticles) to $T_1^{-1} \sim T^3$ (inside the quantum critical regime), to $T$-linear behavior (in the thermal metallic phase) scaling, see Fig.~\ref{simpleweyl_dirty}. 
 
$\bullet \;$ The longitudinal thermal conductivity ($\kappa_{jj}$) can also be a good probe to expose various regimes of the phase diagram. Specifically, $\kappa_{jj}/T$ scales as (a) $T$ when the BdG-Weyl quasiparticles dominate the transport (clean limit), (b) $T^{2/3}$ in the entire quantum critical regime, (c) approaches a constant value in the thermal metallic side as $T \to 0$, see Fig.~\ref{simpleweyl_dirty}. Note that the magnitude of thermal conductivity will in general be different along various crystallographic directions due to the natural anisotropy in the Weyl paired state.

Above, we have discussed the experimental implications of the Weyl-paired superconducting state and how it evolves as a function of impurity scattering strength. If, on other hand, the $T$-linear dependence of the penetration depth in Fig.~\ref{fitlambda:expthr} arises from an underlying line node, then the measurements of the specific heat can be a good tool to pin down such a nodal structure. In the presence of a nodal loop (or double-Weyl node), $T^2$-dependence of the specific heat is expected to occupy a progressively wider window of temperature with a gradual decrease of impurity scattering, see Fig.~\ref{doubleweyl_dirty}. By contrast, with increasing strength of impurity scattering, the $T^2$ dependence will be gradually replaced by the $T$-linear dependence in $C_v$ (dominated by a thermal metal). The inverse of the NMR relaxation time in the presence of a nodal loop in the quasiparticle spectra is then expected to display a smooth crossover from $T_1^{-1} \sim T^3$ 
 (dominated by pseudo-ballistic quasiparticles) to $T$-linear dependence (governed by thermal metallic phase), as the disorder in the system increases. Notice $\kappa_{jj}/T$ is expected to display a $T$-linear scaling, but only when the heat-current flows in the \emph{basal plane} containing the nodal loop. Otherwise, with increasing (decreasing) impurity strength the residual value of $\kappa_{jj}/T$ as $T \to 0$ should increase (decrease)~\cite{Taillefer-review}. By contrast, in the presence of double-Weyl nodes (separated along $\hat{z}$ direction (say), for example), $\kappa_{zz}/T \sim \mbox{constant}$, while $\kappa_{jj}/T \sim T$ at low temperature, where $j=x,y$. 

We realize that it may be very difficult to tune the concentration of impurities experimentally, the task further complicated by the fact that  nodal superconductivity is easily destroyed by (non-magnetic) impurities. Nevertheless, one could still make meaningful conclusions about the pairing by exploring the phase diagram at a fixed impurity concentration, as a function of temperature. For instance, descending in temperature to the left of the disorder controlled diffusive QCP [the red dot in Fig.~\ref{simpleweyl_dirty}], one would expect to see the crossover from the critical regime (where $C_v \sim T^2$, $T_1^{-1} \sim T^3$ and $\kappa_{jj}/T \sim T^{2/3}$) to the pure Weyl regime (with $C_v \sim T^3$, $T_1^{-1} \sim T^5$ and $\kappa_{jj}/T \sim T^1$). Therefore, we believe that it is still conceivable to pin the actual nature of the pairing symmetry in half-Heusler compounds with the presently available experimental tools. We hope our detailed discussion will motivate future experiments in this class of materials.


\section{Strong topological superconductivity: Odd-parity isotropic $p$-wave pairing \label{sec: p-wave}}

In this section we study superconductivity in the Luttinger semimetal (LSM)
with isotropic $p$-wave pairing. This odd-parity pairing is the spin-3/2 generalization 
of the $B$ phase of $^3$He \cite{Volovik2003}. 
The paired state represents a time-reversal invariant, class DIII (strong)
topological superconductor (TSC) \cite{SRFL2008}. 
The topology induces a two-dimensional (2D) gapless Majorana fluid
to appear at the material surface. 
Our goal is to investigate the stability (``topological protection'') of this 2D Majorana fluid to 
perturbations that are inevitable at the surface of a real material: quenched impurities and
residual interparticle interactions.

In a previous work \cite{GDF17}, we investigated exactly these questions in the Luttinger Hamiltonian 
[Eqs.~(\ref{HLDef}) and (\ref{luttinger-spin})]
with isotropic $p$-wave pairing, but with one crucial difference. In Ref.~\cite{GDF17},
we assumed that $m > m_0$, i.e.\ that both bands in Eq.~(\ref{lut_energy}) ``bend together'',
as in the light and heavy hole bands of GaAs~\cite{murakami}. 
Assuming that both bands participate in superconductivity, the bulk winding number $\nu = 4$ in that case,
and the surface Majorana fluid exhibits coexisting linear and cubic dispersing branches \cite{Yang16}.
We showed that interactions can destabilize the clean fluid \cite{GDF17}, inducing spontaneous
time-reversal symmetry breaking and surface thermal quantum Hall 
order\footnote{We note that such surface order can also arise in the presence of $p+is$ pairing in the bulk~\cite{goswami-roy-axionSC}.}
\cite{spinthermal-2,SRFL2008,Ryu2012,Stone2012}.
By contrast, we demonstrated that quenched surface disorder is a strong perturbation  
that induces critical Anderson delocalization, with multifractal surface wave functions 
and a power-law divergence of the disorder-averaged density of states. These results
were obtained numerically via exact diagonalization, and were found to agree very well with the predictions
of a certain 2D conformal field theory (CFT). The CFT is the current algebra SO($n$)${}_\nu$
(with $\nu = 4$), where $n \rightarrow 0$ is a replica index \cite{WZWP4}. 
We concluded that the surface states are governed by this CFT in the presence of arbitrarily weak
disorder. Moreover, in a separate work we established that the class DIII SO($n$)${}_\nu$
theory is stable against the effects of residual quasiparticle-quasiparticle interactions \cite{WZWP4}. 
The main takeaway of Ref.~\cite{GDF17} was that disorder can \emph{enhance} topological protection
at the surface of a higher-spin TSC.

The SO($n$)${}_\nu$ CFT can be ``derived'' via certain conformal embedding rules for 
surface states of model spin-1/2 TSCs \cite{WZWP4}. In the case
of the LSM with $p$-wave pairing studied here and for the closely related model in \cite{GDF17}, these rules do not 
obviously apply. In particular, the conformal embedding argument assumes that the clean
limit is also a CFT, i.e.\ free relativistic fermions (in $2 + 0$ dimensions; in the absence
of interactions, we can study the problem at a fixed single particle energy \cite{WZWP4}). 
By contrast, the clean surface states of higher-spin TSCs typically have
higher (e.g.\ cubic) dispersion \cite{Fang2015,Yang16}, and are not conformally invariant.

Here we consider the problem in the LSM, where electron and hole bands bend oppositely.
This gives rise to a different winding number ($\nu = 3$) and different surface states,
depending on the doping. 
In fact, we invent here a \emph{generalized surface model}
(see Sec.~\ref{sec: DIIIGen}) that allows us to efficiently simulate 
noninteracting
surface states corresponding to a bulk TSC in class DIII 
with arbitrary integer winding number $\nu$. The model has $\nu$-fold dispersion, such that the 
large-$\nu$ limit corresponds to a highly flattened surface band with a strongly diverging clean DoS.

\begin{figure}[b]
\includegraphics[angle=0,width=.45\textwidth]{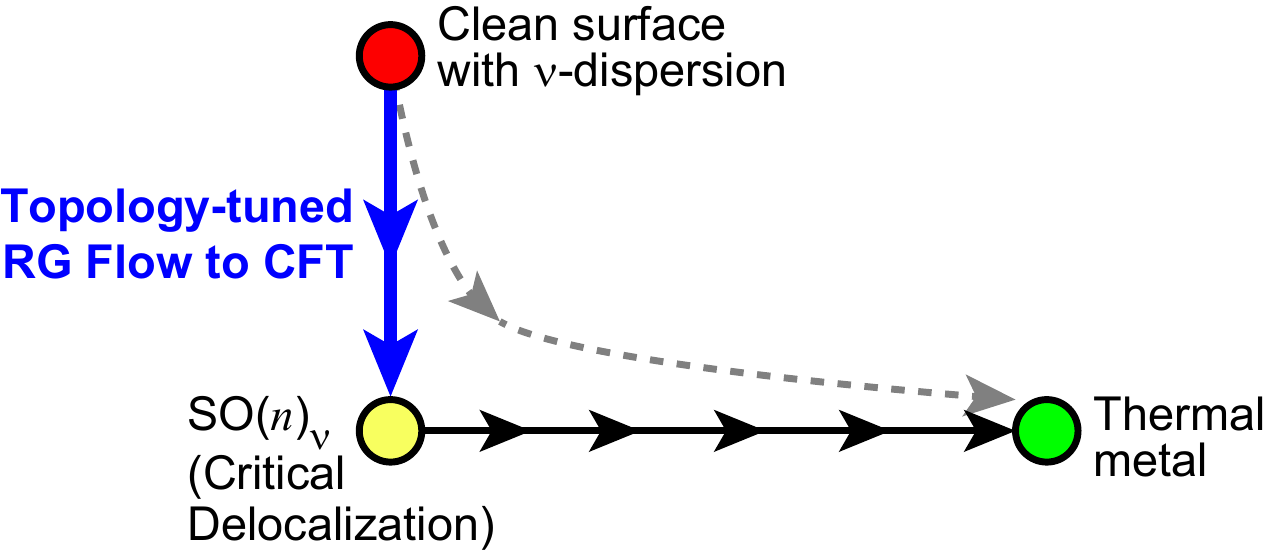}
\caption{
Schematic phase diagram for the noninteracting 2D surface states of a class DIII
bulk topological superconductor. The fixed point representing the clean surface 
band structure 
(red dot)
is unstable in the presence of time-reversal preserving quenched disorder for any $\nu \geq 3$, 
where $\nu$ is the integer bulk winding number. 
The precise form of the clean limit depends on details. 
For a  spin-1/2 bulk, one can have $\nu$ species of massless relativistic Majorana fermions,
with disorder that enters as a nonabelian gauge potential scattering between these \cite{WZWP4}. 
For isotropic $p$-wave pairing in the LSM studied here with winding number $\nu = 3$, the surface states 
in the hole-doped case consist of a single two-component surface fermion with cubic dispersion, see Fig.~\ref{Fig--SurfaceBSCubic} and Eq.~(\ref{hsDef})
\cite{Fang2015}, c.f.\ Refs.~\cite{Yang16,GDF17}.  
Our generalized surface theory
in Eq.~(\ref{hWeff}) has $\nu$-fold dispersion for the corresponding winding number. 
The disordered system should be described by a class DIII non-linear sigma model with a Wess-Zumino-Novikov-Witten (WZNW)
term. The WZNW term prevents Anderson localization \cite{AndRev08,WZWP4}. 
This theory has a stable thermal metal phase (green dot) 
and 
an unstable, critically delocalized fixed point.
The latter (yellow dot) is governed by the SO($n$)${}_\nu$ CFT \cite{WZWP4,WZWP3}. 
Our numerical results are generally consistent with the  SO($n$)${}_\nu$ theory, see Figs.~\ref{Fig--nu=3}--\ref{Fig--nu=7},
implying that the renormalization group trajectory away from the clean limit 
is \emph{fine-tuned by the topology} to flow into the CFT (solid vertical flow), instead of flowing 
into the thermal metal (dashed flow).
The same conclusion was reached for a model with $\nu = 4$ in Ref.~\cite{GDF17}.
}
\label{Fig--CFTFlow}
\end{figure}

On physical grounds, the most general expectation for class DIII in this case would be that disorder induces a 
surface thermal metal \cite{AndRev08}. 
In two spatial dimensions, the thermal metal phase in class DIII is stable due to weak antilocalization. 
Moreover, the SO($n$)${}_\nu$ CFT fixed point, while stable against interactions, 
is technically \emph{unstable} towards flowing into the thermal metal \cite{GDF17};
see Fig.~\ref{Fig--CFTFlow}.   
Despite this, in Ref.~\cite{GDF17} for winding number $\nu = 4$ and 
here for \emph{generic} $\nu \geq 3$, we provide strong numerical evidence that any disorder
induces the quantum critical scaling associated to SO($n$)${}_\nu$, with universal predictions
for experiment that depend only on $\nu$. 
These include power-law scaling for the tunneling density of states, 
a quantized thermal conductivity divided by temperature \cite{WZWP3,GDF17}, 
and a universal multifractal spectrum of local DoS fluctuations.  
These states are also robust against interactions for any $\nu$ \cite{WZWP4}.

Our results suggest a deep connection between the bulk topology of three-dimensional TSC
and the universal physics of the \emph{quench-disordered} two-dimensional Majorana surface fluid, 
despite the fact that key attributes of the clean surface depend on details of the bulk. 
In particular, it suggests a \emph{topological generalization} of the conformal embedding rule 
[$\text{SO}(n \nu)_1 \supset \text{SO}(n)_\nu \oplus \text{SO}(\nu)_n$] 
used to link $\nu$ clean relativistic Majorana fermions to the SO($n$)${}_\nu$ CFT in the presence of disorder \cite{WZWP4}.
This topological generalization should apply 
in the replica limit $n \rightarrow 0$
to any surface band structure for any strong class DIII TSC with winding number $\nu \geq 3$, subject to time-reversal
invariant quenched disorder.

Beyond fundamental interest, the Eliashberg calculations in Ref.~\cite{savary} suggest that 
isotropic $p$-wave pairing gives the dominant non-$s$-wave channel in a hole-doped LSM
due to optical-phonon--mediated pairing. For this reason we focus mainly on the hole-doped model
in the following, which has $\nu = 3$ (see below).

\subsection{Bulk and surface theory}

We write the Luttinger Hamiltonian in terms
of the Nambu spinor defined by Eq.~(\ref{nice-Nambu}),
\begin{align}\label{HDef}
	H
	=
	\frac{1}{2}
	\int \frac{d^3 \vex{k}}{(2 \pi)^3}
	\,
	\Psi_N^\dagger(\vex{k})
	\,
	\hat{h}(\vex{k})
	\,
	\Psi_N(\vex{k}),
\end{align}
where the $8\times8$ Bogoliubov-de Gennes (BdG) Hamiltonian is
\begin{align}\label{hBdG}
	\hat{h}(\vex{k})
	=
	\hat{h}_{L}(\vex{k})
	\,
	\tau_3
	+        
	\Delta_p
	\left(
	\vex{J}
	\cdot
	\vex{k}
	\right)
	\,
	\tau_1.
\end{align}
Here $\hat{h}_L$ is the Luttinger operator from Eq.~(\ref{luttinger-spin}) and 
$\vex{J}$ denotes the vector of spin-3/2 generators [see Eq.~(\ref{JDef})]. 
The Pauli matrices $\{\tau_\mu\}$ act on the particle-hole (Nambu) space. 
The parameter $\Delta_p$ is the real $p$-wave pairing amplitude; 
with this choice, Eq.~(\ref{hBdG}) is time-reversal invariant [see Eq.~(\ref{ChiralTRI})]. 
It also satisfies the particle-hole condition in Eq.~(\ref{MajPH}), using Eq.~(\ref{pseudoreal}).

We assume weak BCS pairing so that $\mu > 0$ ($\mu < 0$) describes
superconductivity in the $|m_s| = 1/2$ conduction ($|m_s| = 3/2$ valence) 
band of the Luttinger Hamiltonian. The physical bulk quasiparticle energy spectrum of Eq.~(\ref{hBdG}) is fully gapped,
\begin{align}
	E_{\pm}(\vex{k}) 
	= 
	\sqrt{
	\left(
	|\lambda_1 
	\pm 
	2 \lambda_2| k^2
	- 
	|\mu|
	\right)^2 
	+ 
	\left[\left({\textstyle{\frac{2 \mp 1}{2}}}\right) \Delta_p k \right]^2
	},
\end{align}
where $2 \lambda_2 > \lambda_1$ is required so that conduction and
valence bands bend oppositely [or $m_0 > m$ in Eq.~(\ref{lut_energy})], 
and $E_+$ ($E_-$) corresponds to superconductivity in 
the conduction (valence) band.  
The assumption of weak BCS pairing around a finite Fermi surface means
that we can project the BdG Hamiltonian into the 
$|m_s| = 1/2$ conduction 
or 
$|m_s| = 3/2$ valence 
band. 
The results are 
\begin{eqnarray}\label{hBdGProj}
	\hat{h}_{1/2}(\vex{k})
	&=&
	\left[(\lambda_1+2\lambda_2)k^2-\mu\right] \tau_3
	+
	\frac{\Delta_p}{2}
	\begin{bmatrix}
		- k_z & -\frac{\kb^2}{\kc} \\
		-\frac{\kc^2}{\kb} & k_z
	\end{bmatrix}
	\tau_1,
\nonumber\\
	\hat{h}_{3/2}(\vex{k})
	&=&
	\left[(\lambda_1-2\lambda_2)k^2-\mu\right] \tau_3	
\nonumber\\
	&&+
	\frac{\Delta_p}{\alpha(\vex{k})}
	\begin{bmatrix}
		-
		k_z \, \beta(\vex{k})	
		& 
		\kb^3
	\\
		\kc^3
		& 
		k_z \, \beta(\vex{k})	
	\end{bmatrix}
	\tau_1,
\end{eqnarray}
where $\kc \equiv k_x - i k_y$, $\kb = \kc^*$, and
\begin{align}
	\alpha(\vex{k})
	=
	\frac{2}{3}
	\left(4 k_z^2  + |\kc|^2 \right),
	\quad
	\beta(\vex{k})
	=
	\left(4 k_z^2 + 3 |\kc|^2 \right).
\end{align}
Here we have diagonalized $(\vex{J}\cdot\vex{k})^2$
but not $(\vex{J}\cdot\vex{k})$, so that the matrix elements
are rational functions of the momentum components. 
This is essential for obtaining a local surface theory, derived below.

\begin{figure}
\subfigure[]{
\includegraphics[angle=0,width=.21\textwidth]{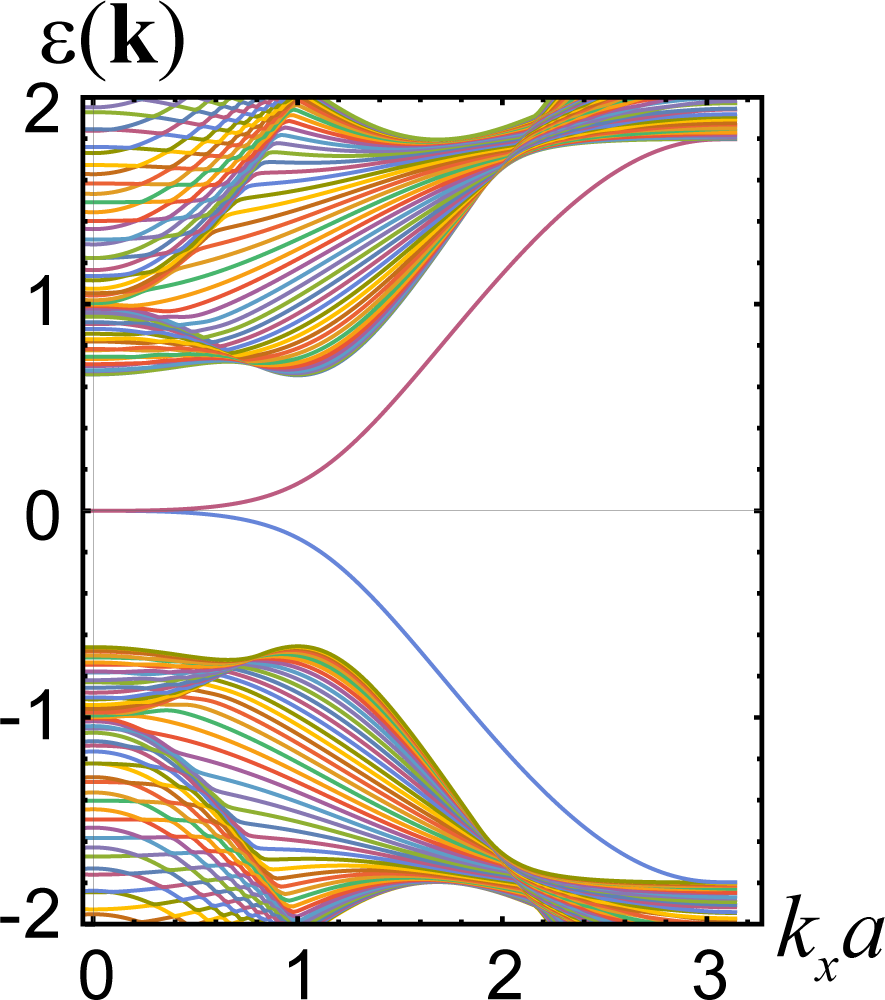}
\label{Fig--SurfaceBSCubic}
}
\subfigure[]{
\includegraphics[angle=0,width=.21\textwidth]{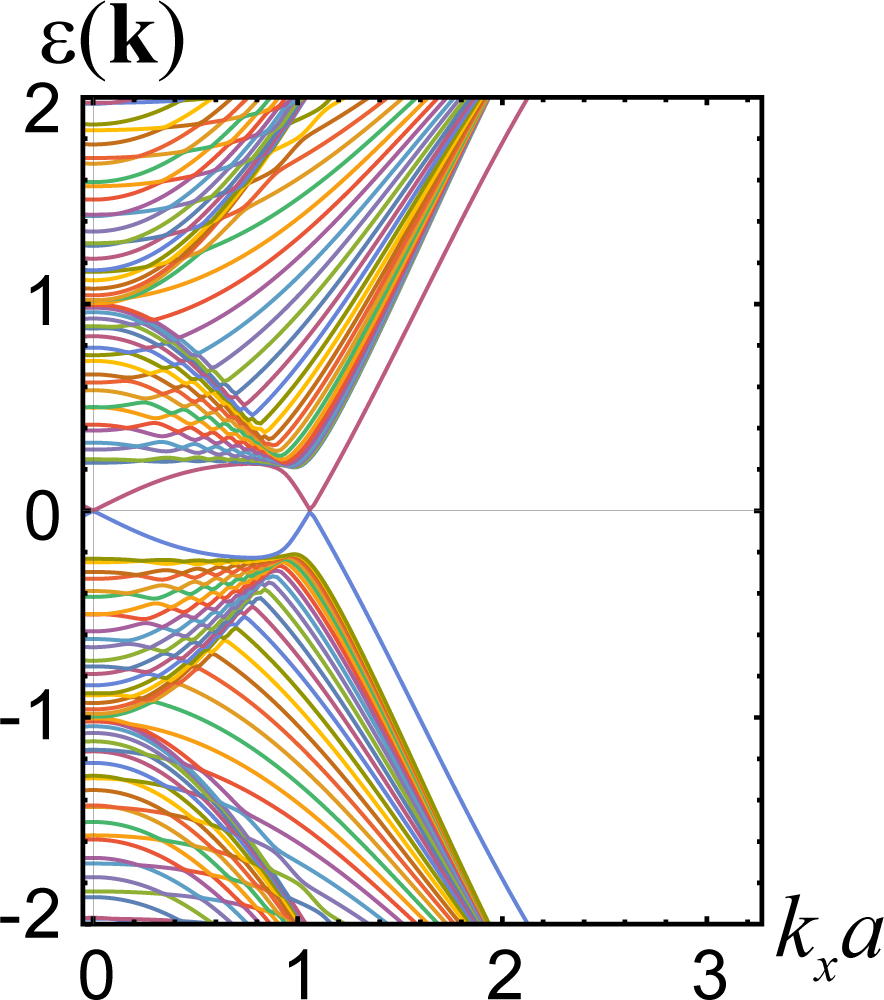}
\label{Fig--SurfaceBSLin}
}
\caption{Surface Majorana fluid band structure for the Luttinger Hamiltonian
with isotropic $p$-wave pairing. This is a class DIII, strong topological superconductor 
with winding number $|\nu| = 3$ for pairing arising from either the conduction or valence bands. 
Results shown here are obtained from a lattice regularization and termination of Eq.~(\ref{hBdG});
the momentum $k_x$ is measured in units of the lattice spacing $a$.
The 
left
panel (a) shows the cubic-dispersing two-dimensional surface states obtained for hole-doping, see Eq.~(\ref{hsDef}).
Only positive $k_x$ is depicted, since the results are symmetric under reflection about $k_x = 0$.
The BdG parameters are 
$\Delta_p=1$, 
$\lambda_1=0.1$, 
$\lambda_2=0.5$ 
and 
$\mu=-1$. 
The 
right
panel (b) shows a relativistic cone centered at $\vex{k} = 0$ and a gapless ring
in the electron-doped case. The parameters are the same as for (a), except that $\mu = + 1$.} 
\label{Fig--SurfaceBS}
\end{figure}

We employ the winding number defined by Schnyder \textit{et al}.\ \cite{SRFL2008} to
characterize the topology of the bulk. 
After rotating $\tau_3 \rightarrow \tau_2$, one introduces the 
matrix $4 \times 4$ matrix $Q(\vex{k}) = U^{-1}(\vex{k}) \, \Lambda \, U(\vex{k})$,
where $U(\vex{k})$ diagonalizes 
$\hat{h}_{|m_s|}(\vex{k})$, 
and $\Lambda = \diag(1,1,-1,-1)$ is the flattened matrix of energy eigenvalues. 
Then $Q$ is off-diagonal, 
\begin{align}
	Q
	=
	\begin{bmatrix}
	0 & q\\
	q^{-1} & 0
	\end{bmatrix},
\end{align}
and the winding number is given by 
\begin{align}
	\!\!
	\nu
	=
	\int 
	\frac{d^3 k}{24\pi^2}
	\,
	\epsilon^{ijk}
	\Tr\left[
		(q^{-1}\parr_i q)
		(q^{-1}\parr_j q)
		(q^{-1}\parr_k q) 
	\right],\!\!
\end{align}
with repeated indices summed. 
We find that $|\nu| = 3$ for the valence and conduction bands.
From here on we ignore $\sgn(\nu)$, which can only be important at an interface
(e.g., a physical surface) 
where this sign flips. 
Our winding number is in agreement with Ref.~\cite{Fang2015} for the $|m_s| = 3/2$ band, but differs
from that obtained for the $|m_s| = 1/2$ band, in which $\nu = 1$ was claimed.  
We show below that surface state calculations support our results.  
We believe that the discrepancy comes from the fact the authors of Ref.~\cite{Fang2015} 
used a Fermi surface winding number method \cite{winsurf}, which gives the correct winding number 
only if the system is non-degenerate.

To obtain the effective surface Hamiltonian we follow the conventional approach of 
terminating in the $z$-direction and diagonalizing $\hat{h}_{|m_s|}(\vex{k},k_z \rightarrow -i \parr_z)$ 
where $\vex{k}={k_x,k_y}$ denotes momentum parallel to the surface. 
For the $|m_s| = 3/2$ valence band, 
applying hard-wall boundary conditions we obtain zero energy surface states at $\vex{k} = 0$ of the form
\begin{align}\label{psi_0}
	\ket{\psi_{0,m_s}}
	=
	\ket{\tau_2 = \sgn(m_s)}
	\otimes
	\ket{m_s}
	\otimes
	\ket{f_{m_s}}.
\end{align}
The particle-hole spin locks along the $+ \tau_2$ ($-\tau_2$) direction
for positive (negative) $m_s$ \cite{Yang16,GDF17}. 
In Eq.~(\ref{psi_0}), $\braless{z \!}\ket{f_{m_s}} = f_{m_s}(z)$ denotes the bound state envelope function.

Using first-order $k\cdot p$ perturbation theory we obtain the surface effective Hamiltonian,
\begin{align}\label{hsDef}
	\hsth(\vex{k})
	\propto
	\frac{\Delta_p}{k_F^2}
	\begin{bmatrix}
	0 & i\kb^3 \\
	-i\kc^3 & 0 \\
	\end{bmatrix}.
\end{align}
Eq.(\ref{hsDef}) satisfies the projected version of the particle-hole symmetry
in Eq.~(\ref{MajPH}), 
\begin{align}\label{PH-Surf}
	-
	\Mps
	\,
	\left[\hsth\right]^\T(-\vex{k})
	\,
	\Mps
	=
	\hsth(\vex{k}),
	\quad
	\Mps = \sigma_1,
\end{align}
and the projected time-reversal symmetry [Eq.~(\ref{ChiralTRI})] 
\begin{align}\label{Chiral-Surf}
	-
	\Mss
	\,
	\hsth(\vex{k})
	\,
	\Mss
	=
	\hsth(\vex{k}),
	\quad
	\Mss = \sigma_3.
\end{align}
Here the matrices $\{\sigma_\mu\}$ act on the components $m_s = \pm 3/2$.

Fig.~\ref{Fig--SurfaceBS}
shows the clean Majorana surface bands obtained numerically 
from a lattice regularization of Eq.~(\ref{hBdG})
for 
(a) $|m_s| = 3/2$ valence-band--hole
and 
(b) $|m_s| = 1/2$ conduction-band--electron
superconductivity. 
Below we focus on the hole-doped case in which the surface fluid has cubic dispersion
[Eq.~(\ref{hsDef})]. 
The surface fluid in the electron-doped case is
depicted in Fig.~\ref{Fig--SurfaceBSLin}, and 
exhibits a linear Majorana cone around $\vex{k}=0$ 
and a zero-mode ring at finite surface momentum; 
the latter structure is inconsistent with $\nu = 1$ \cite{Fang2015,Yang16}.

\subsection{Quenched surface disorder, class DIII SO($n$)${}_\nu$ conformal field theory, and numerical results}

We now turn to perturbations of the surface theory, focusing on the cubic-dispersing Majorana fluid that arises from hole-doped superconductivity. We can write the surface Hamiltonian as
\begin{align}\label{Hs0Def}
		\Hs 
		=	 
		\frac{1}{2} 
		\int d^2 \vex{r} \, 
		\eta^\T \, 
		\Mps \, 
		\left(
		\sigma_- \, \parr^3 - \sigma_+ \, \parb^3
		\right)
		\eta,
\end{align}
where  $\eta \rightarrow \eta_{m_s}$ is a two-component Majorana spinor and $\vex{r}$ is the position vector. 
The chiral derivative operators are $\{\parr,\parb\} \equiv (1/2)(\partial_x \mp i \partial_y)$, 
while $\sigma_\pm \equiv \sigma_1 \pm i \sigma_2$. Here we have set the prefactor of Eq.~(\ref{hsDef}) equal to one.

The simplest class of surface perturbations are constant bilinears. Such an operator can be written as
$\eta^\T \Mps \Lambda \eta$, with $\Lambda$ a $2 \times 2$ Hermitian matrix. The only bilinear that
satisfies particle-hole in Eq.~(\ref{PH-Surf}) (i.e., which does not vanish under Pauli exclusion) is 
the mass term $\Lambda = \sigma_3 \simeq J^z$. This is the projection of the spin operator perpendicular to the surface.
The nonzero expectation value of this term 
(due e.g.\ to a coupling with an external Zeeman field)
would open a surface energy gap and signal time-reversal symmetry breaking. 
The time-reversal broken state would reside in a plateau of a surface thermal quantum Hall effect \cite{spinthermal-2,SRFL2008,Ryu2012,Stone2012}.

These considerations are almost identical to $^3$He-B \cite{Volovik2003,SRFL2008}, which has spin-1/2 and $\nu = 1$. 
The only difference is that the derivatives in Eq.~(\ref{Hs0Def}) appear to the first power for $\nu = 1$, whereas here
we get the $\nu = 3$ power for the spin-3/2 bulk.

Residual quasiparticle-quasiparticle interactions should be short-ranged (due to screening by the bulk superfluid).
Since $\eta$ is a two-component Majorana field, the most relevant interaction that we can write down is 
\begin{align}
		\His = u \int d^2 \vex{r} \; \eta_1 \Nabla \eta_1 \cdot \eta_2 \Nabla \eta_2. 
\end{align}
The coupling strength $u$ has dimensions of length for cubic dispersion and is therefore irrelevant in the sense of the renormalization group (RG) \cite{Fang2015}.

Finally we turn to quenched disorder, which is always present at the surface of a real sample. 
We assume the disorder is non-magnetic, but may arise due to neutral adatoms, charged impurities, grain boundaries, etc. In other words, any time-reversal invariant surface potential perturbation is allowed. 
Since the only bilinear without derivatives is the massive, time-reversal odd $J^z$ operator discussed above, we must broaden the search to include bilinears with derivatives. 
The most relevant possible potential can be encoded in the Hamiltonian
 \begin{align}\label{HdSDef}
		\Hds
		=	 
		-
		\frac{i}{2} 
		\int d^2 \vex{r}
		\left[
		\eta^\T(\vex{r}) \, 
		\Mps \,
		\sigma_\alpha
		\,
		\stackrel{\longleftrightarrow}{\frac{\partial}{\partial x^\beta}}
		\eta(\vex{r})
		\right]
		P_{\alpha \beta}(\vex{r}).
\end{align}
In this equation repeated indices are summed, $\alpha,\beta \in \{x,y\}$.  
We assume that $P_{\alpha \beta}(\vex{r})$ is a white-noise-correlated random potential
with variance $\lambda$. Then $\lambda$ has dimensions of $1/(\text{length})^2$ and is a relevant
perturbation to the clean cubic band structure.

The effects of disorder cannot be treated perturbatively.
The standard procedure would produce a disorder-averaged nonlinear sigma model 
in class DIII, which possesses a stable thermal metal phase \cite{SenthilFisher,AndRev08}. 
Although the thermal metal is perturbatively accessible in the sigma model
with the WZNW term,
the critical SO($n$)${}_\nu$ CFT fixed point is not, except for the limit of large $\nu$.
Therefore we resort to numerics in the remainder of this section. 
The question we want to answer is whether disorder flows into the SO($n$)${}_\nu$ CFT
or the thermal metal, see Fig.~\ref{Fig--CFTFlow}.

The noninteracting BdG Hamiltonian implied by Eqs.~(\ref{Hs0Def}) and (\ref{HdSDef}) has
momentum space matrix elements 
\begin{align}\label{Dirtyhs}
	\!\!
	\left[\hs\right]_{\vex{k},\vex{k'}}
	=&\,
	\begin{bmatrix}
	0 & i \kb^3 \\
	-i \kc^3 & 0 
	\end{bmatrix}
	\!
	\delta_{\vex{k},\vex{k'}}
	+
	(k_x + k'_x)
	\begin{bmatrix}
	0 & 1
    \\
    1 & 0 
	\end{bmatrix}
	\!
	P_x(\vex{k}-\vex{k}')
\nonumber
\\&\,   
     +
    (k_y + k'_y)
    \begin{bmatrix}
	0 & -i
    \\
    i  & 0 
	\end{bmatrix}
	\!
	P_y(\vex{k}-\vex{k}'),
\end{align}
where we have taken $P_{\alpha\beta}(\vex{r})$ to be diagonal in its lower indices. 
Gaussian white noise disorder can be efficiently simulated in momentum space
using a random phase method \cite{YZCP1},
\begin{align}\label{Palpha}
	P_\alpha(\vex{k})
	=
	\frac{\sqrt{\lambda}}{L}
	e^{i \theta_\alpha(\vex{k})}
	\exp\left(- \frac{k^2 \xi^2}{4}\right),
\end{align}
where $\theta_\alpha(-\vex{k}) = - \theta_\alpha(\vex{k})$, but these are otherwise independent,
uniformly distributed random phases. 
The parameters $L$, $\xi$ and $\lambda$ denote the system size, correlation length, and disorder strength respectively. For exact diagonalization, we choose periodic boundary conditions so that $\vex{k} = (2 \pi / L) \vex{n}$, and the components of $\vex{n} \in \{\mathbb{Z},\mathbb{Z}\}$ run over a square with $-N_k \leq n_i \leq N_k$, for $i = 1,2$.
Here $N_k$ determines the size of the vector space in which we diagonalize, which is $2(2 N_k + 1)^2$. 
While the choice of $L$ is arbitrary, we use it to fix the ultraviolet momentum cutoff $\Lambda = 2 \pi N_k / L$. 
The correlation length $\xi$ and the dimensionful disorder strength $\lambda$ are then measured in terms of powers of $\Lambda$. 
The random-phase approach is equivalent to the disorder-average up to finite-size corrections \cite{YZCP1}.
We perform the calculations in momentum space in order to avoid fermion doubling.

\begin{figure}[t!]
\subfigure{
\includegraphics[angle=0,width=0.23\textwidth]{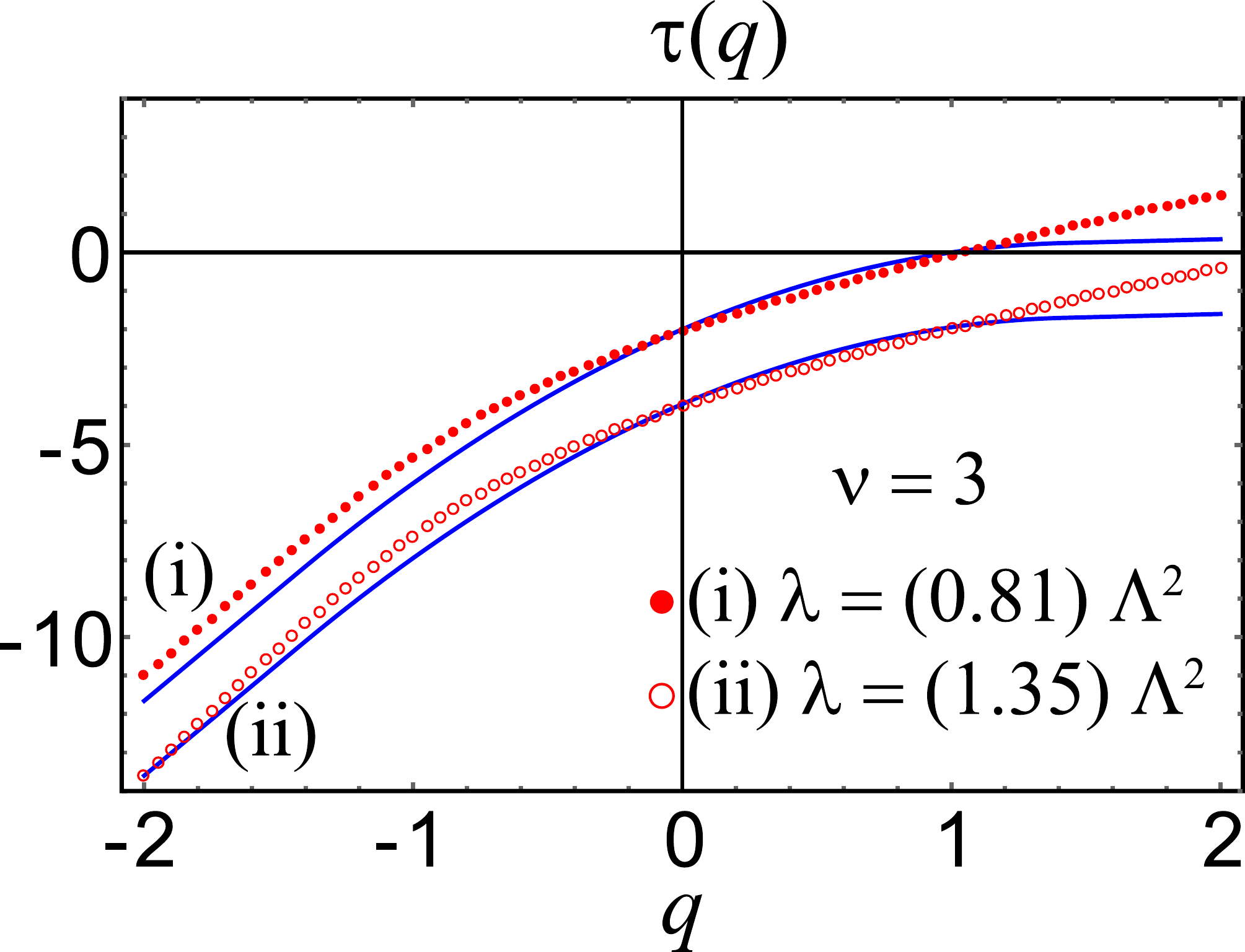}
}
\hspace{-8pt}
\subfigure{
\includegraphics[angle=0,width=0.23\textwidth]{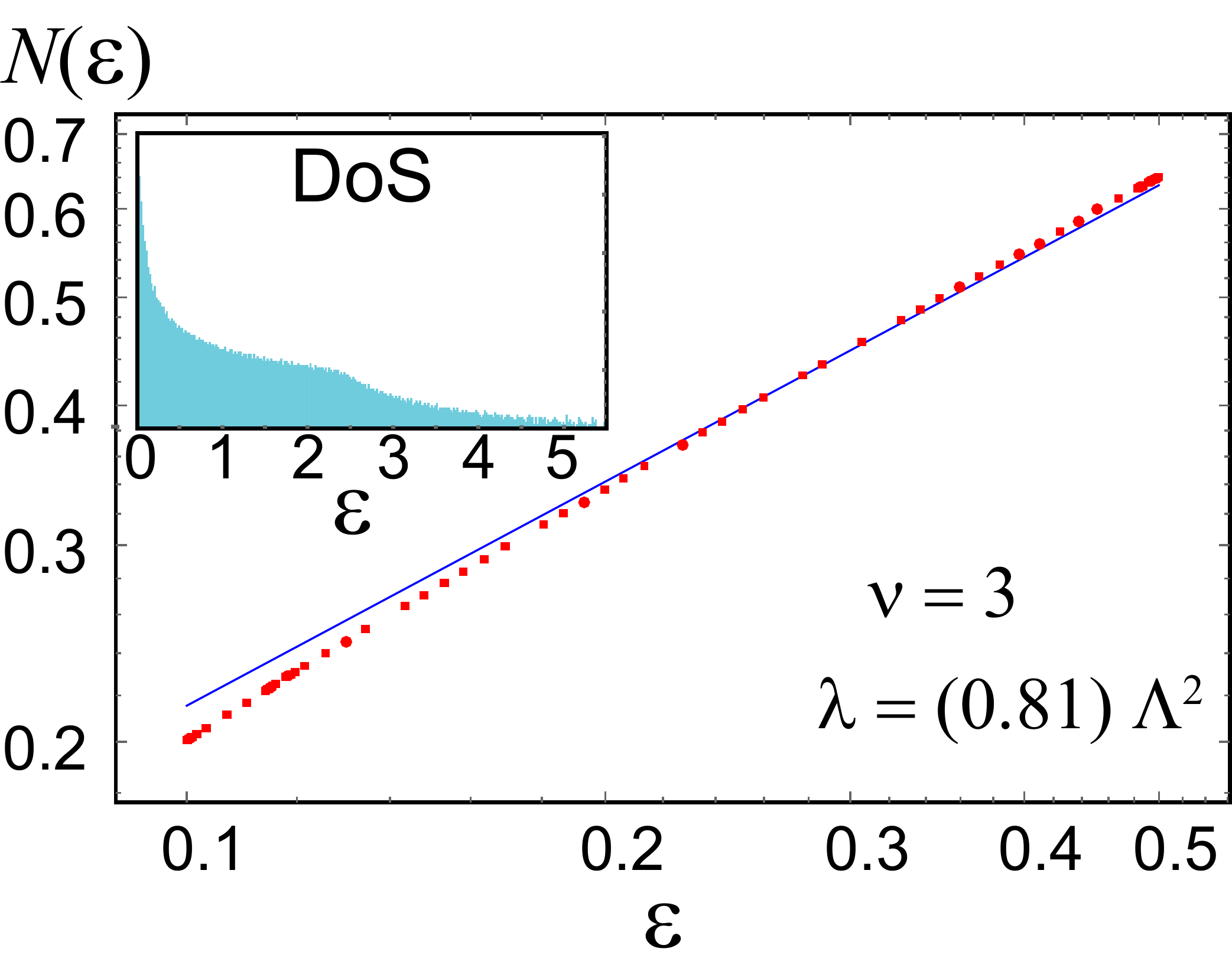}
}
\caption{
Numerical results for the surface states of the hole-doped 
Luttinger semimetal with isotropic $p$-wave pairing in the bulk
and time-reversal symmetry preserving disorder on the surface. 
The winding number of the bulk is $\nu = 3$. 
The
left
plot shows the multifractal spectrum [Eqs.~(\ref{BoxProb})--(\ref{tauq})]
for two typical lowest energy surface wave functions in fixed disorder realizations.
The dotted red curves are the numerical results, while the solid blue curve is the 
analytical prediction from the SO($n$)${}_3$ CFT [Eqs.~(\ref{tauq})--(\ref{theta})]. 
The curves marked (i) and (ii) correspond to two different disorder strengths $\lambda$;
the second one is shifted vertically for clarity.  
Since $\lambda$ has dimensions of 1/$(\text{length})^2$, it is measured in units of
the squared momentum cutoff $\Lambda^2$ (see text). 
The system size is a $109 \times 109$ grid of momenta. 
Box sizes $b$=1 and $b$=5 are used to extract $\tau(q)$ [see Eqs.~(\ref{BoxProb}) and (\ref{IPR})]. 
The 
right
plot shows the integrated density of states $N(\e)$. 
In this case, both the clean limit and the SO($n$)${}_3$ theory predict 
$N(\e) \sim \e^{2/3}$ [Eq.~(\ref{N(e)Def})]. 
The full surface density of states is exhibited in the inset.
For $\nu = 3$ the effects of disorder are strong, as indicated by the 
analytical result for the universal multifractal spectrum (blue curves, 
left
panel). 
It is almost ``frozen''
[a frozen state has $\tau(q) = 0$ for $q > q_c$ \cite{Chamon96,RyuHatsugai01,CarpentierLeDoussal,Fyodorov09,YZCP1}]. 
This means that the typical wave function consists of a few rare peaks with arbitrarily large separation,
see Fig.~1(b) in Ref.~\cite{YZCP1} for an example.  
We expect that finite size effects are quite severe in this case, 
responsible for the deviation between the analytical prediction and numerics.
See Figs.~\ref{Fig--nu=5} and \ref{Fig--nu=7} for higher $\nu$, which give much better agreement. 
}
\label{Fig--nu=3}
\end{figure}

To characterize the disordered surface theory, we study the scaling of the disorder-averaged DoS $\rhos(\e)$ and wave function multifractality, measures that are expected to show universal behavior at the SO($n$)${}_{3}$ fixed point.
The clean surface has $\rhos(\e) \propto |\e|^{-1/3}$ due to the cubic dispersion. 
For winding number $\nu$, in the presence of time-reversal preserving disorder the SO($n$)${}_\nu$ theory predicts the scaling
behavior of the disorder-averaged DoS \cite{LeClair08,WZWP4} to be
$
		\rhos(\e) \propto |\e|^{-1 / (2 \nu - 3)}.
$
In the case of the hole-doped LSM with $\nu = 3$, the clean and dirty CFT predictions coincide. For the generalized surface theory introduced below
[defined via Eq.~(\ref{hWeff})] or the $\nu = 4$ model studied in Ref.~\cite{GDF17}, the clean and dirty predictions differ, so that the DoS provides a useful diagnostic.
We will plot the integrated density of states (IDoS) $N(\e)$. For the SO($n$)${}_\nu$ theory 
\begin{align}\label{N(e)Def}
	N(\e) \equiv \int_0^\e d \e' \, \rhos(\e') \sim |\e|^{(2 \nu - 4)/(2 \nu - 3)}.
\end{align}

The other measure that we will employ here as a numerical test for 
the SO($n$)${}_\nu$ CFT is wave function multifractality. 
The disorder-induced spatial fluctuations of the local DoS $\rhos(\e, \vex{r})$ 
are encoded in the multifractal spectrum $\tau(q)$ \cite{WZWP4,AndRev08}. 
The $\tau(q)$ spectrum measures the sensitivity of extended wave functions to 
the sample boundary. By partitioning a large area $L\times L$ of the surface 
with boxes of small size $b << L$, 
one can define the box probability $\mu_n$ 
and inverse participation ratio (IPR) $\mathcal{P}_q$ 
in terms of a particular wave function $\psi(\vex{r})$ via
\begin{align}\label{BoxProb}
		\mu_n
		=
		\frac{
		\int_{\mathcal{A}_n}d^2\vex{r} \, |\psi(\vex{r})|^2
		}{
		\int_{L^2}d^2\vex{r} \, |\psi(\vex{r})|^2}, 
		\quad
		\mathcal{P}_q\equiv\sum_n\mu^q_n,
\end{align}
where $\mathcal{A}_n$ denotes the $n$th box. 
In the case of a \emph{typical} critically delocalized wave function, 
one expects that
\begin{align}\label{IPR}
		\mathcal{P}_q
		\sim
		\left(b/L\right)^{\tau(q)},
\end{align}
where $\tau(q)$ is self-averaging and universal~\cite{AndRev08}. 
For TSC surface states, the multifractal spectrum $\tau(q)$ is 
expected to have the form \cite{Chamon96,WZWP4,YZCP1}
\begin{align}\label{tauq}
		\tau(q)
		=
		\left\{
		\begin{array}{ll}
		(q-1)(2 - \theta_{\nu} \, q),				\quad 	& q < |q_c|,\\
		(\sqrt{2} - \sqrt{\theta_{\nu}})^2 q,		\quad	& q > q_c, \\
		(\sqrt{2} + \sqrt{\theta_{\nu}})^2 q,		\quad	& q < -q_c,	
		\end{array}
		\right.
\end{align}
where
\begin{align}\label{qc}
		q_c \equiv \sqrt{2 / \theta_{\nu}}.
\end{align}
The spectrum is quadratic below the \emph{termination threshold} $q = \pm q_c$, beyond which it is linear
\cite{Chamon96,AndRev08,MFCP1}. 

For disordered class DIII surface states and winding number $\nu$, the SO($n$)${}_\nu$ theory predicts \cite{WZWP4}
\begin{align}\label{theta}
		\theta_\nu  = 1/(\nu - 2), \quad \nu \geq 3. 
\end{align}
Eqs.~(\ref{N(e)Def}) and (\ref{theta}) are exact results that obtain from the primary field spectrum of SO($n$)${}_\nu$
in the replica $n \rightarrow 0$ limit. 
Isotropic $p$-wave pairing in the Luttinger semimetal gives $\nu = 3$, so that 
$\theta_\nu = 1$ and $q_c = \sqrt{2} \simeq 1.4$. This corresponds to quite strong multifractality, which presents some difficulties as we will see. By contrast, large $\nu$ gives $\theta_\nu \ll 1$ and $q_c \gg 1$, corresponding to weakly multifractal (nearly plane-wave) states.

\begin{figure}
\subfigure{
\includegraphics[angle=0,width=0.23\textwidth]{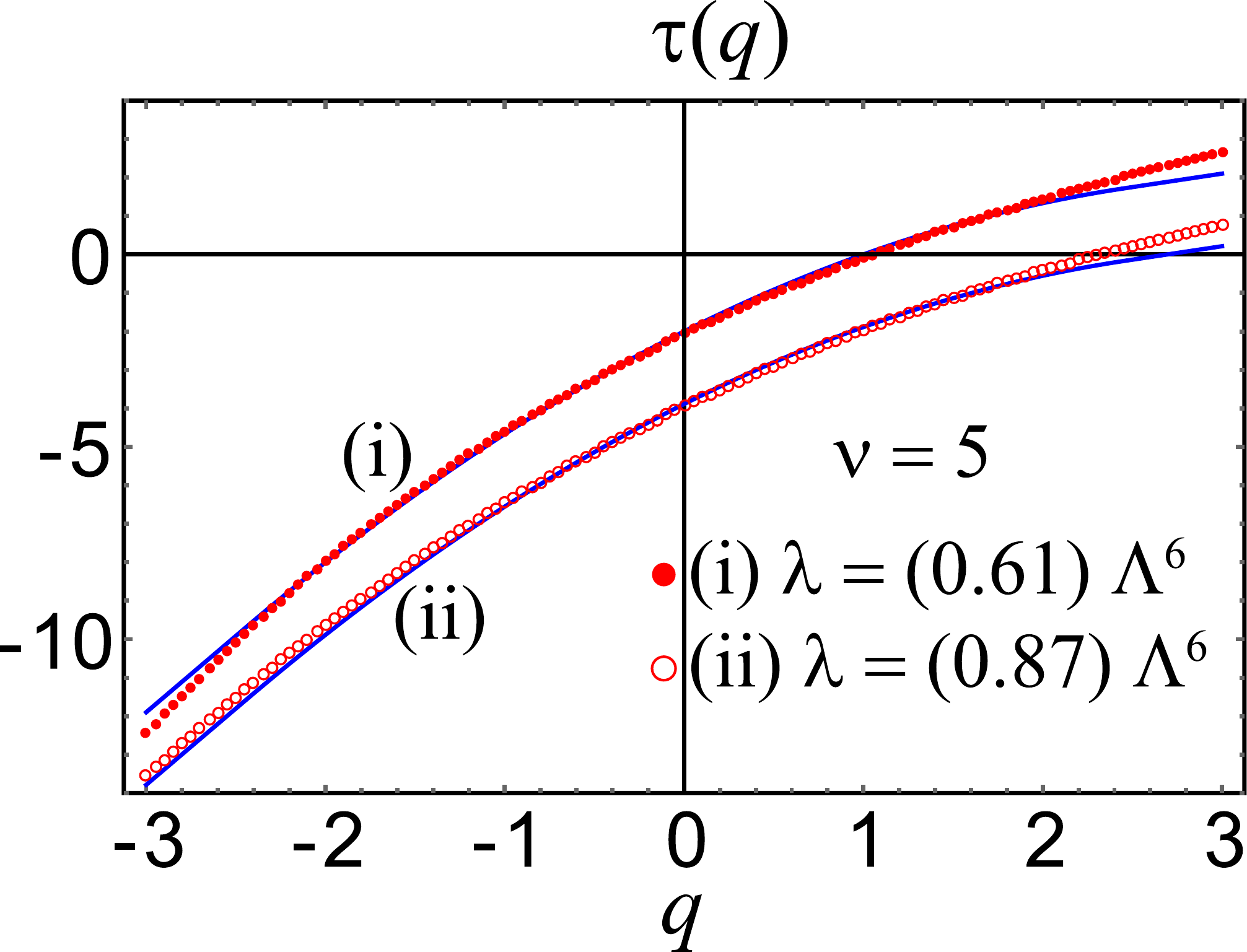}
}
\hspace{-8pt}
\subfigure{
\includegraphics[angle=0,width=0.23\textwidth]{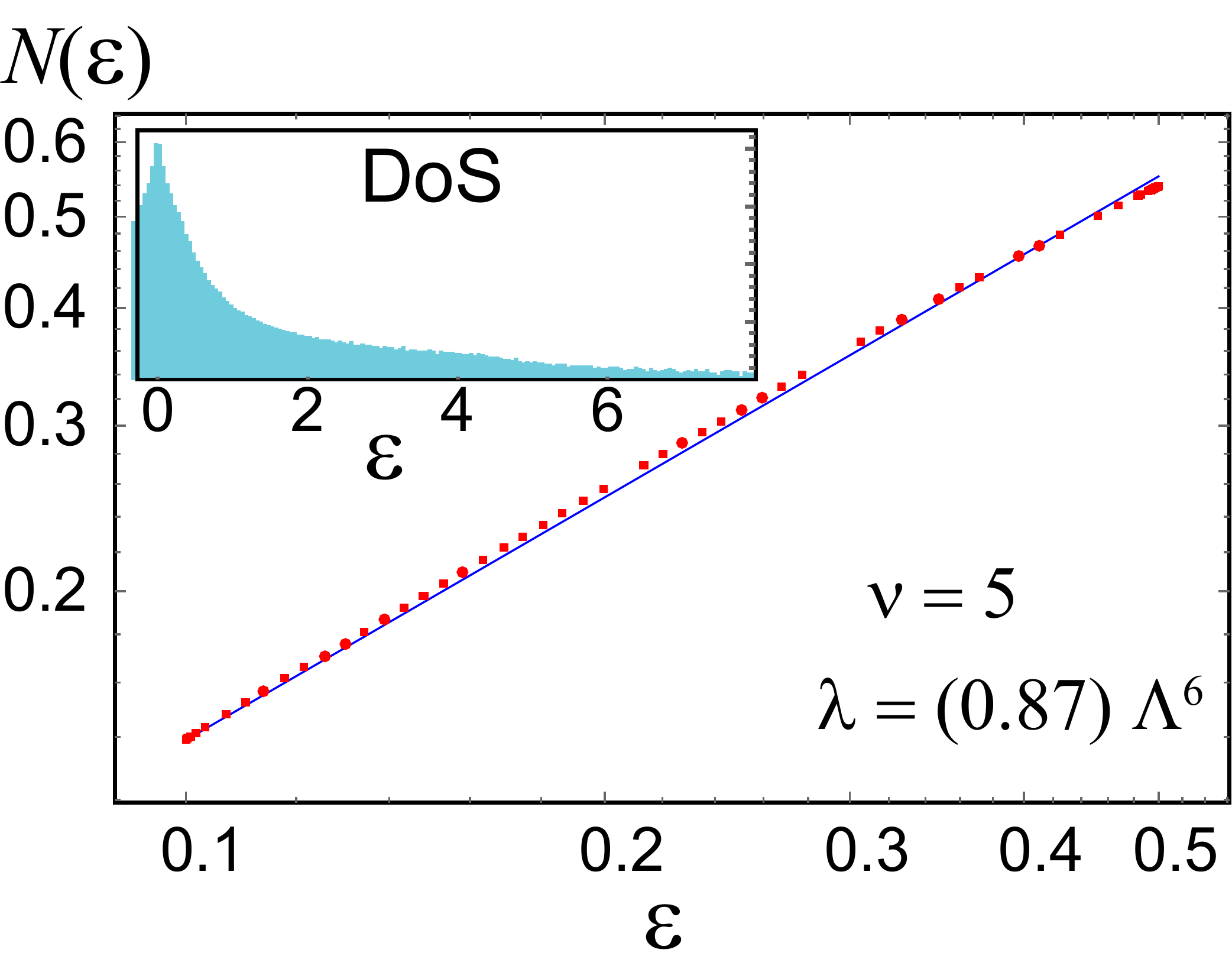}
}
\caption{Same as Fig.~\ref{Fig--nu=3}, but for the generalized surface model in Eq.~(\ref{hWeff}) with $\nu = 5$. 
Numerical results are shown as red dotted curves, 
while analytical predictions (blue solid curves) for $\tau(q)$ and $N(\e)$ obtain from the SO($n$)${}_5$ CFT.
Box sizes $b$=3 and $b$=6 are used to extract $\tau(q)$. 
The disorder strength $\lambda$ formally has units of $1/(\text{length})^6$, hence proportional to the sixth power 
of the momentum cutoff $\Lambda$. The absolute disorder strength is of the same order as in Fig.~\ref{Fig--nu=3},
with the same system size. 
The termination threshold $q_c = \sqrt{6} \simeq 2.45$ [see Eq.~(\ref{qc})].}
\label{Fig--nu=5}
\end{figure}

\begin{figure}
\subfigure{
\includegraphics[angle=0,width=0.23\textwidth]{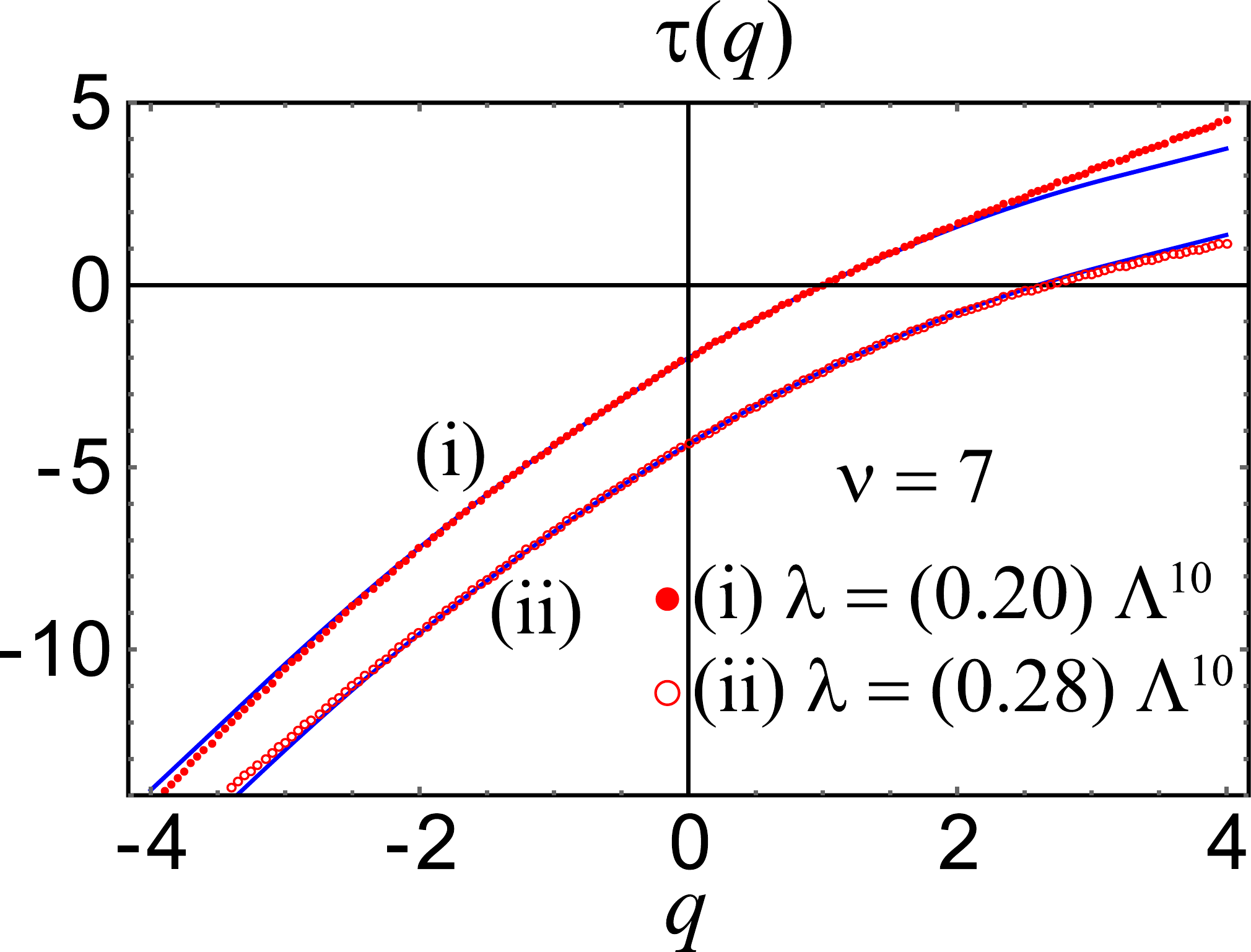}
}
\hspace{-8pt}
\subfigure{
\includegraphics[angle=0,width=0.23\textwidth]{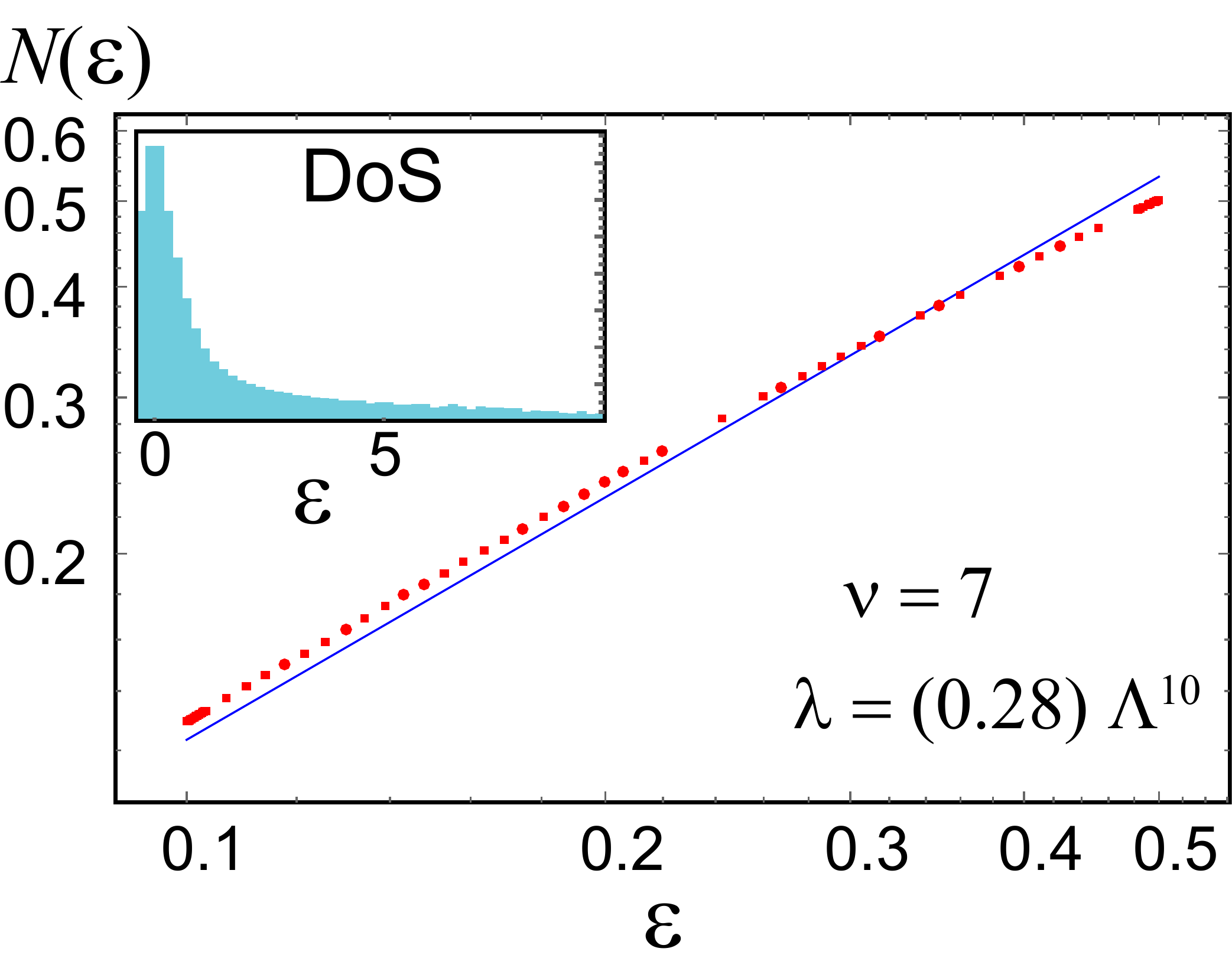}
}
\caption{
Same as Fig.~\ref{Fig--nu=3}, but for the generalized surface model in Eq.~(\ref{hWeff}) with $\nu = 7$. 
Numerical results are shown as red dotted curves, 
while analytical predictions (blue solid curves) for $\tau(q)$ and $N(\e)$ obtain from the SO($n$)${}_7$ CFT.
Box sizes $b$=3 and $b$=6 are used to extract $\tau(q)$. 
The disorder strength $\lambda$ formally has units of $1/(\text{length})^{10}$, hence proportional to the tenth power 
of the momentum cutoff $\Lambda$. The absolute disorder strength is of the same order as in Fig.~\ref{Fig--nu=3},
with the same system size.
The termination threshold $q_c = \sqrt{10} \simeq 3.16$ [see Eq.~(\ref{qc})].}
\label{Fig--nu=7}
\end{figure}

Fig.~\ref{Fig--nu=3}
depicts our numerical results for the $\tau(q)$ spectrum for two different disorder strengths,
and the IDoS $N(\e)$ for one disorder strength. 
As mentioned above, the IDoS is not particularly useful for $\nu = 3$ because the clean
and dirty CFT predictions coincide. Moreover, the strong divergence in the corresponding DoS 
$\varrho(\e)$ makes it difficult to get sufficient resolution in the peak itself.

We find that the multifractal spectrum becomes disorder-independent for sufficiently large disorder
strengths. This is important, because the thermal metal phase should exhibit weak multifractality
and a weak DoS divergence, but both features would be disorder- and scale-dependent 
due to weak antilocalization \cite{SenthilFisher}.
We observe rough agreement between the analytical SO($n$)${}_3$ CFT prediction
[Eqs.~(\ref{tauq})--(\ref{theta})] and the numerics. This should be compared to
the $\nu = 4$ model studied in Ref.~\cite{GDF17}, wherein quite good agreement was obtained.
Even better results are found for the higher-$\nu$ model explicated in the next section,
see Figs.~\ref{Fig--nu=5} and \ref{Fig--nu=7}.

We attribute the relatively poor fit for $\tau(q)$ in Fig.~\ref{Fig--nu=3}
to the \emph{strong multifractality} predicted
by SO($n$)${}_3$. This is indicated by the solid blue curve in the top panel of Fig.~\ref{Fig--nu=3}.
The analytical $\tau(q)$ is almost ``frozen.'' 
A frozen state has $\tau(q) = 0$ for $q > q_c \geq 1$ \cite{Chamon96}.  
A critically delocalized, but frozen state
consists of a \emph{few} rare probability peaks, with arbitrarily large separation between these 
\cite{Chamon96,RyuHatsugai01,CarpentierLeDoussal,Fyodorov09,YZCP1}. 
The peaks are sufficiently rare that their heights do not scale with a power of the system size $L$,
similar to an Anderson localized state. 
(The term ``frozen'' originates via a mapping to the classical glass transition in the random energy model \cite{Chamon96,CarpentierLeDoussal,Fyodorov09}.) 
Frozen states also resemble the ``random singlet'' wave functions of the Jordan-Wignerized
random bond XY model in 1D, which have the quality of random telegraph signals \cite{SUSYInt}. 
In a previous study \cite{YZCP1}, we found that the momentum space method does not scale well 
for frozen states, and we believe this is the source of the relatively poor fit in 
Fig.~\ref{Fig--nu=3}.

Note that it is only meaningful to compare the multifractal spectrum to the analytical prediction over the range $|q| < q_c$, 
since the spectrum becomes linear outside of this. Only the difference in slopes at $q = \pm q_c$ are meaningful \cite{Ghorashi18}.

\subsection{Generalized surface: Higher winding numbers and numerical results \label{sec: DIIIGen}}

If we believe that finite size effects are responsible for
the relatively poor fit between numerics and the SO($n$)${}_3$ CFT
in Fig.~\ref{Fig--nu=3}, the obvious way to improve is to increase
the system size. Instead of doing this (which requires more computer
memory), we take another approach. 

We conjecture that the Majorana surface fluid 
of a class DIII TSC with bulk winding number $\nu \in 2 \mathbb{Z} + 1$ (odd)
can be captured by the generalized $2 \times 2$ surface model 
\begin{align}\label{hWeff}
	\left[\hsnu\right]_{\vex{k},\vex{k'}}
	=&\,
	\begin{bmatrix}
	0 & i \kb^\nu \\
	-i \kc^\nu & 0 
	\end{bmatrix}
	\!
	\delta_{\vex{k},\vex{k'}}
	+
	(k_x + k'_x) \,
	\sigma_1 \,
	P_x(\vex{k}-\vex{k}')
\nonumber
\\&\,   
     +
    (k_y + k'_y) \,
	\sigma_2 \,
	P_y(\vex{k}-\vex{k}').
\end{align}
For $\nu = 1$ we get the spin-1/2 surface states of $^3$He-$B$,
while $\nu = 3$ corresponds to the hole-doped LSM [Eq.~(\ref{Dirtyhs})]. 
Again taking $P_{x,y}$ to be random phase, white noise variables 
as in Eq.~(\ref{Palpha}), the disorder strength $\lambda$ is relevant
for any $\nu \geq 3$, while it is irrelevant for $^3$He-$B$.

How do we know that the surface Hamiltonian in Eq.~(\ref{hWeff})
can be taken to represent a TSC, without connecting it to a bulk model
for general $\nu$? Certainly the clean limit of this model is artificial
and extremely unstable (to both disorder and interactions) for large $\nu$. 
Both attributes follow from the strongly diverging clean DoS,
\begin{align}
	\varrho(\e) \sim |\e|^{-(\nu - 2)/\nu}.
\end{align}
However, if the ``topological tuning'' scenario articulated in Fig.~\ref{Fig--CFTFlow}
is correct, then \emph{any} clean starting point should lead to the same 
disordered fixed point, the SO($n$)${}_\nu$ CFT~\cite{WZWP4}.

We can infer the bulk winding number $\nu$ by computing the \emph{surface winding number}
$\Ws$. This obtains by adding the homogeneous time-reversal symmetry-breaking mass term
$m \, \sigma_3$ to the clean band structure in Eq.~(\ref{hWeff}), and computing~\cite{Volovik2003} 
\begin{align}\label{W}
	\Ws(m)
	=&\,
	\frac{\epsilon_{\alpha \beta \gamma}}{3! (2 \pi)^2}
	\int_{-\infty}^\infty
	d \omega
	\int
	d^2\vex{k}
	\Tr
	\bigg[
	\left(
	\hat{G}^{-1}
	\partial_\alpha
	\hat{G}
	\right)
\nonumber	\\&\,
	\qquad\;\;\;
	\times
	\left(
	\hat{G}^{-1}
	\partial_\beta
	\hat{G}
	\right)
	\left(
	\hat{G}^{-1}
	\partial_\gamma
	\hat{G}
	\right)
	\bigg],
\end{align}
where $\Tr$ denotes the trace over the two spinor components,
and $\alpha,\beta,\gamma \in \{\omega,k_x,k_y\}$ (repeated indices are summed). 
The surface state Green function's $\hat{G}(\omega,\vex{k},m)$ is given by
\begin{align}\label{G}
	\hat{G}(\omega,\vex{k},m)
	\equiv
	\left[
	-
	i \, \omega \, \hat{1}
	+
	\hat{h}_{m}(\vex{k})
	\right]^{-1},
\end{align}
where
$\hat{h}_m = \hsnu|_{P_\alpha = 0} + m \, \sigma_3$ is 
the clean, gapped surface Hamiltonian. 
The surface winding number determines the thermal
Hall conductivity \cite{KaneFisher,spinthermal-2,Capelli02,Zhang2011,Ryu2012,Stone2012} 
\bsub
\begin{align}
	\kappa_{xy} =&\, \Ws \, \kappa_\circ, 
\\
\label{kapcirc}
	\kappa_\circ =&\, \pi^2 k_B^2 T/6 h.
\end{align}
\esub
Here $h$ is Planck's constant. 

It is easy to check that
\begin{align}
	\Ws(m) = (\nu/2) \, \sgn(m).
\end{align}
For $\nu =1$ this is the standard ``half-integer'' (shifted) surface quantum Hall effect familiar from $^3$He-$B$ and topological insulators \cite{Volovik2003,SRFL2008}. For a relativistic Majorana surface fluid, it can be shown that the maximum possible value of $\Ws$ is the bulk winding number divided by two \cite{WZWP4}. We conclude that the surface Hamiltonian in Eq.~(\ref{hWeff}) is a representative surface band structure for a class DIII TSC with winding number $\nu$.

Following the same logic of the previous section, we compare the numerical diagonalization of Eq.~(\ref{hWeff}) in momentum space to the predictions of the SO($n$)${}_\nu$ CFT. 
Results for $\nu = 5$ and $\nu = 7$ are shown in Figs.~\ref{Fig--nu=5} and \ref{Fig--nu=7},
respectively.
In these cases, the multifractal spectrum $\tau(q)$ and the IDoS $N(\e)$ match very well
the corresponding CFT predictions in Eqs.~(\ref{tauq})--(\ref{theta}) and (\ref{N(e)Def}),
respectively. The reason for the better matching is the weaker multifractality of 
the critical wave functions with increasing $\nu$, as predicted by the CFT.

The SO($n$)${}_\nu$ fixed point is stable against residual quasiparticle-quasiparticle interactions \cite{WZWP4}.
In addition to universal energy scaling of the DoS and wave function multifractality (both which could be detected via STM),
the ratio of the thermal conductivity to temperature $T$ is predicted to be quantized in the $T \rightarrow 0$ limit: \cite{SRFL2008,WZWP3}
\begin{align}\label{QuantKappa}
	\lim_{T \rightarrow 0}
	\frac{\kappa_{xx}}{T}
	=
	\frac{|\nu|}{\pi}
	\frac{\kappa_\circ}{T},
\end{align}
where $\kappa_\circ$ was defined by Eq.~(\ref{kapcirc}).


\section{Conclusions and outlook}~\label{conclusions}

To summarize, we have presented possible topological superconducting phases, 
including both gapless and gapped, in a doped Luttinger system (see Sec.~\ref{pairing:luttinger}). 
We showed that while pseudospin singlet $s$-wave pairing yields a trivial fully gapped state, 
the $d$-wave counterparts (belonging to either $T_{2g}$ or $E_{g}$ representations) 
often (if not always) lead to Weyl superconductors at low temperature at the cost of the time-reversal symmetry 
(see Sec.~\ref{Weyl-SC}). 
We argued that the simple Weyl nodes 
(sources and sinks of Abelian Berry curvature, see Figs.~\ref{Berry:Eg}--\ref{Berry:T2g_kxkykz}) 
that arise from complex combinations of 
simple $d$-wave nodal loops 
cause a power-law suppression of the density of states $\varrho(E) \sim |E|\; (\mbox{for nodal-loop}) \to |E|^2 \; (\mbox{for simple Weyl nodes})$ at low energies. 
Therefore, Weyl paired states are generically expected, at least within the framework of weak coupling pairing.

While any Weyl pairing supports one-dimensional (pseudospin degenerate) Fermi arcs as surface Andreev bound states, 
only the $T_{2g}$ paired state can lead to non-trivial anomalous pseudospin and thermal Hall conductivities at 
low temperature (Sec.~\ref{hall:thermal-spin}).
The simple Weyl BdG quasiparticles remain sharp in the presence of weak randomness in the system 
(in contrast to double-Weyl fermions in the $d_{xy}+id_{x^2-y^2}$ phase or nodal-loop states, see Fig.~\ref{doubleweyl_dirty}).
Stronger bulk disorder in the BdG-Weyl system induces a continuous quantum phase transition into a 
thermal metallic phase (Sec.~\ref{Weyl:disorder} and Fig.~\ref{simpleweyl_dirty}). 
The critical regime occupies a large portion of the phase diagram, where $\varrho(E) \sim |E|$, 
as shown in Fig.~\ref{simpleweyl_dirty}, which induces the corresponding scaling of physical observables 
such as specific heat, thermal conductivity, etc.

We demonstrated that nucleation of any $d$-wave pairing always causes a small lattice distortion or nematicity
that in turn gives rise to a non-trivial $s$-wave component in the paired state (see Fig.~\ref{Fig:strain-together}). 
Such symmetry-guaranteed coupling between $d$- and $s$-wave pairing with a lattice distortion may allow one to 
strain-engineer various exotic $s+d$ pairings, specifically in weakly correlated materials 
(Sec.~\ref{externalstrain}). 
We found that, within a simple picture of pairing, time-reversal symmetry breaking $s+id$ order seems to 
be extremely unlikely (with $s$-wave and $d+id$-type pairings being separated by a first order transition, see Fig.~\ref{Fig--s+id}). 
This interesting possibility cannot be completely ruled out (Sec.~\ref{sec:s+id}). 
We also showed that when the pairing interactions in the $T_{2g}$ and $E_g$ channels are of comparable strength, a 
myriad of gapless topological superconductors can be realized in the system 
(see Sec.~\ref{EgT2g:Competition}, Table~\ref{egt2g:competitiontable} and 
Fig.~\ref{Fig--egt2g}),
while only the $d_{xy}+id_{x^2 - y^2}$ paired state, supporting double-Weyl fermions, 
would exhibit non-trivial anomalous thermal and spin Hall conductivities (Sec.~\ref{hall:thermal-spin}). 
However, in the presence of inversion symmetry breaking (the situation in half-Heusler compounds) 
only thermal Hall conductivity remains sharply defined.

In terms of these nodal pairings we also attempted to understand the recent experimental data for 
the penetration depth in YPtBi~\cite{Exp:Paglione-2}, suggestive of the existence of gapless quasiparticles 
inside the paired state. 
We showed that $T$-linear fit, when augmented by a contribution from an ordinary $s$-wave component (always present with any $d$-wave pairing via the aforementioned coupling to the strain), matches extremely well with the experimental penetration depth data in YPtBi~\cite{Exp:Paglione-2}, see Fig.~\ref{fitlambda:expthr}.
We argued that this $T$-linear contribution may originate from either nodal loops in simple $d$-wave pairing for example (see Fig.~\ref{Fig--Eg}), or from the effect of quenched disorder (such that the system gets stuck inside the wide quantum critical regime in Fig.~\ref{simpleweyl_dirty}) on simple Weyl nodes stemming from the $d+id$ pairing.
Although we strongly believe that the former source of $T$-linear dependence 
is most likely, we proposed various experiments on specific heat, thermal conductivity, NMR relaxation time, 
Hall conductivity etc.\ (Sec.~\ref{experiment}), which can possibly pin the actual nature of the pairing in 
half-Heusler compounds~\cite{Exp:Takabatake, Exp:Paglione-1a, Exp:Bay2012, Exp:visser-1, Exp:Taillefer, Exp:Zhang, Exp:visser-2, Exp:Paglione-1, Exp:Paglione-2}.

Finally, we investigated the effects of disorder on the cubically dispersing surface states that arise
from odd-parity, fully gapped $p$-wave pairing (as in $^3$He-$B$). Using a generalized surface
model with $\nu$-fold dispersion for winding number $\nu \geq 3$, we demonstrated excellent agreement
between numerical results and the conformal field theory (CFT) SO($n$)${}_\nu$ for higher $\nu$. The CFT characterizes
the critical delocalization of the surface in the presence of disorder, whilst the naively expected
thermal metal phase is absent in our numerics. This suggests a deep connection between the bulk
topology on one hand, and the \emph{disordered} surface physics on the other, reminiscent of 
key aspects of the integer quantum Hall effect. A key open question is whether there exists a
\emph{topological generalization} of the conformal embedding rule 
[$\text{SO}(n \nu)_1 \supset \text{SO}(n)_\nu \oplus \text{SO}(\nu)_n$], employed to explain
the robustness of CFT results in the case of spin-1/2 topological superconductors \cite{WZWP4}.

Perhaps the most urgent issue in the context of superconductivity in a doped Luttinger semimetal 
is that of pairing mechanisms. Recently, it has been argued that such pairing can in principle be mediated 
by electron-phonon interactions, specifically due to optical phonons~\cite{savary}. 
However, given that promising candidates such as half-Heuslers and 227 pyrochlore iridates also display 
magnetic orders, pairing in these materials may also arise from strong electronic interactions. 
Understanding the effects of magnetic fluctuations on various pairing scenarios is a challenging, 
but crucial question that we leave for a future investigations.

\emph{Note added}: After our paper was posted to the arXiv, another 
preprint~\cite{savary-fu-lee-new} appeared, which also discusses the topology of various paired 
states, using a slightly different language. Qualitatively our conclusions appear to be same.

\acknowledgements

We thank Pallab Goswami and Pavan Hosur for useful discussions. 
This research was supported by the Welch Foundation Grants No.~E-1146 (S.A.A.G.),
No.~C-1809 (B.R. and M.S.F.), and No.~C-1818 (A.H.N.), 
and by NSF CAREER Grants No.~DMR-1552327 (M.S.F.) and No.~DMR-1350237 (A.H.N.). 
We are grateful to J. P. Paglione and Hyunsoo Kim for sharing with us and giving us permission 
to replot the data from Ref.~\cite{Exp:Paglione-2}. A.H.N. and M.S.F. acknowledge the hospitality of the Aspen Center for
Physics, which is supported by National Science Foundation Grant No.
PHY-1607611.


\begingroup

\appendix

\section{Luttinger model components}~\label{Luttinger:details}

In this Appendix we present some essential details of the Luttinger Hamiltonian. 
The ${\mathbf d}$-vector appearing in Eq.~(\ref{luttinger}) 
is a quadratic function of momentum measured from the $\Gamma=(0,0,0)$ point of the Brillouin zone, 
where Kramers degenerate valence and conduction bands touch each other, 
namely ${\mathbf d} = k^2 \, \hat{\vex{d}}$. 
The quantity $\hat{\mathbf d}$ is a five-component unit vector and its components are given by    
\begin{eqnarray}~\label{d-hats}
\hat{d}_1 &=&	\frac{i\left[Y^1_{2} + Y^{-1}_{2}\right]}{\sqrt{2}} = 	\frac{\sqrt{3}}{2} \sin 2 \theta \, \sin \phi = \sqrt{3} \; \hat{k}_y \hat{k}_z, \nonumber \\
\hat{d}_2 &=& \frac{\left[Y^{-1}_{2} + Y^{1}_{2}\right]}{\sqrt{2}} = \frac{\sqrt{3}}{2} \sin 2 \theta \cos \phi = \sqrt{3} \; \hat{k}_x \hat{k}_z, \nonumber \\
\hat{d}_3 &=& \frac{i\left[Y^{-2}_{2} + Y^{-2}_{2}\right]}{\sqrt{2}} = \frac{\sqrt{3}}{2} \sin^2 \theta \sin 2 \phi = \sqrt{3} \; \hat{k}_y \hat{k}_x, \nonumber \\
\hat{d}_4 &=& \frac{\left[Y^{-2}_{2} + Y^{2}_{2}\right]}{\sqrt{2}} = \frac{\sqrt{3}}{2} \sin^2 \theta \cos 2 \phi = \frac{\sqrt{3}}{2} \left[ \hat{k}^2_x-\hat{k}^2_y \right], \nonumber \\
\hat{d}_5 &=& Y^0_2 = \frac{1}{2} \left( 3 \cos^2\theta-1 \right)=\frac{1}{2} \left[ 2 \hat{k}^2_z-\hat{k}^2_x-\hat{k}^2_y \right].
\end{eqnarray}
Note that $\sum^{5}_{j=1} \left( \hat{d}_j \right)^2=1$. In the above expression $Y^m_l \equiv Y^m_l (\theta, \phi)$ are the spherical harmonics with angular momentum $l=2$.

Five mutually anticommuting $\Gamma$ matrices appearing in Eq.~(\ref{luttinger}) are constructed from $J=3/2$ matrices according to 
\begin{gather}
\begin{aligned}
\Gamma_1 =&\, \frac{1}{\sqrt{3}} \{J^y,J^z\},
&\Gamma_2 =&\, \frac{1}{\sqrt{3}} \{J^x,J^z\},
\\
\Gamma_3 =&\, \frac{1}{\sqrt{3}} \{J^x,J^y\},
&\Gamma_4 =&\, \frac{1}{\sqrt{3}} \left[(J^x)^2 - (J^y)^2 \right],
\end{aligned}
\nonumber\\
\Gamma_5 = \frac{1}{3}\left[2 (J^z)^2 - (J^x)^2 - (J^y)^2 \right], 
\label{GammatoTmunu}
\end{gather}
while $\Gamma_0$ denotes the four dimensional identity matrix. 
Here $\{A,B\} = A B + B A$ is the anticommutator. 
The $\{\Gamma_1,\ldots,\Gamma_5\}$ are components of the rank-two symmetric
traceless tensor operator
\begin{equation}~\label{tracelesstensor:definition}
	T^{\mu \nu} 
	=
	\frac{1}{\sqrt{3}}
	\left[
	\{J^\mu,J^\nu\}
	-
	\frac{2}{3}
	\delta^{\mu,\nu}
	\vex{J}^2
	\right],
\end{equation}
which transforms in the $j = 2$ representation of SU(2)
under spin rotations. 
In the basis specified by Eq.~(\ref{spinor-4}), 
the spin-$3/2$ matrices are defined as  
\begin{eqnarray}\label{JDef}
J^x &=& \frac{1}{2} \left[ \begin{array}{cccc}
0 & \sqrt{3} & 0 & 0 \\
\sqrt{3} & 0 & 2 & 0 \\
0 & 2 & 0 & \sqrt{3} \\
0 & 0 & \sqrt{3} & 0 
\end{array} \right],
\,
J^z = \frac{1}{2} \left[ \begin{array}{cccc}
3 & 0 & 0 & 0 \\
0 & 1 & 0 & 0 \\
0 & 0 & -1 & 0 \\
0 & 0 & 0 & -3
\end{array} \right], \nonumber \\
J^y &=& \frac{i}{2} \left[ \begin{array}{cccc}
0 & -\sqrt{3} & 0 & 0 \\
\sqrt{3} & 0 & -2 & 0 \\
0 & 2 & 0 & -\sqrt{3} \\
0 & 0 & \sqrt{3} & 0 
\end{array} \right].
\end{eqnarray}

A full basis for $4 \times 4$ matrices can be formed
by adding the identity and the 10 commutators $\{\Gamma_{a b} = - i \Gamma_a \Gamma_b\}$
to the 5 $\{\Gamma_a\}$ matrices. 
The matrices in the product basis $\{\Gamma_{a b}\}$ do not transform irreducibly under
the spin SU(2). 
An irreducibly decomposed basis of tensor operators
instead uses 
the three $J^{x,y,z}$ generators ($j = 1$) 
and seven components of a rank-three, traceless symmetric tensor formed from products of these,
\begin{align}\label{T3}	
	T^{\mu \nu \gamma}
	\equiv
	J^{(\mu}
	J^{\nu}
	J^{\gamma)}
	-
	\text{(traces)},
	\quad
	\sum_{\mu = 1}^3
	T^{\mu \mu \nu} 
	= 
	0,
\end{align}
where $(\mu \nu \gamma)$ means complete symmetrization of these indices. 
The latter transforms as $j = 3$ under spin SU(2) rotations. 
The tensor operator 
$T^{\mu \nu \gamma}$ 
plays the key role in the proposal for odd-parity, orbital $p$-wave ``septet'' superconductivity in \cite{brydon,brydon-2,Exp:Paglione-2}.


\section{Band Projection in Luttinger system}~\label{bandprojection}

In this 
appendix
we present the band projection method in the Luttinger system. We first focus on the non-interacting part of the theory. In the eight-component spinor basis, defined in Eq.~(\ref{nice-Nambu}) the isotropic Luttinger Hamiltonian reads as 
\begin{equation}
	\hat{h}^{N}_L ({\mathbf k}) =\tau_3 \hat{h}_L ({\mathbf k}),
\end{equation}
where $\hat{h}_L ({\mathbf k})$ is the four-dimensional Luttinger Hamiltonian defined in Eq.~(\ref{luttinger}), with $m_1=m_2=m$. The Luttinger Hamiltonian can be brought into diagonal form under the following unitary transformation ${\mathcal D}^\dagger_N \hat{h}^N_L {\mathcal D}_N$, where ${\mathcal D}_N=U_1 U_2$ is the diagonalizing matrix, with
\begin{eqnarray}
U_1=\tau_0 \otimes {\mathcal D}, \; U_2=\tau_0 \otimes \left[ \begin{array}{cccc}
1 & 0 & 0 & 0 \\
0 & 0 & 0 & 1 \\
0 & 0 & 1 & 0 \\
0 & 1 & 0 & 0
\end{array}
\right],
\end{eqnarray}
with ${\mathcal D}$ is already defined in Eq.~(\ref{diagonalizer}). Upon performing the unitary rotation with the diagonalizing matrix the Luttinger Hamiltonian takes the form 
\begin{eqnarray}
{\mathcal D}^\dagger_N \hat{h}^N_L {\mathcal D}_N &=& \left( \frac{{\mathbf k}^2}{2 m_0}  -\mu \right) \: \mbox{diag.} \left( {\boldsymbol \sigma}_0, \sigma_0,-{\boldsymbol \sigma}_0, -\sigma_0 \right) \nonumber \\
&+& \frac{{\mathbf k}^2}{2 m} \; \left( {\boldsymbol \sigma}_0, -\sigma_0,-{\boldsymbol \sigma}_0, \sigma_0 \right).
\end{eqnarray}
In the last expression the bold (normal) entries correspond to conduction (valence) band, which then leads to the kinetic energy in the conduction band reported in Eq.~(\ref{kinetic-band}). The kinetic energy in the valence band also assumes the form of Eq.~(\ref{kinetic-band}), but with the modification that effective mass parameter reads as $m_\ast=m m_0/(m-m_0)$. Therefore, valence and conduction bands bend in opposite directions when the effective mass parameters in the Luttinger model satisfy $m>m_0$.

The diagonalization process on pairing operators yields
\begin{eqnarray}~\label{Pairing:BP-Luttinger}
&&{\mathcal D}^\dagger_N \left[ \left( \tau_1 \cos \phi_{sc} + \tau_2 \sin \phi_{sc} \right) \otimes \hat{M}_{4 \times 4} \right] {\mathcal D}_N \nonumber \\
&=& \left( \tau_1 \cos \phi_{sc} + \tau_2 \sin \phi_{sc} \right) \otimes \left[ \begin{array}{c|c}
 \hat{a}_{2 \times 2} & \hat{b}_{2 \times 2} \\
\hline
\hat{c}_{2 \times 2} & \hat{d}_{2 \times 2}
\end{array} 
\right].
\end{eqnarray}
In the final expression $\hat{a}_{2 \times 2} (\hat{d}_{2 \times 2})$ captures the form of a given pairing ($\hat{M}_{4 \times 4}$) on conduction (valence) bands, while two other entries namely $\hat{c}_{2 \times 2}, \hat{b}_{2 \times 2}$ capture the coupling between the valence and conduction bands. Throughout this work we neglect such coupling assuming that the pairing only takes place in the close proximity to the Fermi surface.


\section{Band Projection in massive Dirac system}~\label{Dirac:bandprojection}

In this appendix
we present the band diagonalization procedure and
transformation of six local pairings in a Dirac semiconductor.
In the basis of a four-component Dirac-spinor defined as $\Psi^\top=\left( c^{+}_{\uparrow}, c^{+}_{\downarrow}, c^{-}_{\uparrow}, c^{-}_{\downarrow} \right)$, where $c^{\kappa}_{\sigma}$ is the fermionic annihilation operator with parity $\kappa=\pm$ and spin-projection $\sigma=\uparrow, \downarrow$, the massive Dirac Hamiltonian reads as
\begin{equation}~\label{massiveDirac}
H_D=v i\gamma_0 \gamma_j k_j + m \gamma_0 -\mu,
\end{equation}
where $j=1,2,3$, and summation over repeated indices are assumed. Here, $k_j$s are three spatial component of momentum, measured from the $\Gamma=(0,0,0)$ point of the Brillouin zone, $m$ is the Dirac mass and $\mu$ is the chemical potential. Fermi velocity $v$ is assumed to be isotropic for the sake of simplicity, which from now onward we set to be unity. Mutually anti-commuting four-component Hermitian $\gamma$ matrices are defined as
\begin{eqnarray}~\label{Diracmatrix}
\gamma_0=\kappa_3 \sigma_0, \; \gamma_1= \kappa_2 \sigma_1, \; \gamma_2= \kappa_2 \sigma_2, \nonumber \\
\gamma_3= \kappa_2 \sigma_3, \; \gamma_5=\kappa_1 \sigma_0.
\end{eqnarray}
Two sets of Pauli matrices $\left\{ \kappa_\mu \right\}$ and $\left\{ \sigma_\mu \right\}$, with $\mu=0,1,2,3$ respectively operate on parity and spin index. The above Hamiltonian is invariant under the following discrete symmetry operations: (a) reversal of time ($\mathcal T$), generated by ${\mathcal T}=i \gamma_1 \gamma_3 K$, where $K$ is the complex conjugation, (b) parity or inversion ($\mathcal P$), under which ${\bf r} \to -{\bf r}$ and ${\mathcal P}=\gamma_0$, (c) charge conjugation (${\mathcal C}$), under which ${\mathcal C} \Psi {\mathcal C}^{-1}= \gamma_2 \Psi^{\ast}$. In the massless limit ($m \to 0$), describing a quantum critical point between two topologically distinct insulating phases, the Dirac Hamiltonian also enjoys an emergent continuous $U(1)$ chiral symmetry, generated by 
$\gamma_5$~\cite{roy-sau}.

Since all four dimensional representation of five mutually anti-commuting matrices are unitarily equivalent, we can express the above five matrices from Eq.~(\ref{Diracmatrix}) in term of $\Gamma_j$s and $\Gamma_{jk}$s introduced in Eq.~(\ref{Gamma:definition}) according to
\begin{equation}
\gamma_0=\Gamma_{43}, \; \gamma_1=\Gamma_{24},\; \gamma_2=\Gamma_{14},\; \gamma_3=\Gamma_5,\; \gamma_5=\Gamma_4,
\end{equation}
or equivalently
\begin{equation}
\Gamma_1=\gamma_{25},\; \Gamma_2=\gamma_{15},\; \Gamma_3=\gamma_{05},\; \Gamma_4=\gamma_5, \; \Gamma_5=\gamma_3,
\end{equation}
where $\gamma_{lm}=i \gamma_l \gamma_m$. Therefore, all possible, namely \emph{six}, local-pairings in this system are also captured by the effective single-particle Hamiltonian $H^{local}_{pp}$ in Eq.~(\ref{local-pairing-all}), which in terms of the $\gamma$ matrices can be expressed as
\begin{eqnarray}
H^{local}_{pp} &=& -\int d{\bf r} \bigg\{ \bigg[ \Delta_s \Psi \gamma_{13} \Psi + \Delta_p \Psi \gamma_{02} \Psi + \Delta_1 \Psi \gamma_3 \Psi \nonumber \\
&+& \Delta_2 \Psi \gamma_{05} \Psi + \Delta_3 \gamma_1 + \Delta_0 \Psi \gamma_{25} \Psi \bigg] \bigg\}.
\end{eqnarray}
where $\Psi$ is the four-component Dirac-spinor, introduced earlier. Also notice that we have changed the notations for the pairing amplitudes $\Delta_a$s from Eq.~(\ref{local-pairing-all}). The purpose will be clear in a moment.

We now conveniently define an eight-component Nambu spinor to cast all local pairing in a massive Dirac system in a compact form
\begin{equation}~\label{Nambu:niceDirac}
\Psi_N = \left[ \begin{array}{c}
\Psi \\
\gamma_{13}  \; \left( \Psi^\dagger \right)^\top
\end{array}
\right].
\end{equation}
Note that $\gamma_{13}$ is the unitary part of the time-reversal operator. In this eight-component basis, the Nambu-doubled massive Dirac Hamiltonian from Eq.~(\ref{massiveDirac}) takes the form
\begin{equation}~\label{nambudoubleddirac}
H^{Nam}_D = \tau_3 \otimes \left[ \gamma_{01} k_x + \gamma_{02} k_y + \gamma_{03} k_z + \gamma_0 m -\mu \right].
\end{equation}
The newly introduced set of Pauli matrices $\left\{\tau_\mu \right\}$ operate on the Nambu index.  In the same basis the single-particle Hamiltonian in the presence of all local pairings is given by
\begin{eqnarray}~\label{pairing:LT}
H^{local}_{pp} &=& \left( \tau_1 \cos \phi + \tau_2 \sin \phi \right) \; \bigg[ \Delta_s + \Delta_p \gamma_5 \nonumber \\
&+& \sum^{3}_{\mu=0} \Delta_\mu \gamma_\mu \bigg],
\end{eqnarray}
where $\phi$ is the superconducting phase. Now we can take the advantage of the spinor basis introduced in Eq.~(\ref{Nambu:niceDirac}), to classify all six local pairings according to their transformation under the \emph{Lorentz transformation} (LT). Respectively the pairing proportional to $\Delta_s$ and $\Delta_p$ transforms as scalar and pseudo-scalar Majorana mass under the LT. The set of pairings $\left\{ \Delta_\mu \right\}$, with $\mu=0,1,2,3$, transforms as a ``four"-vector under the LT. While the pseudo-scalar and the spatial-components ($\mu=1,2,3$) of the vector pairings are odd under the parity (${\mathcal P}$), the scalar and the temporal-component of the vector pairing are even under parity [see Table~\ref{Pairing:Dirac-projected}].

The unitary operator that diagonalizes the massive Hamiltonian [see Eq.~(\ref{nambudoubleddirac})] is given by $\tau_0 \tilde{{\mathcal D}} \left( {\bf k}, m \right)$, where ${\bf k}=\left(k_x,k_y,k_z \right)$,
\begin{align}
&\tilde{{\mathcal D}} \left( {\bf k}, m \right) \nonumber \\
&= \begin{bmatrix}
       \frac{k_x-ik_y}{\sqrt{2\lambda(\lambda-m)}} & \frac{k_z}{\sqrt{2\lambda(\lambda-m)}} & \frac{-k_x+ik_y}{\sqrt{2\lambda(\lambda+m)}}           &\frac{-k_z}{\sqrt{2\lambda(\lambda+m)}} \\[0.3em]
       \frac{-k_z}{\sqrt{2\lambda(\lambda-m)}} & \frac{k_x+i k_y}{\sqrt{2\lambda(\lambda-m)}}           & \frac{k_z}{\sqrt{2\lambda(\lambda+m)}} &\frac{-k_x-i k_y}{\sqrt{2\lambda(\lambda+m)}} \\[0.3em]
       0           & \frac{\lambda-m}{\sqrt{2\lambda(\lambda-m)}} & 0 &\frac{\lambda+m}{\sqrt{2\lambda(\lambda+m)}} \\[0.3em]
       \frac{\lambda-m}{\sqrt{2\lambda(\lambda-m)}}           & 0 & \frac{\lambda+m}{\sqrt{2\lambda(\lambda+m)}} &0
     \end{bmatrix},
\end{align}
and $\lambda=\left[ {\bf k}^2 + m^2 \right]^{1/2}$. After performing a unitary rotation with diagonalizing matrix, the Nambu-doubled Massive Dirac Hamiltonian from Eq.~(\ref{nambudoubleddirac}) becomes
\begin{eqnarray}
&&\tau_3 \tilde{\mathcal D}^\dagger H^{Nam}_D \tau_3 \tilde{\mathcal D} = -\mu \;{\rm diag.} \left( {\bm \sigma}_{\bm 0}, \sigma_0, -{\bm \sigma}_{\bm 0}, -\sigma_0 \right) \nonumber \\
&+& \sqrt{{\bf k}^2+m^2} \;{\rm diag.} \left( {\bm \sigma}_{\bm 0}, -\sigma_0,,-{\bm \sigma}_{\bm 0}, \sigma_0 \right).
\end{eqnarray}
In the above expression, the quantities in the bold (normal) font correspond to the conduction (valence) band. We here restrict ourselves with the situation when the chemical potential is placed in the conduction band, i.e., when $\mu>0$. We now make a \emph{large mass} expansion of the kinetic energy term, yielding
\begin{equation}
\sqrt{{\bf k}^2+m^2} -\mu=\frac{{\bf k}^2}{2m}-\left( \mu-m\right) \equiv \frac{{\bf k}^2}{2m}- \tilde{\mu},
\end{equation}
where $\tilde{\mu}=\mu-m$ is the renormalized chemical potential, measured from the bottom of the conduction band, and $k_F=\sqrt{2 m \tilde{\mu}}$ is the Fermi vector. Hence, the kinetic energy for massive Dirac fermions in the close proximity to the Fermi surface assumes the form
\begin{equation}
H_0= \left( \frac{{\bf k}^2}{2m}- \tilde{\mu} \right) \tau_3 \sigma_0.
\end{equation}

Next we perform the same band projection on six local pairings shown in Eq.~(\ref{pairing:LT}). After this transformation the pairing assumes the schematic form shown in Eq.~(\ref{Pairing:BP-Luttinger}). 
Since, we have assumed that chemical potential lies within the conduction band, we are only interested in the two-component representation of the pairings, namely $\hat{a}_{2 \times 2}$. The band-projected version of all six local pairings, quasiparticle spectra etc. are displayed in Table~\ref{Pairing:Dirac-projected}.

\begin{table*}[ht!]
\begin{tabular}{|c|c|c|c|c|c|c|}
\hline
{\bf Pairing} 
& 
{\bf 
$\stackrel{\displaystyle{\text{Transformation}}}{\displaystyle{\text{under LT}}}$  
} 
& 
{\bf 
$\stackrel{\displaystyle{\text{IREP}}}{\displaystyle{D_{3d}}}$ 
}
& 
${\mathcal T}$ & ${\mathcal P}$ & {\bf Pairing near FS} & {\bf Quasiparticle spectrum} \\
\hline \hline
$\Delta_s $ & Scalar mass & $A_{1g}$ & $\checkmark$ & $\checkmark$ & $\Delta_s \sigma_0$ & Fully gapped \\
\hline
$\Delta_p \; \gamma_5$ & Pseudo-scalar mass & $A_{1u}$ & $\checkmark$ & $\times$ & $\Delta_p {\mathbf d} \cdot \boldsymbol{\sigma}$, ${\mathbf d}=\left( \hat{k}_x, -\hat{k}_y, \hat{k}_z \right)$ & Fully gapped \\
\hline
$\Delta_0 \; \gamma_0$ & $0^{\rm th}$-vector component  & $A_{1g}$ & $\checkmark$ & $\checkmark$ & $\Delta_0 \sigma_0 \; \left( m/k_F\right)$ & Fully gapped \\
\hline
$\Delta_1 \; \gamma_1$ & $1^{\rm st}$-vector component & $E_u$ & $\checkmark$ & $\times$ & $\Delta_1 {\mathbf d} \cdot \boldsymbol{\sigma}$, ${\mathbf d}=\left( 0, -\hat{k}_z, \hat{k}_y \right)$ 
	& Gapless: 
	$\displaystyle{    
	\left\{ 
		\begin{aligned}
		&\mbox{2 Dirac points at}
		\\
		&k_x =\pm k_F, k_y=0=k_z 
		\end{aligned}
	\right\}  
	}$  
\\
\hline
$\Delta_2 \; \gamma_2$ & $2^{\rm nd}$-vector component & $E_u$ & $\checkmark$ & $\times$ & $\Delta_2 {\mathbf d} \cdot \boldsymbol{\sigma}$, ${\mathbf d}=\left( -\hat{k}_z, 0, -\hat{k}_x \right)$ 
	& Gapless: 
	$\displaystyle{    
	\left\{ 
		\begin{aligned}
		&\mbox{2 Dirac points at}
		\\
		&k_y =\pm k_F, k_x=0=k_z 
		\end{aligned}
	\right\}  
	}$  
\\
\hline
$\Delta_3 \; \gamma_3$ & $3^{\rm rd}$-vector component & $A_{2u}$ & $\checkmark$ & $\times$ & $\Delta_3 {\mathbf d} \cdot \boldsymbol{\sigma}$, ${\mathbf d}=\left( \hat{k}_y, \hat{k}_x,0 \right)$ 
	& Gapless: 
	$\displaystyle{    
	\left\{ 
		\begin{aligned}
		&\mbox{2 Dirac points at} 
		\\
		&k_z =\pm k_F, k_x=0=k_y 
		\end{aligned}
	\right\}  
	}$  
\\
\hline \hline
\end{tabular}
\caption{
Classification of six local pairing operators for a 4-component, massive three-dimensional Dirac semiconductor.
The purpose of this table is to show how the parent band structure is crucial to determine the
character of local pairing operators; in particular, the results are completely 
different from the Luttinger semimetal, shown in Table~\ref{table_projection}.  
First column: Possible local pairings for Dirac fermions. 
Second column: Transformation of each such local pairing under the Lorentz transformation (LT)~\cite{ohsaku,goswami-roy}. 
Third column: Transformation of each such pairing under $D_{3d}$ point group (relevant for Cu$_x$Bi$_2$Se$_3$)~\cite{fu-berg}. 
Fourth and fifth columns: Transformation of each pairing under time-reversal ($\mathcal T$) and parity (${\mathcal P}$), respectively. 
Sixth column: Form of each local pairing close to the Fermi surface (FS). 
Seventh Column: Quasiparticle spectra and the nodal topology close to the Fermi surface. 
The four-component Hermitian $\gamma$ matrices satisfy the anti-commutation Clifford algebra $\left\{ \gamma_\mu, \gamma_\nu\right\}=2\delta_{\mu \nu}$ and the exact representation of $\gamma$ matrices are given in Eq.~(\ref{Diracmatrix}). 
Note that the $0^{\rm th}$ or temporal component of the vector pairing yields a gapless Fermi surface in a doped Dirac semimetal 
(i.e., when $m \to 0$). 
Otherwise, the gap on the Fermi surface is suppressed by a factor $k^{-1}_F$, where 
$k_F =\sqrt{2 m \tilde{\mu}}$ is the Fermi vector
and
$\tilde{\mu}$ is the chemical potential measured from the bottom of the conduction band. 
The Dirac points on the Fermi surface in the presence of spatial components of the vector 
pairing produces the quasiparticle density of states 
$\varrho(E) \sim |E|^2$. 
The details of the band projection method are presented in 
Sec.~\ref{Dirac:bandprojection}.
}~\label{Pairing:Dirac-projected}
\end{table*}


\section{Superconducting condensation energy}~\label{Append:free-energy}

{We hereby present the calculation of the zero-temperature free energy of a $d$-wave superconductor. The derivation is standard, and we only provide it here for the sake of completeness, since the final result is already quoted in Eq.~(\ref{eq:free1}) in the main text and is also  used to compute the free energy in the case of the ($s+id$) pairing considered below in Section~\ref{Append:s+id}.

The free energy $F$ is given by Eq.~(\ref{eq:free0}) in the main text, and can be expressed in dimensionless units $f=F/[\mu^2\rho(\mu)]$, where $\mu$ is the chemical potential and $\rho(\mu) = 2a^3m^{\ast}\sqrt{2m^{\ast}\mu}/(2\pi^2)$ is the density of states at the Fermi level. We then obtain:
\begin{eqnarray}
f &= & \frac{|\hat{\Delta}_d|^2}{2\lambda_d} \\
&-& \frac{1}{2}\int\limits_{-\omega_D}^{\omega_D} \ud y \sqrt{1+y}\int\frac{\ud \Omega}{4\pi} \sqrt{y^2 + |\hat{\Delta}_d|^2(1+y)^2\hat{d}(\Omega)^2}, \nonumber
\label{eq:freeS}
\end{eqnarray}
where, using the notation introduced in the main text, $\hat{\Delta} = \Delta/\mu$ is the dimensionless order parameter, $\omega_D = \Omega_D/\mu$ is the dimensionless Debye frequency (in the units of $\mu$), and $y=\xi_k/\mu = (k/k_F)^2 - 1$ is the dimensionless energy of the normal quasiparticles relative to the Fermi level. Note that the factor of $\sqrt{1+y}$ in the integrand appears because of the square-root dependence of the DoS in a 3D normal metal. All the angle dependence of the superconducting order parameter is contained in the factor $\hat{d}(\Omega)$, normalized in a standard way such that the cubic harmonics form an orthonormal basis. A resolution of identity is obtained when summed over all harmonics:
\be
1 = \sum_j \langle d_j| d_j \rangle = \sum_j \int\frac{\ud \Omega}{4\pi} |d_j(\Omega)|^2.
\label{eq:identity}
\ee

We now use the weak-coupling approximation, \mbox{$\omega_d \ll 1$} (i.e. $\Omega_d \ll \mu$), allowing us to approximate $1+y \approx y$ in the integrand, thus obtaining
\be
f = \frac{|\hat{\Delta}_d|^2}{2\lambda_d} 
- \int\frac{\ud \Omega}{4\pi} \int\limits_{0}^{\omega_D} \ud y \sqrt{y^2 + |\hat{\Delta}_d|^2\hat{d}(\Omega)^2}.\label{eq:free2}
\ee
Attempting to expand the square root in powers of $(|\hat{\Delta}_d|^2/y^2)$ would result in an infra-red divergence, with the first term yielding an expression of order $|\hat{\Delta}_d|^2 \ln(y)$, and the subsequent terms even more divergent.  This is to be expected, since the zero-temperature free energy is a non-analytic function of $\Delta_d$. Instead, we proceed by formally evaluating the integral over $y$ in Eq.~(\ref{eq:free2}):
\begin{eqnarray}
\label{eq:free3}
f &=& \frac{|\hat{\Delta}_d|^2}{2\lambda_d} 
       - \frac{1}{2} \int\frac{\ud\Omega}{4\pi} \omega_D \sqrt{\omega_D^2 + |\hat{\Delta}_d|^2 \hat{d}(\Omega)^2} \\
       &-& \frac{|\hat{\Delta}_d|}{2} \int\frac{\ud\Omega}{4\pi} \hat{d}(\Omega)^2 \ln\left(\frac{\omega_D + \sqrt{\omega_D^2 + |\hat{\Delta}_d|^2 \hat{d}(\Omega)^2}}{|\hat{\Delta}_d|\, \hat{d}(\Omega)} \right). \nonumber       
\end{eqnarray}

The last term is a non-trivial integral, however we have encountered it before, namely in the zero-temperature gap equation (\ref{eq:gap}):
\be
\frac{1}{\lambda_d} = \int\frac{\ud\Omega}{4\pi} \hat{d}(\Omega)^2 \ln\left(\frac{\omega_D + \sqrt{\omega_D^2 + |\hat{\Delta}_d|^2 \hat{d}(\Omega)^2}}{|\hat{\Delta}_d|\, \hat{d}(\Omega)} \right).
\label{eq:gapS}
\ee
We recognize that the last integral in Eq.~(\ref{eq:free3}) is therefore nothing else but $|\hat{\Delta}_d|^2/2\lambda_d$. This term cancels the first term in Eq.~(\ref{eq:free3}), thus resulting in the final expression for the free energy
\be
f_{SC} = - \frac{1}{2} \int\frac{\ud\Omega}{4\pi} \omega_D \sqrt{\omega_D^2 + |\hat{\Delta}_d|^2 \hat{d}(\Omega)^2}.
\ee
This is the same expression as quoted in Eq.~(\ref{eq:free1}) in the main text.
The normal state free energy is obtained by setting $\Delta_d = 0$, so that $f_N = -\omega_D^2/2$ (it contains the Debye frequency because this was our choice of the ultraviolet cutoff in the initial Eq.~(\ref{eq:freeS})). The Cooper pair condensation energy is thus
\be
f_{SC} - f_{N} = - \frac{\omega_D^2}{2} \int\frac{\ud\Omega}{4\pi}  \left[\sqrt{1+\left(\frac{\hat{\Delta}_d^2}{\omega_D^2}\right)\hat{d}_j^2(\Omega)} -1 \right].
\ee
Now we can expand the square root in the powers of the small parameter $\hat{\Delta}_d/\omega_D$, resulting in
\be
f_{SC} - f_{N} = -\frac{\hat{\Delta}_d^2}{4} \int\frac{\ud\Omega}{4\pi} \hat{d}_j^2(\Omega) + \mathcal{O}\left(|\hat{\Delta}_d|^4\right)
\label{eq:free5}
\ee
Because all the cubic harmonics are normalized to form a complete basis [see Eq.~(\ref{eq:identity})], the solid-angle integral $\int\frac{\ud\Omega}{4\pi} \hat{d}_j^2(\Omega) = 1/5$ is the same for each of the five harmonics, yielding Eq.~(\ref{eq:free1a}) in the main text.

We note that although the condensation energy now has a well defined expansion in powers of $|\Delta|^2$, the gap itself is a non-analytic function of the coupling strength $\lambda_d$, as is standard. We emphasize that  Eq.~(\ref{eq:free5}) should \textit{not} be thought of as the Landau free energy; rather, it is the zero-temperature condensation energy expressed in terms of the self-consistent solution $\Delta_d$ of the gap equation, shown in Eq.~(\ref{eq:gapS}).

}


\section{$d+id$ pairing: energetics and competition within $E_g$ channel}~\label{Append:Eg}

In Sec.~\ref{phaselocking:eg}, 
we have established that there are three inequivalent ways of choosing the basis functions for the two-dimensional representation $E_g$, namely:
\bea
\text{A: } d_1(\kk)&\!=\!&\frac{\sqrt{3}}{2}(k_x^2-k_y^2),\; d_2(\kk)=\frac{1}{2}(2k_z^2-k_x^2 -k_y^2); \nonumber\\
\text{B: } d_3(\kk)&\!=\!&\frac{\sqrt{3}}{2}(k_z^2-k_x^2),\; d_4(\kk)=\frac{\sqrt{3}}{2}(k_z^2 -k_y^2); \label{eq:basis} \\
\text{C: } d_5(\kk)&\!=\!&\frac{1}{2}(2k_x^2 - k_y^2-k_z^2),\; d_6(\kk)=\frac{1}{2}(2k_y^2 - k_x^2-k_z^2). \nonumber
\eea 
In this section, we provide the details of the derivation and solution of the  gap equation for these cases.

The zero-temperature gap equation, Eq.~(\ref{eq:gap0}), acquires the following form in the case of $d_m+id_n$ pairing:
\be
\frac{1}{\lambda_d} = \frac{1}{2}\int\limits_{-\omega_D}^{\omega_D} \ud y \sqrt{1+y} \int\frac{\ud \Omega}{4\pi} \frac{|d_m|^2 + |d_n|^2}{\sqrt{y^2 + |\Delta_d|^2(|d_m|^2 + |d_n|^2)}},
\ee
where $y=(k/k_F)^2-1$. The general structure of the form-factors is as follows:
\bea
|d_m(\kk)|^2 &+& |d_n(\kk)|^2 = \left(\frac{k}{k_F}\right)^4[a^2(\theta)+b^2(\theta)\cos^2(2\phi)] \nonumber \\
&=&(1+y)^2 [a^2(\theta)+b^2(\theta)\cos^2(2\phi)].
\eea
In order to make progress analytically, we shall adopt the weak-coupling approximation $\omega_D\ll 1$, leading to $y \ll 1$ and allowing us to simplify $1+y \approx 1$. The integral over $y$ can then be computed analytically, resulting in
\bea
\frac{1}{\lambda_d} &=& \int\frac{\ud \Omega}{4\pi} [a^2(\theta)+b^2(\theta)\cos^2(2\phi)] \times \\
&\times& \ln\left(\frac{\omega_D + \sqrt{\omega_D^2 + |\Delta_d|^2(a^2 + b^2\cos^2(2\phi))}}{\Delta_d\sqrt{a^2 + b^2\cos^2(2\phi)}}\right).\nonumber
\eea
Under the assumption of weak coupling, $|\Delta_d| \ll \omega_D$, the numerator under the logarithm can be approximated by $2\omega_D$, allowing one to make further progress analytically.
%

The integration over $\phi$ can then readily be performed using the following two identities
\bea
I_0(a,b) &\equiv& \frac{1}{2\pi}\int_0^{2\pi}\ud \phi \ln(\sqrt{a^2+b^2\cos^2(2\phi)}) \nonumber \\ 
&=& \ln\left(\frac{a+\sqrt{a^2+b^2}}{2}\right) \nonumber,
\eea
\bea
I_1(a,b) &\equiv& \frac{1}{2\pi}\int\limits_0^{2\pi}\ud \phi \cos^2(2\phi)\ln(\sqrt{a^2+b^2\cos^2(2\phi)}) \nonumber\\
&=&\frac{1}{4} + \frac{1}{2}\ln\left(\frac{a+\sqrt{a^2+b^2}}{2} \right) - a\frac{\sqrt{a^2+b^2}-a}{2b^2}. \nonumber
\eea
Further denoting 
$\langle f \rangle = \frac{1}{2}\int_{-1}^1\ud (\cos \theta) f(\theta)$
to lighten the notation, we can write down the  gap equation as follows:
\be
\frac{1}{\lambda_d} = \left(\av{a^2} + \frac{\av{b^2}}{2} \right)\ln\!\left(\frac{2\omega_D}{\Delta_d}\right) -  \av{a^2 I_0(a,b)} - \av{b^2 I_1(a,b)}\nonumber
\ee
Note that the angle average
\be
\int \frac{\ud \Omega}{4\pi} \left[ \hat{d}_m(\theta,\phi)^2 + \hat{d}_n(\theta,\phi)^2\right]  = \av{a^2} + \frac{\av{b^2}}{2} = \frac{2}{5}\nonumber,
\ee
since each of the 5 $d$-wave harmonics averages to 1/5. This allows us to finally write down the solution for the gap in a closed form:
\be
\Delta_d = 2\omega_D \exp\left[ -\frac{5}{2} (P_0 + P_1)  \right] \exp\left(-\frac{5}{2}\lambda_d\right),
\ee
where $P_0\equiv \av{a^2 I_0(a,b)}$ and $P_1\equiv \av{b^2 I_1(a,b)}$ are simple c-numbers that can be computed explicitly for each of the three choices of the basis, shown in Eq.~(\ref{eq:basis}), yielding
\bea
\text{A: }&& P_0 = -0.1008,\;  P_1 = -0.0199, \\
\text{B: }&& P_0 = -0.0487,\;  P_1 = -0.0326, \\
\text{C: }&& P_0 = -0.0099,\;  P_1 = -0.0181.
\eea
We then arrive at the final expression for the gap amplitudes for each choice of the basis
 \bea
 \Delta_{E_g}^{(A)}(T=0) &=& 2.705\, \omega_d \exp\left(-\frac{5}{2\lambda_d}\right), \\
 \Delta_{E_g}^{(B)}(T=0) &=& 2.451\, \omega_d \exp\left(-\frac{5}{2\lambda_d}\right), \\
 \Delta_{E_g}^{(C)}(T=0) &=& 2.145\, \omega_d \exp\left(-\frac{5}{2\lambda_d}\right).
 \eea
As anticipated, different choices of bases result in different zero-temperature values of the $(d+id)$ gap, with the basis A ($d_{x^2-y^2} + id_{3z^2-r^2}$) having the largest gap value and therefore the lowest energy. But any $d+id$ paired state from same basis produces identical nodal structure in the ordered phase.

We note that the choice of the basis does not however affect the value of the superconducting transition temperature, with all three choices resulting in the same value of $T_c$ given by the solution of the equation similar to Eq.~(\ref{eq:gap-Tc}):
\allowdisplaybreaks[4]
\bea
\frac{1}{\lambda_d} &=& \frac{1}{2} \int_{-\omega_D}^{\omega_D}  \ud y \frac{(1+y)^{\frac{5}{2}}}{y} \tanh\left( \frac{y}{2k_B T_c}\right) \times \nonumber \\
&\times &  \int \frac{\ud \Omega}{4\pi} \left[\hat{d}_m^2(\Omega) + \hat{d}_n^2(\Omega) \right]. 
\eea
Using the fact that $1+y\approx 1$ in the weak-coupling approximation, we obtain
\be
k_BT_c^{(d+id)} = \frac{2e^\gamma}{\pi}\omega_D e^{-\frac{5}{2\lambda}} \approx 1.134\, \omega_D e^{-\frac{5}{\lambda}}, 
\ee
where $\gamma \approx 0.577$ is the Euler's number.
Therefore, despite possessing the same transition temperature, $d_{x^2-y^2} + id_{3z^2-r^2}$ has the lowest energy among three time-reversal broken candidates for the $d+id$ paired states in the $E_g$ channel, and thus always wins at low temperature.


\section{$s+id$ pairing and competition with $d+id$ phase}\label{Append:s+id}

In this section,
 we provide technical details of the derivation for the $s+id$ phase, which we studied in Sec.~\ref{sec:s+id}. Our starting point is the system of coupled gap equations Eqs.(\ref{eq:gap_s})--(\ref{eq:gap_d}), valid in the weak-coupling approximation $\hat{\Delta}_d \ll \omega_D \ll 1$. We stress that the weak coupling approximation is justified here, since we are concerned with solving the coupled gap equations in the vicinity of the second-order phase transition at $r=r_{c1}$ ($r\equiv \lambda_d/\lambda_s$), where $\hat{\Delta}_d$ vanishes.
In the following, we consider the most general case when the Debye frequencies for the $s$-wave and $d$-wave components, $\omega_D^{(s)}$ and $\omega_D^{(d)}$ respectively, are not necessarily the same. This would be the case if, for instance, the origin of the $s$-wave component is due to electron-phonon coupling, whereas the $d$-wave pairing is mediated by electronic interactions.

To avoid the unnecessary complications associated with the integration over the angle $\phi$, we hereby consider the case of $s+id_{3z^2-r^2}$ pairing, since the form-factor only depends on the polar angle $\theta$. 
We then have the following coupled gap equations
\bea
\frac{1}{\lambda_s} &\!\!=\!\!& \ln(2\omega_D^{(s)}) - \frac{1}{2} \int\limits_{-1}^1 \ud(\cos\theta) \ln\big(\sqrt{\hat{\Delta}_s^2 + \hat{\Delta}_d^2 \hat{d}^2(\theta)} \big),~\label{s+id:s-wave} \\
\frac{1}{\lambda_d} &\!\!=\!\!& \ln(2\omega_D^{(d)}) - \frac{1}{2} \int\limits_{-1}^1 \ud(\cos\theta) \hat{d}^2(\theta) \ln\big(\sqrt{\hat{\Delta}_s^2 + \hat{\Delta}_d^2 \hat{d}^2(\theta)} \big)~\label{s+id:d-wave}. \nonumber \\
\eea
The integration cannot be completed in the closed form, however it is possible to obtain an approximate solution in the vicinity of the transition $r_{c1}$ by expanding in the powers of the small parameter $\hat{\Delta}_d/\hat{\Delta}_s$. We then obtain 
\bea
\frac{1}{\lambda_s} &=& \ln\left(\frac{2\omega_D^{(s)}}{\hat{\Delta}_s}\right) 
    - \frac{3}{20}\left(\frac{\hat{\Delta}_d}{\hat{\Delta}_s}\right)^2 + \mathcal{O}\left(\frac{\hat{\Delta}_d}{\hat{\Delta}_s}\right)^4,~\label{s:apmitude} \\ 
    \frac{1}{\lambda_d} &=& \frac{1}{5}\ln\left(\frac{2\omega_D^{(d)}}{\hat{\Delta}_s}\right) 
    - \frac{27}{280}\left(\frac{\hat{\Delta}_d}{\hat{\Delta}_s}\right)^2 + \mathcal{O}\left(\frac{\hat{\Delta}_d}{\hat{\Delta}_s}\right)^4.~\label{d:apmitude}
\eea
From the first equation, we get $\hat{\Delta}_s\approxeq 2\omega_D^{(s)} \exp(-1/\lambda_s)$, since the  vanishing $d$-wave component does not alter the pure $s$-wave solution at $r_{c1}$ to the leading order.
Substituting $\hat{\Delta}_s$ into the second equation, we find 
\be~\label{s-d:apmplitude}
\left( \frac{\hat{\Delta}_d}{\hat{\Delta}_s} \right)^2 = \frac{56}{27}\left[ 
 \frac{1}{\lambda_s} - \ln\left( \frac{\omega_D^{(s)}}{\omega_D^{(d)}} \right) - \frac{5}{\lambda_d} 
\right],
\ee
from which it follows that  for $\hat{\Delta}_d$ to have a non-trivial solution, $\lambda_d$ must have a lower bound:
\be
\frac{\lambda_d}{\lambda_s} > r_{c1} = \frac{5}{1 - \lambda_s\ln\left( \frac{\omega_D^{(s)}}{\omega_D^{(d)}} \right)}, \label{eq:lower-bound}
\ee
and the results in Sec.~\ref{sec:s+id} are quoted for $\omega_D^{(s)}=\omega_D^{(d)}$. In particular, if the two Debye frequencies are the same, we have $r_{c1}=5$, as verified by direct numerical calculation [see Fig.~\ref{Fig--s+id}(a) and Sec.~\ref{sec:s+id}]. Eq.~(\ref{eq:lower-bound}) also tells us that the ratio of the Debye frequencies cannot be chosen arbitrarily and must satisfy $\omega_D^{(s)}/\omega_D^{(d)} < e^{\frac{1}{\lambda_s}}$
in order for the $(s+id)$ solution to exist. Put alternatively, it requires that the $s$-wave coupling constant is not too large
\be
 \lambda_s<1/\ln\left( \omega_D^{(s)}/\omega_D^{(d)} \right), \label{eq:bound-omegaD}
\ee
otherwise the pure $s$-wave will dominate and the $d$-wave component will never have a chance to develop.

As noted in Sec.~\ref{sec:s+id}, the mere existence of the $s+id$ solution does not guarantee that it will be realized in nature, unless it is lower in energy than the competing $d+id$ phase.
We must therefore compare the free energy of the $s+id$ solution (which is essentially a pure $s$-wave in the vicinity of $r_{c1}$) to that of the $d+id$ phase from Eq.~(\ref{eq:free-d+id}), leading to
\bea
f_s - f_{N} &=& -\frac{\big(\omega_D^{(s)}\big)^2}{2}  \int\frac{\ud \Omega}{4\pi} 
   \left[
   \sqrt{1+\left(\frac{\hat{\Delta}_s}{\omega_D^{(s)}}\right)^2} - 1
   \right] \nonumber\\  
   &=& -\frac{\hat{\Delta}_s^2}{4} + \mathcal{O}\left(\hat{\Delta}_s^4\right),  \\
f_{d+id} - f_{N} &=& -\frac{\hat{\Delta}_{d+id}^2}{10} + \mathcal{O}\left(\hat{\Delta}_{d+id}^4\right).
\eea
Substituting the zero-temperature gap values $\hat{\Delta}_s = 2\omega_D^{(s)}\exp(-1/\lambda_s)$ and  \mbox{$\hat{\Delta}_{d+id} = 2.705\,\omega_D^{(d)}\exp(-2.5/\lambda_d)$} from Eq.~(\ref{eq:gap-d+id}) into the free energies, we find
\be
f_{s} - f_{d+id} = - \big(\omega_D^{(s)}\big)^2 e^{-\frac{2}{\lambda_s}} + 0.734 \big(\omega_D^{(d)}\big)^2 e^{-\frac{5}{\lambda_d}}.
\ee
For a non-trivial $s+id$ wave solution to exist, $\lambda_d$ must exceed the minimal value given in Eq.~(\ref{eq:lower-bound}). Substituting it into the above expression for the free energy, we find
\be
f_{s} - f_{d+id} > \big(\omega_D^{(s)}\big)^2 e^{-\frac{2}{\lambda_s}} \left[
    0.734\, e^{\frac{1}{\lambda_s}} \left(\frac{\omega_D^{(d)}}{\omega_D^{(s)}}\right)  - 1
\right].
\ee 
Therefore, for $s$-wave to be more stable than $d+id$ at $r_{c1}$, the necessary condition is 
$\exp(1/\lambda_s) < 1.362\, \omega_D^{(s)}/\omega_D^{(d)}$, or equivalently
\be
\frac{1}{\lambda_s} < 0.309 + \ln\left( \frac{\omega_D^{(s)}}{\omega_D^{(d)}}   \right).
\label{eq:bound-s}
\ee 
Combining this expression with Eq.~(\ref{eq:bound-omegaD}), we see that $\lambda_s^{-1}$ must belong to a rather narrow interval, given by
\be
\ln\left( \frac{\omega_D^{(s)}}{\omega_D^{(d)}} \right) < \frac{1}{\lambda_s} < 0.309 + \ln\left( \frac{\omega_D^{(s)}}{\omega_D^{(d)}} \right),
\ee
for the $(s+id)$ phase to have a chance to exist. Substituting Eq.~(\ref{eq:bound-s}) into the inequality Eq.~(\ref{eq:lower-bound}), we conclude that for $s+id$ to be  energetically stable, the following lower bound on $\lambda_d$ is necessary
\be
\lambda_d > \frac{5}{0.309} \approx 16.2.\label{eq:lower-bound2}
\ee
Note that this bound is universal, independent of the ratio of the Debye frequencies. We reiterate that the above derivation is rigorous since the weak-coupling approximation is always justified near the $r=r_{c1}$ transition (since $\hat{\Delta}_d$ vanishes at that point).

As remarked in Sec.~\ref{sec:s+id}, the condition in Eq.~(\ref{eq:lower-bound2}) is extremely unlikely to be realized in nature, and if true, it would certainly lie outside the realm of weak-coupling approximation at the apogee of $s+id$ phase (it would imply that $\Delta_{s+id}^{max} > \omega_D^{(d)}$). Therefore for all practical applications, we can safely conclude that $(s+id)$ order is always energetically inferior to its rival $(d+id)$ phase. This conclusion is corroborated by the direct numerical evaluation of the free energies, an example of which is shown in Fig.~\ref{Fig--s+id}(b). Invariably, we find a direct first-order phase transition from a pure $s$-wave into the $d+id$ phase as the ratio of the coupling constants $r=\lambda_d/\lambda_s$ is increased.

We can shed more light on the reason why the $s+id$ solution is less energetically stable by considering the gap equations (\ref{eq:gap_s})--(\ref{eq:gap_d})  in the weak-coupling approximation $\hat{\Delta}\ll\omega_D \ll 1$. 
The gap equations [see Eqs.~(\ref{eq:gap_s}) and (\ref{eq:gap_d})] can further be simplified in the vicinity of the second-order phase transition at $r=r_{c1}$, where the magnitude of  $\hat{\Delta}_d$ can be made arbitrarily small. 
In this limit we can approximate $\hat{\Delta}_s\approxeq 2\omega_D \exp(-1/\lambda_s)$ and the $d$-wave pairing amplitude can then be obtained from [follows from Eqs.~(\ref{s+id:s-wave})--(\ref{s-d:apmplitude})]
\be
\left(\frac{\hat{\Delta}_d}{\hat{\Delta}_s}\right)^2 \approxeq \frac{280}{27}\left[\frac{1}{5\lambda_s} - \frac{1}{\lambda_d}\right].\label{eq:d-component}
\ee
The non-trivial solution for $\hat{\Delta}_d$ is only possible provided $\lambda_d/\lambda_s > 5$, explaining the value $r_{c1} \approx 5$ obtained numerically, see Fig.~\ref{Fig--s+id}(a).
We now must compare the free energy of this solution (which is essentially a pure $s$-wave in the vicinity of $r_{c1}$) to that of $d+id$ from Eq.~(\ref{eq:free-d+id}). 
In order for $s+id$ phase to have lower energy than the $d+id$, in the vicinity of $r_{c1}$, we require that $\lambda^{-1}_s < -\ln\left(C_{d+id}^2/10\right) \approx  0.309$, where $C_{d+id}=2.705$, see Eq.~(\ref{eq:free-d+id}).
Substituting this into Eq.~(\ref{eq:d-component}), we see that for a non-trivial $s+id$ solution to exist, one must ensure the lower bound on the pairing strength $\lambda_d > 5\lambda_s \gtrsim 16.2$. Such a huge value of $\lambda_d$ is unphysical, and moreover, the assumption of the weak coupling would break down in this case. 
Even allowing for different Debye frequencies for the $s$- and $d$-wave components -- justified if the origin of $s$-wave pairing is not the same, for instance due to conventional electron-phonon mechanism -- does not alter the outcome, with the lower bound on $\lambda_d \gtrsim 16.2$ remaining roughly same as before. 
This conclusion is corroborated by numerics, where we certainly have not been able to find an energetically viable $s+id$ solution for any $\lambda_d \lesssim 8$.


\begin{figure}
\subfigure[]{
\includegraphics[width=0.2\textwidth]{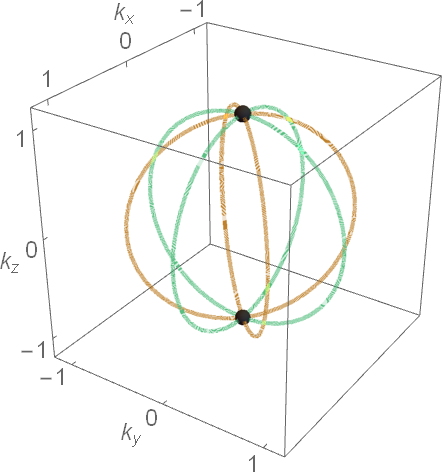}
\label{d3+id4}
}
\subfigure[]{
\includegraphics[width=0.2\textwidth]{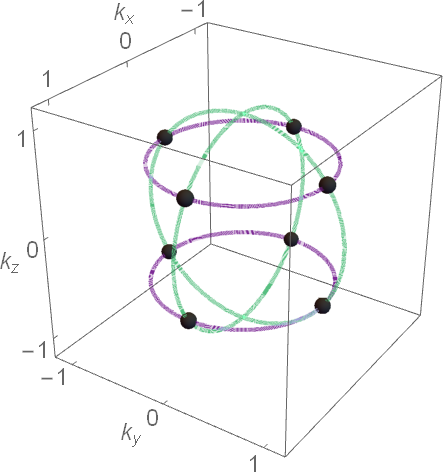}
\label{d3+id5}
}
\\
\subfigure[]{
\includegraphics[width=0.2\textwidth]{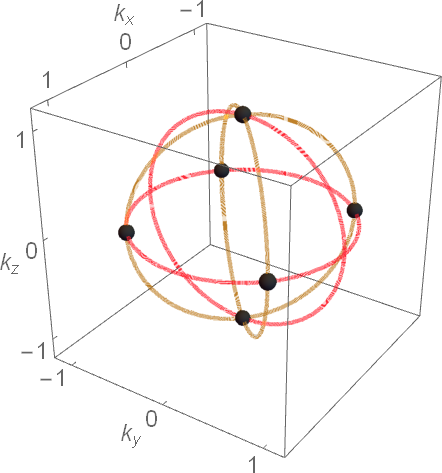}
\label{d2+id4}
}
\subfigure[]{
\includegraphics[width=0.2\textwidth]{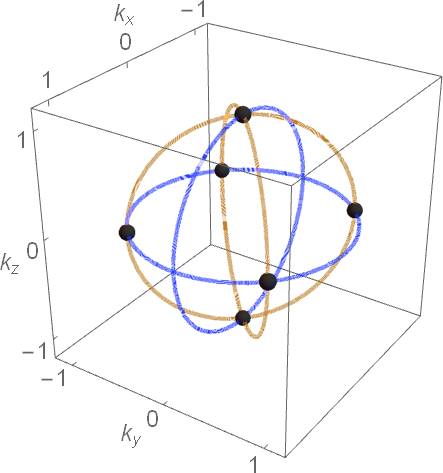}
\label{d1+id4}
}
\\
\subfigure[]{
\includegraphics[width=0.2\textwidth]{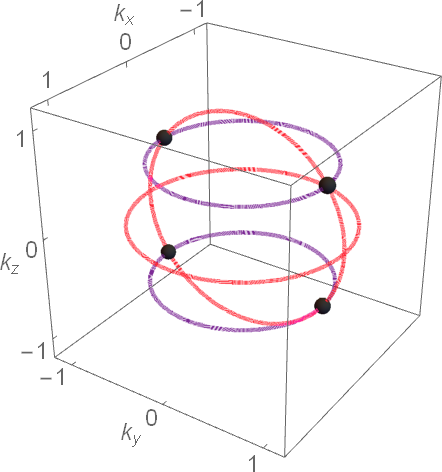}
\label{d2+id5}
}
\subfigure[]{
\includegraphics[width=0.2\textwidth]{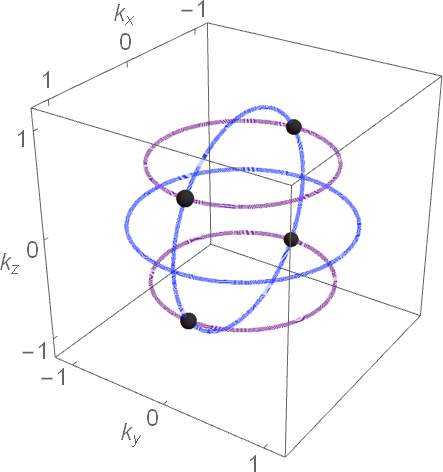}
\label{d1+id5}
}
\caption{Weyl superconductors that obtain via $d+id$ combinations of $T_{2g}$ and $E_g$ local pairings.
Each figure shows the double- or single-Weyl nodes (block dots) that arise from the nodal-loop intersections. 
The pairings are 
(a) $d_{xy}+i d_{x^2-y^2}$,
(b) $d_{xy}+i d_{3 z^2-r^2}$,
(c) $d_{xz}+i d_{x^2-y^2}$,
(d) $d_{yz}+i d_{x^2-y^2}$,
(e) $d_{xz}+i d_{3z^2-r^2}$,
and
(f) $d_{yz}+i d_{3z^2-r^2}$.
Properties of these states are enumerated in Table~\ref{egt2g:competitiontable}. 
Nodal loops shown in 
green, 
red, 
blue, 
brown, 
and 
purple respectively corresponds to the ones 
associated with the 
$d_{xy}$, 
$d_{xz}$, 
$d_{yz}$, 
$d_{x^2-y^2}$, 
and 
$d_{3z^2-r^2}$ pairings, respectively. 
Equations for these loops appear in the rightmost column of Table~\ref{table_projection}. 
Notice that 
$d_{xz/yz}+i d_{x^2-y^2}$ also supports a pair of gapless points at the north and south pole of the Fermi surface. 
These nodes do not possess any topological invariant and thus their existence is purely accidental. 
Here all momentum axes are measured in units of the Fermi momenta $k_F$.} 
\label{Fig--egt2g}
\end{figure}

\section{Nodal topology of $d+id$ pairing: competing $T_{2g}$ and $E_g$ channels}~\label{Append:t2geg}

In Sec.~\ref{EgT2g:Competition} we have highlighted six possible time-reversal symmetry breaking paired states when the pairing interaction in the $T_{2g}$ and $E_g$ channels are of comparable strength. Nodal topology of these paired states were announced in Table~\ref{egt2g:competitiontable}. In Sec.~\ref{hall:thermal-spin} we have discussed the nodal topology of $d_{x^2-y^2}+id_{xy}$ paired state in details. We here present the computation of nodal topology of other five paired states, shown in Table~\ref{egt2g:competitiontable}.

$\bullet$ $d_{xy}+id_{3 z^2-r^2}$ pairing: The location of the nodal points for this paired state are given in the second row of Table~\ref{egt2g:competitiontable}. To extract the nodal topology of these isolated points on the Fermi surface, we first focus on the Weyl node located at $k_x=0, k_y= k_F \sqrt{2/3}, k_z= k_F \sqrt{1/3}$, and conveniently define a set of new momentum variables 
\begin{equation}
p_z=\frac{k_z}{\sqrt{3}}+\frac{\sqrt{2}k_y}{\sqrt{3}},\; p_y=\frac{\sqrt{2}k_z}{\sqrt{3}} -\frac{k_y}{\sqrt{3}},\; p_x=k_x.
\end{equation}
The Weyl nodes are now located at ${\mathbf p}=(0,0,k_F)$. Expanding the kinetic energy around the Weyl node we obtain the following reduced BCS Hamiltonian
\begin{equation}
	\hat{h}_{d_{xy}+id_{3z^2-r^2}}= \tau_3 v_z \delta p_z + v_x  \tau_1 p_x + v_y \tau_2 p_y,
\end{equation} 
where $\delta p_z=p_z-k_F$, $v_x=\sqrt{2}|\Delta_{T_{2g}}|/k_F$, $v_y=|\Delta_{E_g}|/(\sqrt{2} k_F)$, and $v_z=k_F/m$. The above nodal point, as well as the one located at ${\bf k}=-\left( 0, \sqrt{2/3}, \sqrt{1/3} \right) k_F$ are characterized by monopole charge $W_n=+1$ [follows from Eq.~(\ref{monopolecharge})]. By contrast, $W_n=-1$ for the Weyl nodes at ${\bf k}=\pm \left( 0, \sqrt{2/3},- \sqrt{1/3} \right)k_F$. We denote the above four Weyl nodes as ``(a)". For the Weyl nodes located at ${\bf k}=\pm \left( k_F \sqrt{2/3},0, k_F \sqrt{1/3} \right)$ the monopole charge is $W_n=+1$, and Weyl nodes located at ${\bf k}= \pm \left( \sqrt{2/3},0,- \sqrt{1/3} \right)k_F$ have $W_n=-1$. The last four Weyl nodes are denoted as ``(b)".

$\bullet$ $d_{xz}+id_{x^2-y^2}$ pairing: Let us first focus on the Weyl nodes located at ${\bf k}=\pm(1, 1,0)k_F/\sqrt{2}$ [see Table~\ref{egt2g:competitiontable}], and conveniently rotate the momentum axis according to 
\begin{equation}
p_x=\frac{k_x+k_y}{\sqrt{2}}, p_y=\frac{k_x-k_y}{\sqrt{2}}, p_z=k_z. 
\end{equation}
Weyl nodes are now located at ${\bf p}=\pm (k_F,0,0)$. Expanding the kinetic energy around the Weyl nodes we obtain the following reduced BCS Hamiltonian 
\begin{equation}
	\hat{h}_{d_{xz}+id_{x^2-y^2}}= \pm v_x \delta p_x \tau_3 + v_y \tau_1 p_z + v_z \tau_2 p_y,
\end{equation}
where $\delta p_x=p_x \pm k_F$, $v_y=\sqrt{3}| \Delta_{E_{g}}|/ k_F$, $v_z=\sqrt{3}|\Delta_{T_{2g}}|/\left(\sqrt{2} k_F\right)$ and $v_x=k_F/m$. Therefore, two Weyl nodes located at ${\bf k}=\pm(1, 1,0)k_F/\sqrt{2}$ have the monopole charge $W_n=+1$. By contrast, the Weyl nodes located at ${\bf k}=\pm(-1, 1,0)k_F/\sqrt{2}$ have monopole charge $W_n=-1$.

$\bullet$ $d_{yz}+id_{x^2-y^2}$ pairing: Analysis of the nodal topology for this paired state is identical to the previous one. We find that Weyl nodes at ${\bf k}=\pm (1,1,0)k_F/\sqrt{2}$ have monopole charge $W_n=-1$ and Weyl nodes at ${\bf k}=\pm(1,-1,0)k_F/\sqrt{2}$ have monopole charge $W_n=-1$.

$\bullet$ $d_{xz}+id_{3 z^2-r^2}$ pairing: Following the analysis of nodal topology for $d_{xy}+id_{3 z^2-r^2}$ pairing, denoted as ``(a)", we find that Weyl nodes at ${\bf k}=\pm\left( 0,  \sqrt{2/3}, \sqrt{1/3} \right)k_F$ have monopole charge $W_n=+1$, while $W_n=-1$ for the ones located at ${\bf k}=\pm \left( 0,- \sqrt{2/3}, \sqrt{1/3} \right)k_F$.

$\bullet$ $d_{yz}+id_{3 z^2-r^2}$ pairing: From the analysis of nodal topology for $d_{xy}+id_{3 z^2-r^2}$ pairing, denoted as ``(b)",  we find that Weyl nodes at ${\bf k}=\pm \left( \sqrt{2/3}, 0, \sqrt{1/3} \right)k_F$ have monopole charge $W_n=+1$, while $W_n=-1$ for the ones located at ${\bf k}=\pm \left(- \sqrt{2/3}, 0, \sqrt{1/3} \right)k_F$.


\section{Coupling between $s$-wave and $d$-wave pairing with electronic nematicity in doped LSM}\label{strain:LSMfull}

\begin{figure}[t!]
\includegraphics[width=7cm,height=4.0cm]{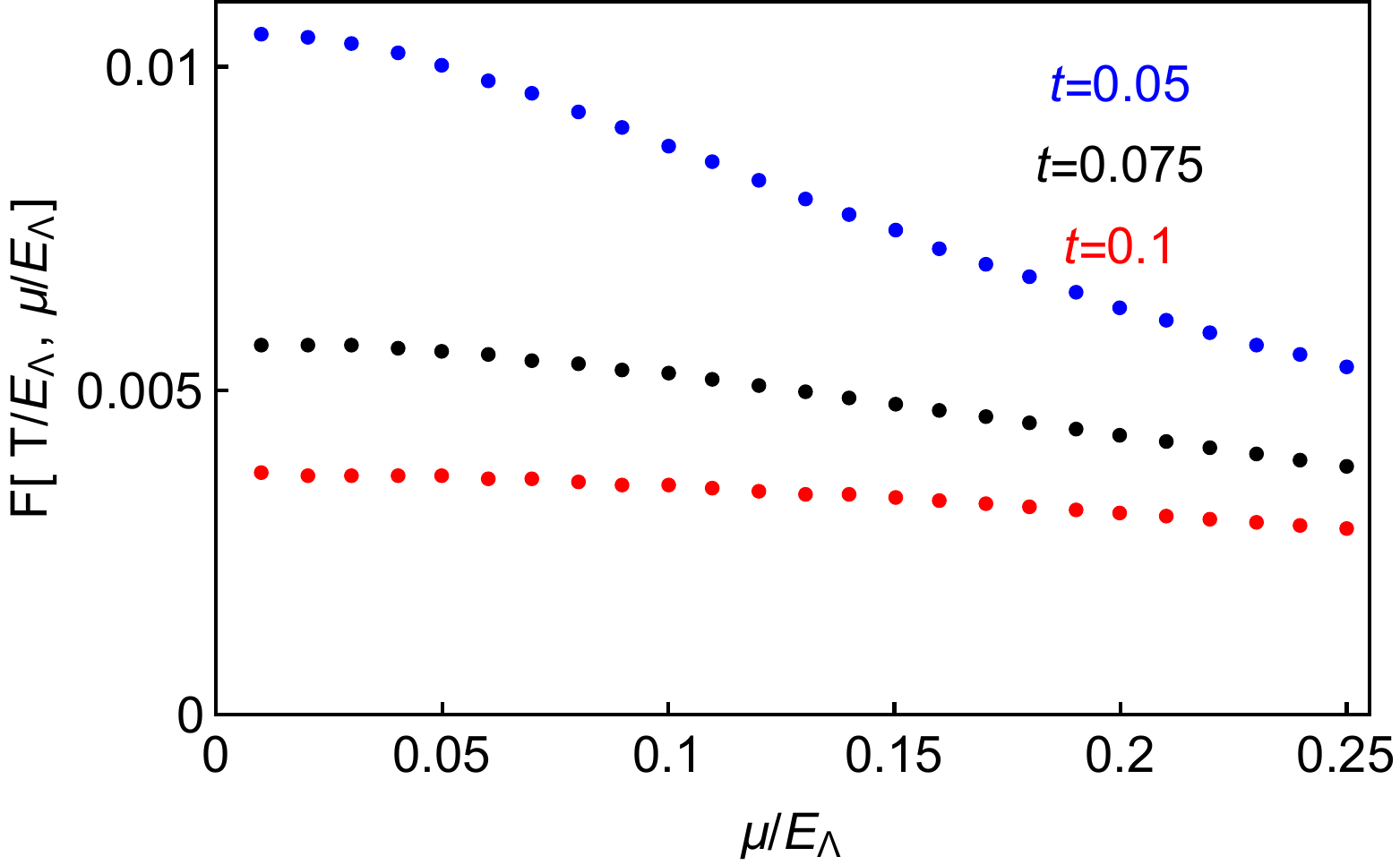}
\caption{ Scaling of the function $F(x,y)$, defined in Eq.~(\ref{function_strain-Full}), with $x$ for various choices of $y$. Here, $x=\mu/E_\Lambda$ and $y=T/E_\Lambda$ are respectively dimensionless chemical potential and temperature, with $E_\Lambda=\Lambda^2/(2m)$, and $t \equiv y$. The function $F$ determines the strength of the coupling between the $s$-wave and $d$-wave pairing with lattice distortion or electronic nematicity, determined through $f^{Lutt}_{str}$ shown in Eq.~(\ref{function_str-fullband}). }~\label{strain_fullband}
\end{figure}


 In Sec.~IV of the main paper, we established a non-trivial coupling between the $s$-wave and $d$-wave pairings with electronic nematicity, which can also be induced by applying a weak external strain. In Sec.~IV we presented the calculation upon projecting all these ingredients onto the Fermi surface, assuming that the external strain is sufficiently weak and that pairing predominantly takes place in the close proximity of the Fermi surface. However, such non-trivial coupling does not depend on this approximation, and we here demonstrate that it is non-trivial even if we take into account the entire Luttinger band of spin-3/2 fermions, with the chemical potential ($\mu$) placed away from the band touching point. The general form of external strain in the Luttinger model has already been displayed in Eq.~(4.1). The contribution from the Feynman diagram shown in Fig.~7(a) now takes the form
\begin{eqnarray}
&& F^{Lutt}_{str} = -\frac{1}{8} \Phi_j \Delta^\mu_l \Delta^\nu_0 \frac{1}{\beta} \sum^{\infty}_{n=-\infty} \int \frac{d^3 {\bf k}}{(2 \pi)^3}
{\bf Tr} \bigg[ \left( \tau_\mu \Gamma_l \right) \nonumber \\
&\times& G(i \omega_n, \mu, {\bf k}) \left( \tau_\nu \Gamma_0 \right) G(i \omega_n, \mu, {\bf k}) \left( \tau_3 \Gamma_j \right) G(i \omega_n, \mu, {\bf k}) \bigg], \nonumber \\
\end{eqnarray}
where 
\begin{eqnarray}
G(i \omega_n, \mu, {\bf k})= \frac{i \omega_n + \mu + \hat{H}_L}{(i \omega_n+\mu)^2 -E^2} \oplus 
\frac{i \omega_n - \mu + \hat{H}_L}{(i \omega_n-\mu)^2 -E^2}, \nonumber \\
\end{eqnarray}     
is the Greens function of spin-3/2 fermions in the Luttinger model, with $E=k^2/(2m)$ and 
\begin{equation}
\hat{H}_L= \frac{k^2}{2m} \sum^{5}_{j=1} \Gamma_j \hat{d}_j. 
\end{equation}
First of all note that for non-vanishing ${\bf Tr}$ we require $j=l$ and $\mu=\nu$, implying that (1) the component of the $d$-wave pairing and of the external strain \emph{must} break the cubic symmetry in an identical fashion (since $j=l$), and (2) external strain supports a time-reversal invariant superposition of $s$-wave and $d$-wave pairing (since $\mu=\nu$). These conclusions are identical to the one we found by computing the triangle diagram [see Fig.7(a)] in the close proximity to the Fermi surface after the band projection. In what follows we compute the above expression from the entire Luttinger band for spin-3/2 fermions. 

After some straightforward algebra and completing the integral over the solid angle we arrive at the following expression (setting $\hbar=k_B=1$) 
\begin{widetext}
\begin{eqnarray}
F^{Lutt}_{str} = \frac{\mu T}{10\pi^2} \sum^{\infty}_{n=-\infty} \int^{\Lambda}_0 k^2 dk  
 \frac{\left[ E^4+ \left[ (11-10\pi^2) \omega^2_n - (1+10\pi^2)\mu^2 \right] E^2  + 10 \pi^2 (\omega^2_n +\mu^2)^2 \right]}{\left[ E^4 + 2E^2 (\omega^2_n -\mu^2) + (\omega^2_n +\mu^2)^2 \right]^2},
\end{eqnarray} 
\end{widetext} 
where $\Lambda$ is the ultraviolet momentum cut-off up to which the spectra of conduction and valence band scale quadratically with momentum. The above contribution to the free energy can now be written compactly as 
\begin{eqnarray}
F^{Lutt}_{str}= \frac{\mu m^3}{\Lambda^3} F\left( \frac{\mu}{E_\Lambda}, \frac{T}{E_\Lambda} \right),
\end{eqnarray} 
where $E_\Lambda=\Lambda^2/(2m)$ is the ultraviolet energy cut-off and the two parameters inside the function $F$ are therefore dimensionless. Also note that in natural units ($\hbar=k_B=1$), mass ($m$) is a dimensionless parameter and $\Lambda$ bears the dimension of inverse length. Hence, we can suitably define a free-energy density $f^{Lutt}_{str}$ associated with $F^{Lutt}_{str}$ according to $f^{Lutt}_{str}=F^{Lutt}_{str} \Lambda^3$, where 
\begin{equation}~\label{function_str-fullband}
f^{Lutt}_{str}=\mu m^3 F\left( \frac{\mu}{E_\Lambda}, \frac{T}{E_\Lambda} \right),
\end{equation}  
and
\begin{widetext}
\begin{eqnarray}~\label{function_strain-Full}
F(x,y)= y \sum^{\infty}_{n=-\infty} \int^1_0 \frac{z^2 dz}{10\pi^2} 
\frac{\left[ z^8+ \left[(11-10\pi^2) \Omega^2_n -(1+10\pi^2) x^2 \right]z^4  + 10 \pi^2 (\Omega^2_n +x^2)^2 \right]}{\left[ z^8 + 2z^4 (\Omega^2_n -x^2) + (\Omega^2_n +x^2)^2 \right]^2}, 
\end{eqnarray}
\end{widetext}
with $z=k/\Lambda$, $x=\mu/E_\Lambda$, $y=T/E_\Lambda$, and $\Omega_n=\omega_n/E_\Lambda$. Here, $\omega_n=(2n+1)\pi T$ is the fermionic Matsubara frequency. The scaling of the function $F(x,y)$ with $x$ for various choices of $y$ are shown in Fig.~\ref{strain_fullband}.

Therefore, the non-trivial coupling between the $s$-wave and $d$-wave pairing with lattice distortion or electronic nematicity in a Luttinger semimetal solely arises from the spin-3/2 nature of the quasiparticles. Note that such coupling exists only in the presence of a Fermi surface, as $f^{Lutt}_{str} \to 0$ when $\mu \to 0$. 

Finally we show that the induced $s$-wave component ($\Delta_s$) is finite due to the lattice distortion ($\Phi$) in the presence of a dominant $d$-wave pairing ($\Delta_d$). To this end we write down the phenomenological Landau potential to the quartic order, given by 
\begin{eqnarray}
f_{\rm quar.} &=& r_1 \Delta^2_d + u_1 \Delta^4_d + r_2 \Delta^2_s + u_2 \Delta^4_s + u_{12} \Delta^2_s \Delta^2_d \nonumber \\
              &+& a \; \Phi \; \Delta_s \; \Delta_d,
\end{eqnarray}
where $r_1, r_2, u_1, u_2, u_{12}, a$ are phenomenological (and nonuniversal) parameters. Earlier we showed the scaling of the coefficient of cubic term, namely $a$, with various band parameters. All the terms appearing in the first line of $f_{\rm quar.}$ are the standard one, while the cubic coupling is very specific to spin-3/2 fermions in cubic environment. In what follows, we work in the regime where $r_1<0$, but $r_2>0$. Hence, the $d$-wave pairing is spontaneously developed in the system, while the $s$-wave component can only be induced. In this regime $\Delta_s, \Phi \ll \Delta_d$ and we can obtain analytic solutions for $\Delta_d$ and $\Delta_s$ by minimizing $f_{\rm quar.}$, yielding   
\begin{equation}
\Delta_d=\pm \sqrt{\frac{2 u_1}{r_1}}\equiv \pm \Delta^0_d, \quad 
\Delta_s= - \frac{a \Phi \Delta^0_d}{2 \left[ r_2 + U_{12} \left( \Delta^0_d \right)^2 \right]}.
\end{equation}
Therefore, cubic coupling between the $s$-wave and $d$-wave pairings with lattice distortion or electronic nematicity is responsible for non-trivial solution of induced $s$-wave pairing. By contrast, the standard quartic coupling between two pairing channels, proportional to $u_{12}$ reduces the strength of the the induced $s$-wave pairing. 

This analysis can immediately be generalized when $r_1>0$, but $r_2<0$ so that the $s$-wave pairing is spontaneously generated in the system. Under that circumstance presence of an external strain ($\Phi$) can induce a $d$-wave pairing as we argued the main text.


\section{Nodal topology of $s+d$-wave pairings}~\label{Append:sdNodes}

In this appendix, we briefly review the nodal topology of various $d$-wave and $d+id$-type pairings in the presence of an accompanying $s$-wave pairing, which can be induced by lattice deformation (due to dominant $d$-wave pairings, see Sec.~IV). We show that when the $s$-wave component is sufficiently small it only shifts the location of (a) nodal-loops of an individual $d$-wave pairing or (b) point nodes arising from $d+id$-type pairings. 

1. $s+d_{x^2-y^2}$ pairing: The effective single-particle Hamiltonian for $s+d_{x^2-y^2}$ pairing reads as 
\begin{eqnarray}
H= \tau_3 \left[ \frac{{\bf k}^2}{2 m} -\mu \right] + \tau_1 \left[ \Delta_s - \frac{\sqrt{3} \Delta_d}{2k^2_F} \left( k^2_x-k^2_y \right) \right], 
\end{eqnarray} 
where $\Delta_s$ and $\Delta_d$ are respectively the amplitude of $s$-wave and $d_{x^2-y^2}$ pairing, and conveniently we chose the relative phase between these two pairings to be $\pi$ (hence preserves time-reversal symmetry). The spectra of the above single-particle Hamiltonian are given by 
\begin{eqnarray}
E=\pm \left\{ \left[\frac{{\bf k}^2}{2 m} -\mu \right]^2 + \left[ \Delta_s - \frac{\sqrt{3}\Delta_d}{2k^2_F} \left( k^2_x-k^2_y \right) \right]^2 \right\}^{1/2}. \nonumber \\
\end{eqnarray}  
The equations for two nodal-loops are then given by 
\begin{eqnarray}
2 k^2_x+k^2_z=k^2_F, \quad  
k_y=\pm \left[ k^2_x -k^2_F \frac{2 \Delta_s}{\sqrt{3} \Delta_d}\right]^{\frac{1}{2}}. 
\end{eqnarray}
Therefore, $s+d_{x^2-y^2}$ pairing continues to support two nodal-loop as long as $\Delta_s/\Delta_d<\sqrt{3}/4 \approx 0.43$, similar to the situation in pure $d_{x^2-y^2}$ pairing. Following the same strategy one can show that $s+d_{xy}$, $s+d_{xz}$, $s+d_{yz}$ and $s+d_{3 z^2-r^2}$ pairings continue to support two nodal-loops as long as the accompanying $s$-wave component is sufficiently small.

2. $s+d_{3z^2-r^2}+i d_{x^2-y^2}$ pairing: Recall the pure $d_{3z^2-r^2}+id_{x^2-y^2}$ pairing supports eight Weyl nodes at $\pm k_x=\pm k_y = \pm k_z =k_F/\sqrt{3}$. With the addition of the $s$-wave pairing eight Weyl nodes get shifted to 
\begin{eqnarray}
\pm k_x &=& \pm k_y= \frac{k_F}{\sqrt{3}} \left[ 1+ \frac{\Delta_s}{\Delta_d} \right]^{\frac{1}{2}}, \nonumber \\
k_z &=& \pm \frac{k_F}{\sqrt{3}} \left[1-2 \frac{\Delta_s}{\Delta_d} \right]^{\frac{1}{2}},
\end{eqnarray}
The reason why $s+d_{3z^2-r^2}+i d_{x^2-y^2}$ pairing continues to support eight Weyl nodes is the following (see also main text). Note that individually $s+d_{3 z^2-r^2}$ and $s+d_{x^2-y^2}$ pairing supports two nodal-loops for weak enough $s$-wave component. Therefore, eight Weyl nodes in the $s+d_{3z^2-r^2}+i d_{x^2-y^2}$ paired state are located at the intersection points of four nodal-loops. Following the same approach we find that for a specific phase locking, namely $\left(\phi_{xy}, \phi_{xz}, \phi_{yz} \right)=\left(0,2\pi/3, 4 \pi/3 \right)$, among three $d$-wave pairings belonging to the $T_{2g}$ representation supports eight Weyl nodes when accompanied by a small $s$-wave component (only shifted from the ones reported in Sec.~IIIc).


\section{Inversion symmetry breaking and two gap structure}~\label{Append:Inversionassymetry}

In this appendix we present the detailed analysis of the quasiparticle spectra 
for the $s+d+id$ type paired state, but in the absence of the \emph{inversion symmetry}. 
We showed in the previous appendix that addition of a small $s$-wave pairing to either 
$d$-wave or $d+id$-type pairing does not change the nodal topology, only shifts the location 
of the nodal-loops or point nodes. In this appendix, we will show that if we add an 
inversion asymmetric term to the kinetic energy 
(1) Kramers degeneracy of the pseudo-spin degenerate Fermi surface is lost and we end up with 
two Fermi surfaces, and then 
(2) in the presence of an $s$-wave component to the dominant $d+id$-type pairing, 
it is conceivable to keep only one of the Fermi surfaces gapless, while the other one 
becomes fully gapped. By contrast, in the absence of $s$-wave pairing, a pure 
$d+id$ pairing continues to support eight Weyl nodes on both Fermi surfaces. We substantiate 
this statement by focusing on the $d_{x^2-y^2}+id_{3z^2-r^2}$ pairing. Our analysis can be 
generalized to other pairings (such as the one with specific phase locking 
$\left(\phi_{xy}, \phi_{xz}, \phi_{yz} \right)=\left(0,2\pi/3, 4 \pi/3 \right)$ 
within the $T_{2g}$ sector), discussed in Sec.~IIIc.

The non-interacting Hamiltonian in the absence of the inversion symmetry (with its simplest realization) is given by 
\begin{eqnarray}
	H^{\rm Inv.}_0 = \left( \frac{k^2}{2 m} -\mu \right) \tau_3 + \tau_3 v \left({\boldsymbol \sigma} \cdot {\bf k}\right),
\end{eqnarray}
where $v$ bears the dimension of Fermi velocity and measures the strength of inversion symmetry breaking. 
Three Pauli matrices $\left\{ \sigma_\mu \right\}$ for $\mu=1,2,3$ operate on the pseudo-spin index. 
The spectra of above Hamiltonian is $  k^2/(2m)-\mu + \tau v |{\bf k}|$ for $\tau=\pm 1$, 
confirming the lack of Kramers degeneracy (due to inversion asymmetry).

Now in the presence of $d_{x^2-y^2}+id_{3z^2-r^2}$ pairing the quasiparticle spectra of BdG fermions 
are given by $\pm E_{\tau, \bf k}$, where 
\begin{eqnarray}
E_{\tau,\bf k}= \bigg[ \left( \frac{k^2}{2m}-\mu \right)^2 + v^2 k^2 + 2 \tau v k \left( \frac{k^2}{2m}-\mu \right)  \nonumber \\
+ \frac{\Delta^2}{k^4_F} \left\{ \frac{3}{4}\left( k^2_x-k^2_y \right)^2 + \frac{1}{4} \left( 2 k^2_z- k^2_x-k^2_y \right)^2 \right\} 
\bigg]^{1/2},
\end{eqnarray}
for $\tau=\pm$. Hence, the Weyl nodes are now located at  
\begin{equation}
\pm k_x=\pm k_y =\pm k_z = k^\tau_0,
\end{equation}
where for $\tau=\pm$
\begin{equation}
k^\tau_0 = \sqrt{2 m \mu + m^2 v^2} + \tau m v. 
\end{equation}
Both Fermi surfaces are gapless (irrespective of the strength of 
inversion symmetry breaking) and each of them accommodates eight Weyl nodes. 
Note that in the above expression $k_F \neq \sqrt{2 m \mu}$ due to the presence of 
two Fermi surfaces. Here $k_F$ should be considered as a large 
momentum scale such that $k_x,k_y,k_z \ll k_F$.

With the addition of the $s$-wave component the energy spectra inside the 
$s+d_{x^2-y^2}+id_{3z^2-r^2}$ pairing in the absence of inversion symmetry breaking are given by 
\begin{eqnarray}
E^\tau_{\bf k} &=& \bigg[ \left\{ \left( \frac{k^2}{2m}-\mu \right) + \tau v k \right\}^2  + \frac{\Delta^2_d}{4 k^4_F} \left(2 k^2_z-k^2_x-k^2_y \right)^2 \nonumber \\
&+& \left( \Delta_s - \frac{\sqrt{3}\Delta_d}{2 k^2_F} \left( k^2_x-k^2_y\right)\right)^2
\bigg]^{1/2}.
\end{eqnarray}
The locations of the Weyl nodes on two Fermi surfaces (denoted by index $\tau=\pm$) 
are given by $\left(\pm k^\tau_x, \pm k^\tau_y ,\pm k^\tau_z \right)$, where
\begin{eqnarray}
k^\tau_x &=& \frac{1}{\sqrt{3}} \left[ \left( \sqrt{2 m \mu + m^2 v^2} + \tau m v \right)^2 + \sqrt{3} k^2_F \frac{\Delta_s}{\Delta_d} \right]^{1/2}, \nonumber \\ 
k^\tau_y &=& \frac{1}{\sqrt{3}} \left[ \left( \sqrt{2 m \mu + m^2 v^2} + \tau m v \right)^2 - \sqrt{3} k^2_F \frac{\Delta_s}{\Delta_d} \right]^{1/2}, \nonumber \\ 
k^\tau_z &=& \frac{1}{\sqrt{3}} \left[ \sqrt{2 m \mu + m^2 v^2} + \tau m v \right].
\end{eqnarray} 
Note that real solutions for $k^\tau_x$ and $k^\tau_z$ exist for any strength of 
inversion symmetry breaking and $s$-wave component. However, both $k^{\pm}_y$ are real-valued only when $v<v_0$, where 
\begin{equation}
v_0 =\frac{1}{2 m} \left[ \sqrt{3} k^2_F \frac{\Delta_s}{\Delta_d} + \frac{4}{\sqrt{3}} \frac{m^2 \mu^2}{k^2_F} \frac{\Delta_d}{\Delta_s} - 4 m \mu \right]^{1/2}.
\end{equation}    
On the other hand, for $v>v_0$ only one Fermi surface (the bigger one, with $\tau=+1$) hosts eight 
Weyl nodes, while the other one (the smaller one, with $\tau=-1$) becomes fully gapped.

Therefore, depending on the strength of the inversion symmetry breaking it is conceivable 
to realize a situation when the $s+d_{x^2-y^2}+i d_{3z^2-r^2}$ pairing gives rise to one 
nodal and one fully gapped Fermi surfaces. This situation stands as a possible microscopic 
origin for the proposed two gap structure in the penetration depth [see Eq.~(6.1) of main text]. 
It must be noted that the accompanying $s$-wave component in the presence of dominant $d+id$ 
type pairing is not induced by the lack of the inversion symmetry, as both of them are \emph{even} 
under the spatial inversion. The $s$-wave component is induced by lattice distortion mediated 
by the dominant $d$-wave pairings, discussed in Sec.~IV of the main paper. Nonetheless, once the 
$s$-wave component is established in the system by lattice distortion (due to dominant $d$-wave 
pairing) it can receive further assistance from electron-phonon interaction which is always 
finite in real system.

We note that the actual inversion symmetry breaking term in YPtBi for example is more complex 
than the one we discussed in this appendix~\cite{brydon}. The sole purpose of the present analysis 
is to demonstrate that in the absence of inversion symmetry, $s+d+id$ pairing can give rise 
to two gap structure: \emph{one of the Fermi surfaces remains gapless, supporting Weyl nodes, while the other one becomes fully gapped}. 
The last observation provides a microscopic justification for the two-gap structure we subscribe in Sec.~VI 
to compare our theoretical predictions with the experimental observation. However, due to lack of microscopic details 
in a correlated system, such as actual strength of inversion asymmetry, 
and its renormalization due to electronic interactions and disorder, 
strength of electron-phonon coupling and elastic constants (required to estimate the actual strength of induced $s$-wave component), 
further theoretical justification of our proposed scenario in a specific material such as YPtBi is 
very difficult (a common limitation in any correlated system). Thus, we have to rely on complimentary 
experiments (apart from the penetration depth measurement), discussed in depth in Sec.~VIA, 
to test the validity of the our proposed scenario in superconducting half-Heusler compounds.


\end{document}